\documentclass[a4paper,11pt]{article}
\usepackage{amsmath,amsfonts,amssymb,mathrsfs,theorem,calc,graphicx,color,enumerate,tabularx,bbm,mathtools,rotating}
\usepackage{titling}
\usepackage{geometry}
\usepackage[round,longnamesfirst]{natbib}
\definecolor{dblue}{rgb}{0,0,0.3}
\usepackage[hypertexnames=false]{hyperref} 
\hypersetup{
  colorlinks=true,
  linkcolor=dblue,
  anchorcolor=dblue,
  citecolor=dblue,
  filecolor=dblue,
  menucolor=dblue,
  urlcolor=dblue,
  pdftitle={iMaxEntBw},
  pdfauthor={Vitaliy Oryshchenko},
  pdfcreator={\LaTeX\ with package \flqq hyperref\frqq}, 
  final=true,
  breaklinks=true
  }

\providecommand{\Vol}[2][]{\mathop{\mathrm{Vol}}\nolimits_{#1}\left(#2\right)}
\providecommand{\ConvexHull}[1]{\mathop{\mathrm{conv}}\left(#1\right)}
\providecommand{\ones}[1]{\boldsymbol{\iota}_{#1}}                       
\providecommand{\I}[1]{\mathbbm{1}\left\{#1\right\}}                     
\newcommand{\transp}{^{\mathsf{T}}}                                      
\newcommand{\ceil}[1]{\left\lceil#1\right\rceil}                         
\newcommand{\norm}[1]{\lVert#1\rVert}                                    
\newcommand{\abs}[1]{\left\vert#1\right\vert}                            
\newcommand{\littleO}[2][]{\mathit{o}_{#1}\left(#2\right)}               
\newcommand{\bigO}[2][]{\mathit{O}_{#1}\left(#2\right)}                  
\newcommand{\R}{\mathbbm{R}}
\DeclareMathOperator{\E}{E}
\DeclareMathOperator{\Var}{Var}
\DeclareMathOperator{\MISE}{MISE}

\DeclareMathOperator{\IVar}{IVar}
\DeclareMathOperator{\corr}{corr}
\DeclareMathOperator*{\argmin}{argmin}
\DeclareMathOperator*{\argmax}{argmax}
\DeclareMathOperator{\CV}{CV}

\theoremstyle{plain} \newtheorem{lemma}{Lemma}
\theoremstyle{plain} \newtheorem{conjecture}{Conjecture}
\theoremstyle{plain} {\theorembodyfont{\upshape} \newtheorem{assumption}{Assumption}}

\begin{document}
\title{Indirect Maximum Entropy Bandwidth}
\author{Vitaliy Oryshchenko\thanks{Address for correspondence: 2.068 Arthur Lewis Building, Economics DA, School of Social Sciences, University of Manchester, Oxford Road, Manchester M13 9PL, United Kingdom. E-mail: \href{mailto:vitaliy.oryshchenko@manchester.ac.uk}{vitaliy.oryshchenko@manchester.ac.uk}.} \\
University of Manchester }
\date{\today}
\maketitle
\begin{abstract}
This paper proposes a new method of bandwidth selection in kernel estimation of density and distribution functions motivated by the connection between maximisation of the entropy of probability integral transforms and maximum likelihood in classical  parametric models. 
The proposed estimators are designed to indirectly maximise the entropy of the leave-one-out kernel estimates of a distribution function, which are the analogues of the parametric probability integral transforms. 

The estimators based on minimisation of the Cram\'{e}r-von Mises discrepancy, near-solution of the moment-based estimating equations, and inversion of the Neyman smooth test statistic are discussed and their performance compared in a simulation study. The bandwidth minimising the Anderson-Darling statistic is found to perform reliably for a variety of distribution shapes and can be recommended in practice. 

The results will also be of interest to anyone analysing the cross-validation bandwidths based on leave-one-out estimates or evaluation of nonparametric density forecasts. 


\end{abstract}
\section{Introduction}
Let $X_{1},\ldots,X_{n}$ be a sample of independent, identically distributed random variables with an absolutely continuous distribution function (d.f.) $F$ and density $f$. The kernel estimators of $F$ (KDFE) and $f$ (KDE) at a point $x$ are
\begin{equation}\label{Eq:KDFE}
\widehat{F}(x;b)=n^{-1}\sum_{i=1}^{n}K_{b}(x-X_{i}), \quad b\geq0,
\end{equation}
\citep{nadaraya1964b,watson1964} and 
\begin{equation}\label{Eq:KDE}
\hat{f}(x;b) = n^{-1}\sum_{i=1}^{n}k_{b}(x-X_{i}), \quad b>0,
\end{equation}
\citep{rosenblatt1956,parzen1962} respectively, 
where $k_{b}(z)=k(z/b)/b$ and $K_{b}(z)=K(z/b)=\int_{-\infty}^{z/b}k(v)dv$ are the kernels, with $k$ being symmetric about the origin and integrating to unity, and $b=b_{n}$ is the bandwidth sequence. The empirical distribution function (EDF) can be obtained as a special case of \eqref{Eq:KDFE} with $b=0$, viz. $F_{n}(x)=\widehat{F}(x;0)=n^{-1}\sum_{i=1}^{n}\I{X_{i}\leq x}$. 
It is well known that under mild conditions $\widehat{F}$ and $\hat{f}$ are uniformly strongly consistent and asymptotically normal estimators of $F$ and $f$, respectively \citep{yamato1973,silverman1978}.

By far the most common measure of fit in kernel smoothing is the mean integrated squared error (MISE). 
Other less commonly applied criteria include the $L_{1}$ norm  \citep{devroye1985}, the Kullback-Leibler (KL) divergence \citep{hall1987}, and the Hellinger distance \citep{kanazawa1993}. One consequence of using the curve-fitting framework to evaluate the quality of the estimates \eqref{Eq:KDFE} and \eqref{Eq:KDE} is that in general the bandwidths optimal for KDE and KDFE differ (intuitively, the d.f. is easier to estimate than the density). For example, the bandwidths optimal in the MISE sense for $\widehat{F}$ and $\hat{f}$ are asymptotically of order $n^{-1/5}$ and $n^{-1/3}$, respectively \citep[see e.g.][]{wand1995}. Thus the ability to recover an estimate of a d.f. from that of a density, and vice versa, using the identity $\widehat{F}(x;b) = \int_{-\infty}^{x}\hat{f}(z;b)dz$ is lost. 

This paper posits an alternative framework for bandwidth selection in kernel estimation of density and distribution functions. 
To motivate the proposed methods consider the classical parametric setup where $Z_{i}$, $i=1,\ldots,n$, are independent random variables with common d.f. $G$ and density $g$, and the model is given by the family of d.f.'s $\mathscr{F}=\{F_{\theta}(z):\theta\in\Theta\subseteq\R^{p}\}$ with densities $f_{\theta}$. 
The model $\mathscr{F}$ may or may not contain the true distribution $G$. 
The probability integral transform (PIT) of $Z_{i}$ \emph{under the model} $F_{\theta}$ is defined as 
$V_{i}(\theta) = F_{\theta}(Z_{i})$.
The vector of PITs, $V(\theta)=\left(V_{1}(\theta),\ldots,V_{n}(\theta)\right)\transp$, is distributed on $I^{n} = [0,1]^{n}$,
the $n$-dimensional unit hypercube placed in the positive orthant of $\R^{n}$ with one vertex at the origin;  
the components of $V(\theta)$ are independent and have a common d.f., $Q_{\theta}(u)$, $u\in[0,1]$, say, with density $q_{\theta}$. 

It is well known that under appropriate regularity conditions the following equalities hold: 
\begin{equation}
\label{Eq:MLE.MaxEnt.Equivalence}
   \argmin_{\theta\in\Theta}\; D\left(g\Vert f_{\theta}\right) 
 = \argmax_{\theta\in\Theta}\; \E_{G}\left[\ln f_{\theta}(Z)\right]
 = \argmax_{\theta\in\Theta}\; h\left(U(\theta)\right)
 = \argmin_{\theta\in\Theta}\; D\left(q_{\theta}\Vert r\right),
\end{equation}
where $r$ denotes the uniform density on $[0,1]$, $D(g\Vert f) = \int g(z)\ln\left(\left.g(z)\right/f(z)\right)dz$ is the relative entropy of $g$ with respect to $f$ (the KL between $g$ and $f$), and $h(Z) = -\int f(z)\ln f(z)dz$ is the Shannon differential entropy of a random vector $Z$ with density $f$, hereinafter simply the entropy. 
The first two equalities underpin the quasi- (or pseudo-) maximum likelihood \citep{akaike1973,white1982} and the maximum spacings estimators \citep{shao1994}. 
The third equality in \eqref{Eq:MLE.MaxEnt.Equivalence} can be used to motivate testing the goodness of fit or evaluation of density forecasts by checking the uniformity of PITs \citep{gneiting2007,diebold1998}, although such tests are usually motivated by the fact that PITs are uniform if $F_{\theta}=G$ \citep{neyman1937,rosenblatt1952}, thus bypassing the maximum entropy interpretation. 

It is thus natural to consider an alternative class of estimators for $\theta$ defined as minimisers of KL or other distance or  discrepancy measure between $Q_{\theta}$ ($q_{\theta}$) and the uniform d.f. (density).
In view of \eqref{Eq:MLE.MaxEnt.Equivalence} such estimators can be called the {\it indirect maximum entropy} (iMaxEnt) estimators.
The idea of fine-tuning the model to achieve uniformity of PITs (combined with independence of $V_{i}$'s in a time series context) has been put forward in e.g. \citet{dawid1984}, but to the best of my knowledge no such estimators have been formally studied before. 
Clearly, if $G=F_{\theta_{0}}\in\mathscr{F}$, then in all cases in \eqref{Eq:MLE.MaxEnt.Equivalence} the optimum is achieved with $f_{\theta}=g$ and $V_{i}(\theta)$ independent, uniformly distributed on a unit interval.
Under the appropriate regularity conditions the iMaxEnt estimators will then also be consistent for $\theta_{0}$. 
When $G\notin\mathscr{F}$ the pseudo-true values of iMaxEnt estimators will generally differ from the maximiser of the entropy of $V(\theta)$. However, if the model provides a close approximation to the true distribution, one can expect all such pseudo-true values to be similar. 

While the iMaxEnt estimators for parametric models are not further pursued here, the ideas carry over to kernel estimation of density and distribution functions. Specifically, for a fixed kernel the estimators \eqref{Eq:KDFE} and \eqref{Eq:KDE} can be viewed as a model indexed by the bandwidth parameter, and the natural analogue of the PIT under this model is the leave-one-out kernel estimate of a d.f. at a sample value $x_{i}$. 
Then a bandwidth which results in the  joint distribution of the leave-one-out estimates being close to uniform over their support can be seen as the bandwidth indirectly maximising the Shannon entropy of PITs. 

The properties of the leave-one-out estimates are examined in Section \ref{Sec:main.results}, and several estimation procedures for the indirect maximum entropy bandwidth are proposed. Performance of selected estimators is evaluated in a small simulation study reported in Section \ref{Sec:Monte.Carlo}. 
Section \ref{Sec:Conclusions} concludes.
Proofs and other technical results are given in the appendices and the supplement. 

\vspace*{0.5\baselineskip}

\noindent\textbf{Notation}.\; 
For a nonzero vector $w\in\R^{n}$ and a real number $t$, $H_{w,t}$ denotes the hyperplane $H_{w,t}=\{x\in\R^{n}:\left\langle w,x\right\rangle=t\}$. 
For a point $x = (x_{1},\ldots,x_{n})$ in $\R^{n}$  the permutohedron $P_{n}(x)$ is the convex hull of all vectors obtained from $x$ by permutations of the coordinates, 
$P_{n}(x) = \ConvexHull{(x_{\sigma(1)},\ldots,x_{\sigma(n)})\vert\sigma\in S_{n}}$,
where $S_{n}$ is the symmetric group. 
$P_{n}(n-1,n-2,\ldots,0)$ is called the  regular permutohedron.
The dimension of $P_{n}(x)$ is at most $n-1$, and if not all the coordinates of $x$ are equal, then $\dim P_{n}(x)=n-1$. 
$\Vol[n-1]{P_{n}}$ denotes the usual $(n-1)$-dimensional volume of the projection of $P_{n}$ onto $x_{n}=0$.

\section{Main results} \label{Sec:main.results}
\subsection{Assumptions}
\begin{assumption}\label{Ass:DGP} 
(a) $X_{1},\ldots,X_{n}$ are independent, identically distributed random variables from an absolutely continuous distribution $F$ with a bounded density function $f$; \\
(b) $f$ possesses a fourth derivative which is continuous and square integrable. 
\end{assumption}
\begin{assumption}\label{Ass:kernel}
(a) The kernel, $k:\R\mapsto[0,\infty)$ is a bounded function, $k(-x)=k(x)$, $\int_{-\infty}^{\infty} k(x)dx=1$, and $\mu_{2}(k)=\int_{-\infty}^{\infty} x^{2}k(x)dx<\infty$;\\
(b) $\mu_{4}(k)=\int_{-\infty}^{\infty} x^{4}k(x)dx<\infty$, and $\psi_{2,1}(K)=2\int_{-\infty}^{\infty}xK(x)k(x)dx<\infty$. 
\end{assumption}
\begin{assumption}\label{Ass:bandwidth}
The bandwidth $b>0$ is a non-stochastic sequence such that as $n\to\infty$, $b\to0$ and $bn\to\infty$.
\end{assumption}

Assumptions \ref{Ass:DGP}(a), \ref{Ass:kernel}(a), and \ref{Ass:bandwidth} are standard in the smoothing literature. The requirement that $bn\to\infty$ in Assumption \eqref{Ass:bandwidth} is needed for consistency of the KDE \eqref{Eq:KDE}. 
Assumption \ref{Ass:kernel}(a) is the definition of a second order kernel. 
Kernels of orders higher than two are not considered here because they necessarily take negative values and thus the resultant density estimates are not necessarily non-negative and the d.f. estimates not necessarily non-decreasing. This would both undermine the connection with parametric models and cause some of the subsequent results to break down.

Assumptions \ref{Ass:DGP}(b) and \ref{Ass:kernel}(b) are used for the asymptotic expansions. 

\subsection{The leave-one-out estimates}
Considering the KDFE \eqref{Eq:KDFE} (and KDE \eqref{Eq:KDE}) as a family of models indexed by a parameter $b>0$ for a given $k$, the natural analogue of the probability integral transform is the leave-one-out estimate\footnote{\label{Fn:KCDFE.PIT.Ordering}$V_{i}(b)$ can also be written as $V_{i}(b) = \frac{n}{n-1}\widehat{F}(X_{i};b)-K_{b}(0)/(n-1)$. If $k$ is a probability density function symmetric around zero, then $K_{b}(0)=K(0)=1/2$. Furthermore, $K(x)$, and hence $\widehat{F}(x;h)$ are non-decreasing in $x$, and therefore $V_{j}\geq V_{i}$ whenever $X_{j}\geq X_{i}$.} of $F$ at $x_{i}$, $i=1,\ldots,n$, viz. 
\begin{equation}\label{Eq:KCDFE.PIT.def}
V_{i}(b) = \widehat{F}_{(-i)}(X_{i};b)=\frac{1}{n-1}\sum_{\substack{j=1 \\ j\neq i}}^{n}K_{b}(X_{i}-X_{j}), \qquad n\geq2.
\end{equation}
Equivalently, $V_{i}(b)=\int_{-\infty}^{x_{i}}\hat{f}_{(-i)}(z;b)dz$, where $\hat{f}_{(-i)}(z;b)$ is the KDE \eqref{Eq:KDE} constructed using all but the $i^{th}$ observation. 
Under Assumptions \ref{Ass:DGP}(a) and \ref{Ass:kernel}(a), the components of $V(b)=\left(V_{1}(b),\ldots,V_{n}(b)\right)\transp$ are identically distributed on the $[0,1]$ interval with a marginal CDF $G(v) = G(v;n,b)$, say, but are dependent by construction, viz. for $i\neq j$, $\corr(V_{i},V_{j})=-1/(n-1)$ by symmetry. 

\begin{lemma}
\label{Lemma:KCDFE.PIT.Support}
If Assumptions \ref{Ass:DGP}(a) and \ref{Ass:kernel}(a) hold, then the support of the leave-one-out estimates $V(b)$, $b\geq0$, is the the regular permutohedron scaled by $(n-1)^{-1}$, viz.
\begin{equation}
\Pi_{n} = P_{n}\left(1,(n-2)/(n-1),(n-3)/(n-1),\ldots,1/(n-1),0\right). \tag*{$\square$}
\end{equation}
\end{lemma}
$\Pi_{n}$ is an $(n-1)$-dimensional polytope that lies in the central section of a unit hypercube, $\Pi_{n}\subset \left(I^{n}\cap H_{\ones{n},n/2}\right)$, and its barycentre is $c = \ones{n}/2$, where $\ones{n}$ is an $n$-vector of ones. 
In particular, $\sum_{i=1}^{n}V_{i}=n/2$ and the joint distribution of $V(b)$ is singular. 
$\E V_{j}=1/2$ and all the higher moments of $V_{j}$  approach the respective moments of the uniform distribution on $[0,1]$ as $n\to\infty$ and $b\to0$ at the rate $\max(b^{2},n^{-1})$. 

\begin{lemma}\label{Lemma:KCDFE.PIT.Moments}
If Assumptions \ref{Ass:DGP}(a,b), \ref{Ass:kernel}(a,b), and \ref{Ass:bandwidth} hold, then as $n\to\infty$, for $r\geq2$, $j=1,\ldots,n$,
\begin{equation*}
\E V_{j}^{r} = \frac{1}{r+1} - \frac{1}{2}\mu_{2}(k)\xi_{2,r}(F)b^{2}   +\frac{r-1}{2(r+1)}n^{-1}  -\psi_{2,1}(K)\xi_{1,r}(F)bn^{-1} 
  + \bigO{\max(b^{2}n^{-1},b^{4},n^{-2})},
\end{equation*}
where for $r\geq2$, $\xi_{2,r}(F) = \frac{r(r-1)}{2}\int_{-\infty}^{\infty}F^{r-2}(x)f^{3}(x)dx$, $\xi_{1,r}(F) = \int_{-\infty}^{\infty}\frac{r(r-1)}{2}F^{r-2}(x)f^{2}(x)dx$,  and $0<\xi_{r}(F)\leq\frac{r}{2}\sup_{x}f^{2}(x)$, $0<\xi_{1,r}(F)\leq \frac{r}{2}\sup_{x}f(x)$. \hfill$\square$
\end{lemma}

\subsection{The indirect maximum entropy bandwidth}\label{SSec:iMaxEnt.BW.Estim}
Following the ideas outlined in the introduction it is natural to seek a bandwidth which maximises the entropy of $V(b)$.
By the Maximum Entropy Principle, the (possibly infeasible) MaxEnt bandwidth, $b^{\circ}$, would be such that $V(b^{\circ})$ is uniformly distributed over its support, $\Pi_{n}$. Since the components of $V(b)$ are identically distributed, the marginal distribution of $V_{i}(b)$, $G(v)$, and the density, $g(v)$, are estimable. The MaxEnt bandwidth can therefore be estimated \emph{indirectly} by minimising an appropriate distance or discrepancy measure between $G$ (or $g$) and the marginal d.f. (density) of the components of a random vector distributed uniformly over $\Pi_{n}$. 

\begin{lemma}\label{Lemma:Unif.Pn.Marg}
Suppose an $n$-vector $(u_{1},\ldots,u_{n})\transp$ has a (singular) uniform distribution on a permutohedron $\Pi_{n}$, then the marginal distributions of $u_{i}$, $i=1,\ldots,n$, are identical with probability density function given by
\begin{equation}
\label{Eq:Marginal.Finite.n}
l_{n}(u) = \frac{n-1}{n^{n-2}}\Vol[n-2]{P_{n-1}\left(n-1,n-2,\ldots,j+1,\;2j-1-(n-1)u,\;j-2,\ldots,0\right)}, 
\end{equation}
for $u\in\left[\frac{j-1}{n-1},\frac{j}{n-1}\right]$, $j=1,\ldots,n-1$. 
The marginal density  at $0$ and $1$ is $l_{n}(0) = l_{n}(1) = \left(1-1/n\right)^{n-2}$. 
\hfill$\square$
\end{lemma}
The density \eqref{Eq:Marginal.Finite.n}, the corresponding d.f. $L_{n}(u)=\int_{0}^{u}l_{n}(t)dt$, $u\in[0,1]$, moments or other quantities of interest can be computed exactly or approximated as described in Appendix  \ref{Sec:Permutohedron}. 
Figure \ref{Fig:PnMarg} shows the exact density $l_{n}(u)$ for $n=2,\ldots,11$ (panel A) and 
the approximated density for $n=100,\ldots,1000000$ (panel B).
The values of the first five even central moments of a random variable with density $l_{n}(u)$ for $n$ between $2$ and $1,000,000$ are tabulated in the Supplement.

\begin{figure}[htbp]\centering
\begin{tabular}{cc}
{\small (A) Exact density \eqref{Eq:Marginal.Finite.n} for $n=2,\ldots,11$ }&{\small (B) Estimated density for $n=100,\ldots,1,000,000$}\\
\includegraphics[width=0.49\linewidth]{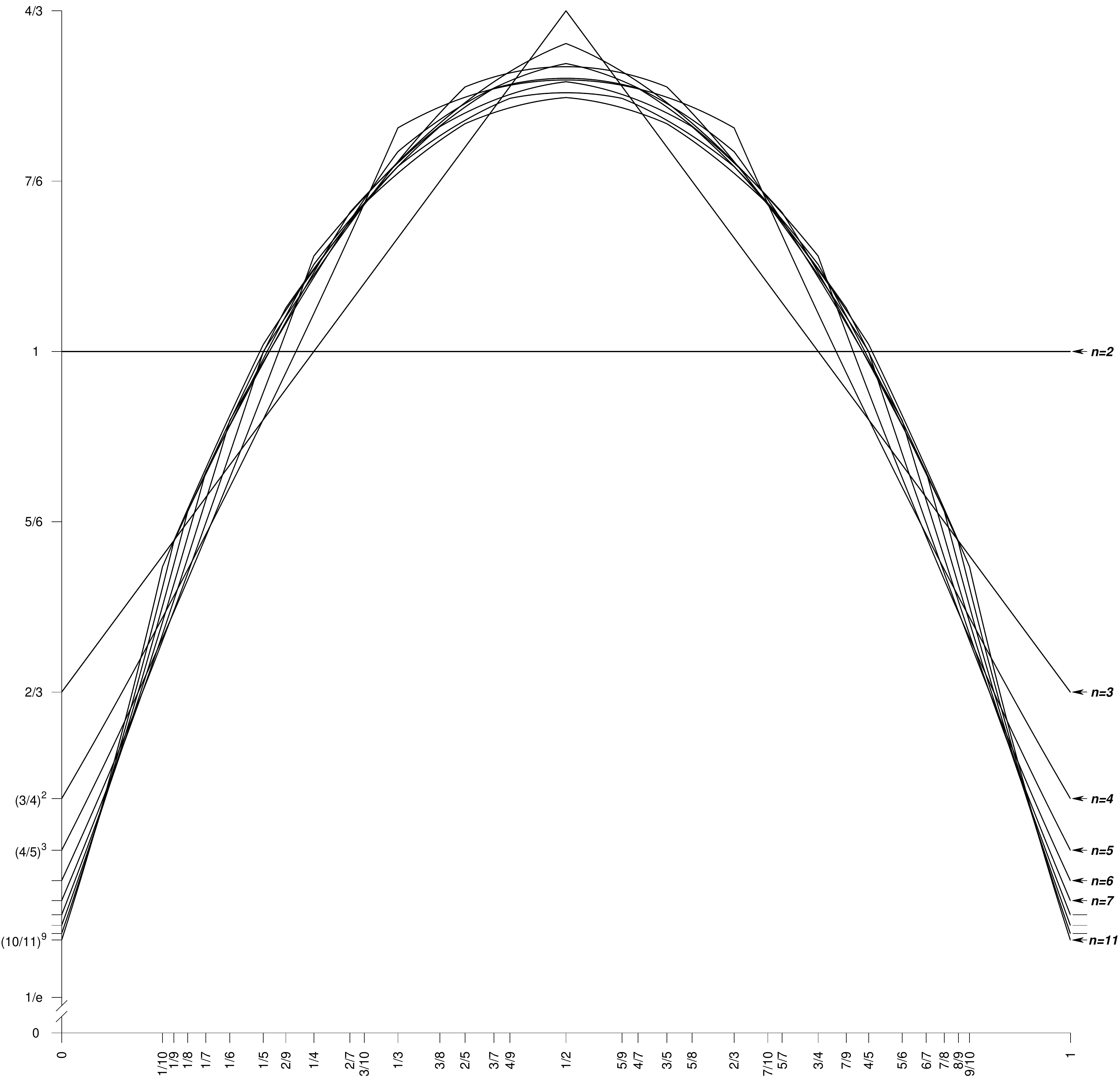} & 
\includegraphics[width=0.49\linewidth]{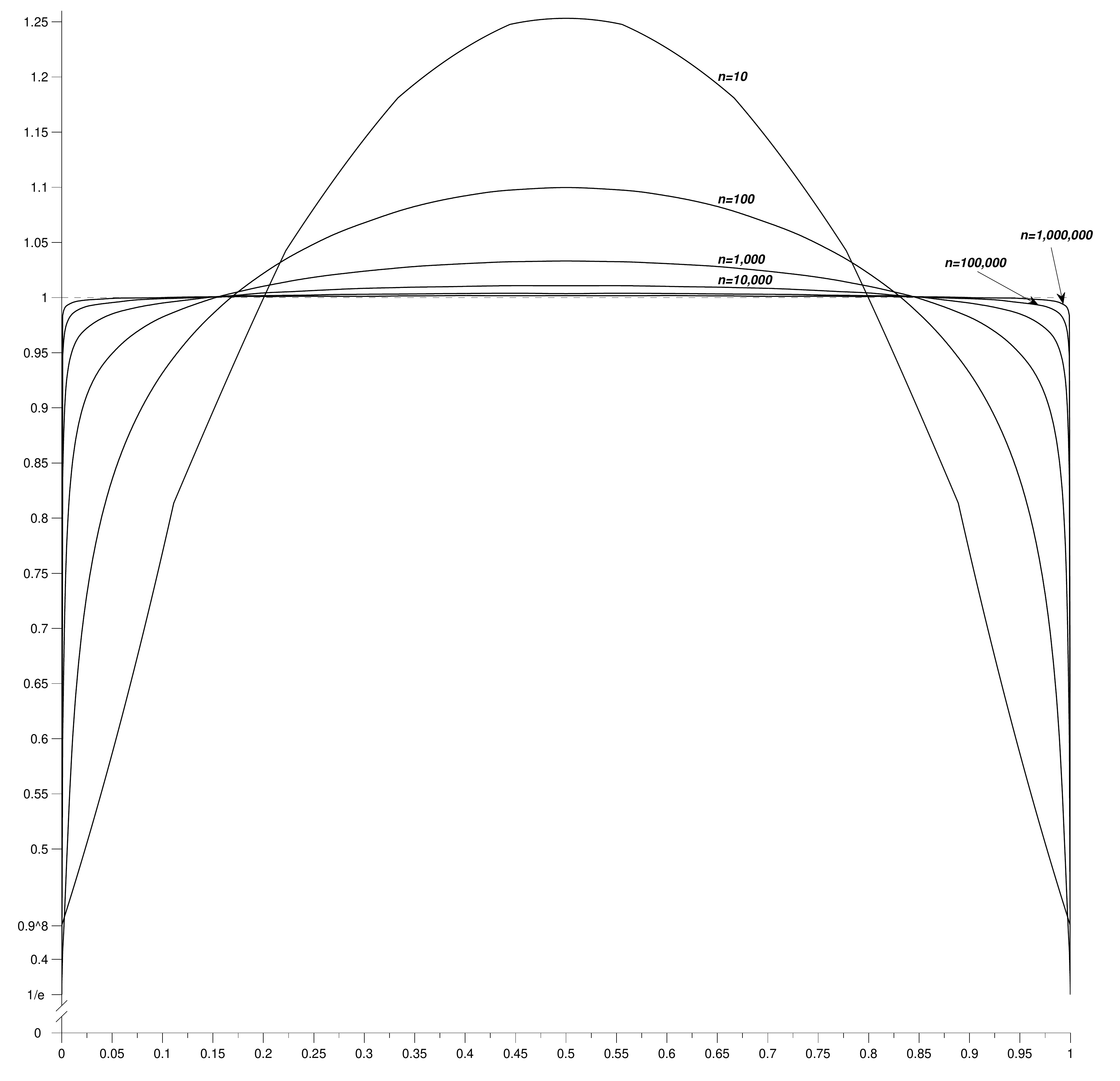} 
\end{tabular}
\begin{flushleft}
{\footnotesize Legend: $u$ on the horizontal axes, density on the vertical axes.\newline
Panel (B): Beta kernel density estimates \citep{chen1999} on an equispaced grid of $1000$ points from $5e8/n$ random uniform draws on $\Pi_{n}$ and bandwidth $0.001$. Estimates are symmetrised, adjusted at the endpoints, and rescaled to integrate to unity.}
\vspace*{-\baselineskip}
\end{flushleft}
\caption{The marginal density $l_{n}(u)$}
\label{Fig:PnMarg}
\end{figure}

\begin{figure}[htbp]\centering
\begin{tabular}{ccc}
\multicolumn{3}{c}{\small $5,000$ random draws}\\
\includegraphics[width=0.3\linewidth]{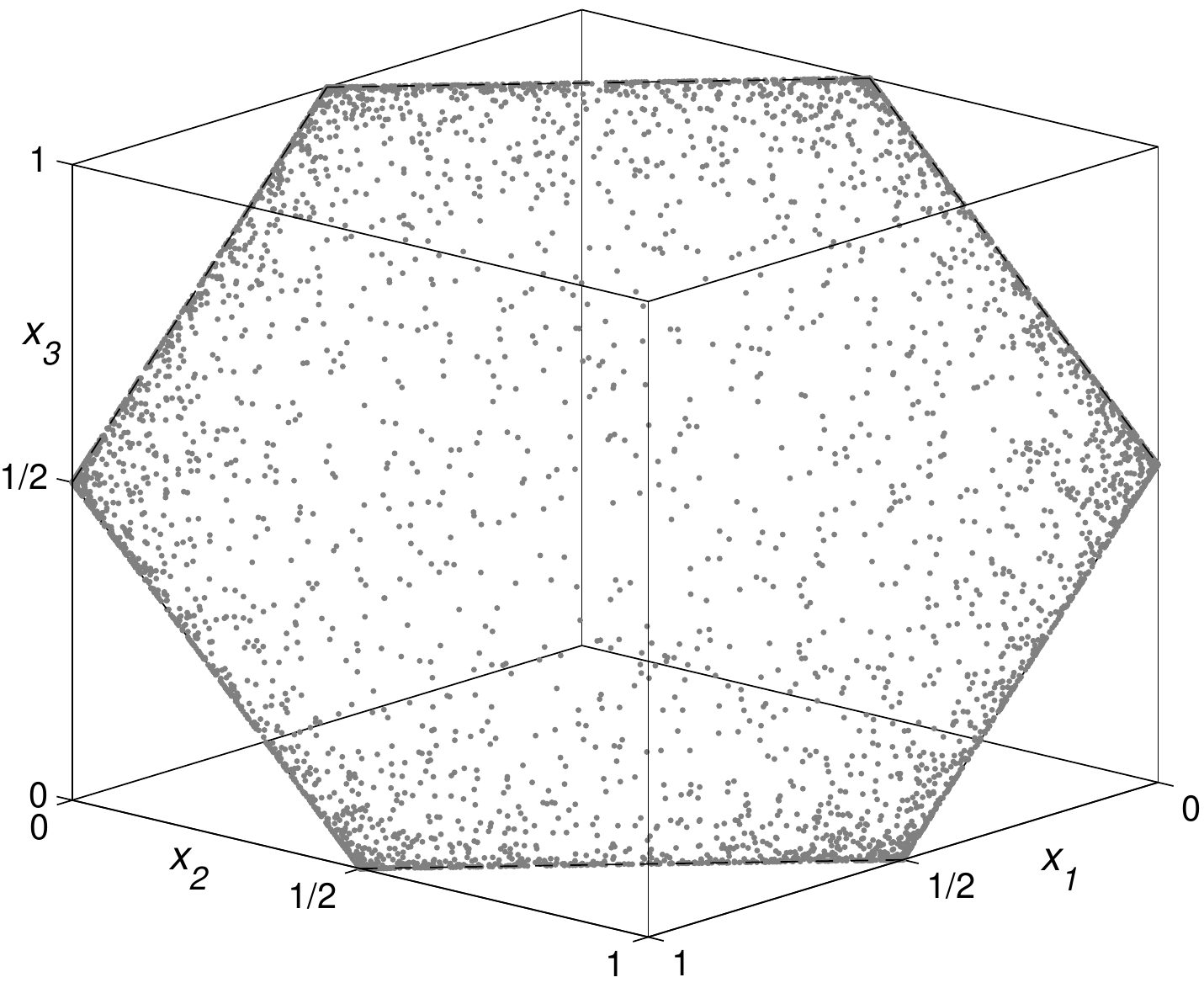} & 
\includegraphics[width=0.3\linewidth]{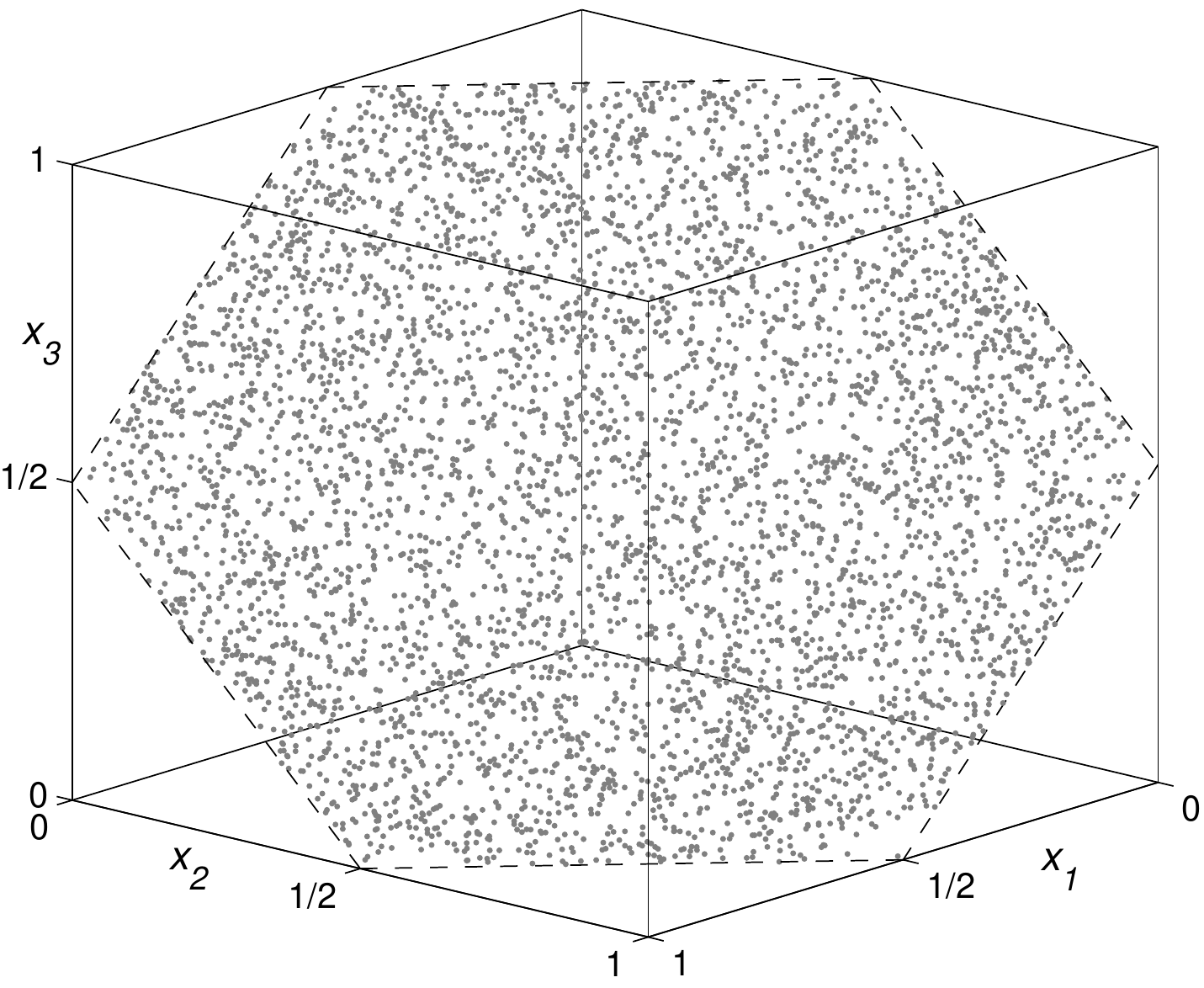} & 
\includegraphics[width=0.3\linewidth]{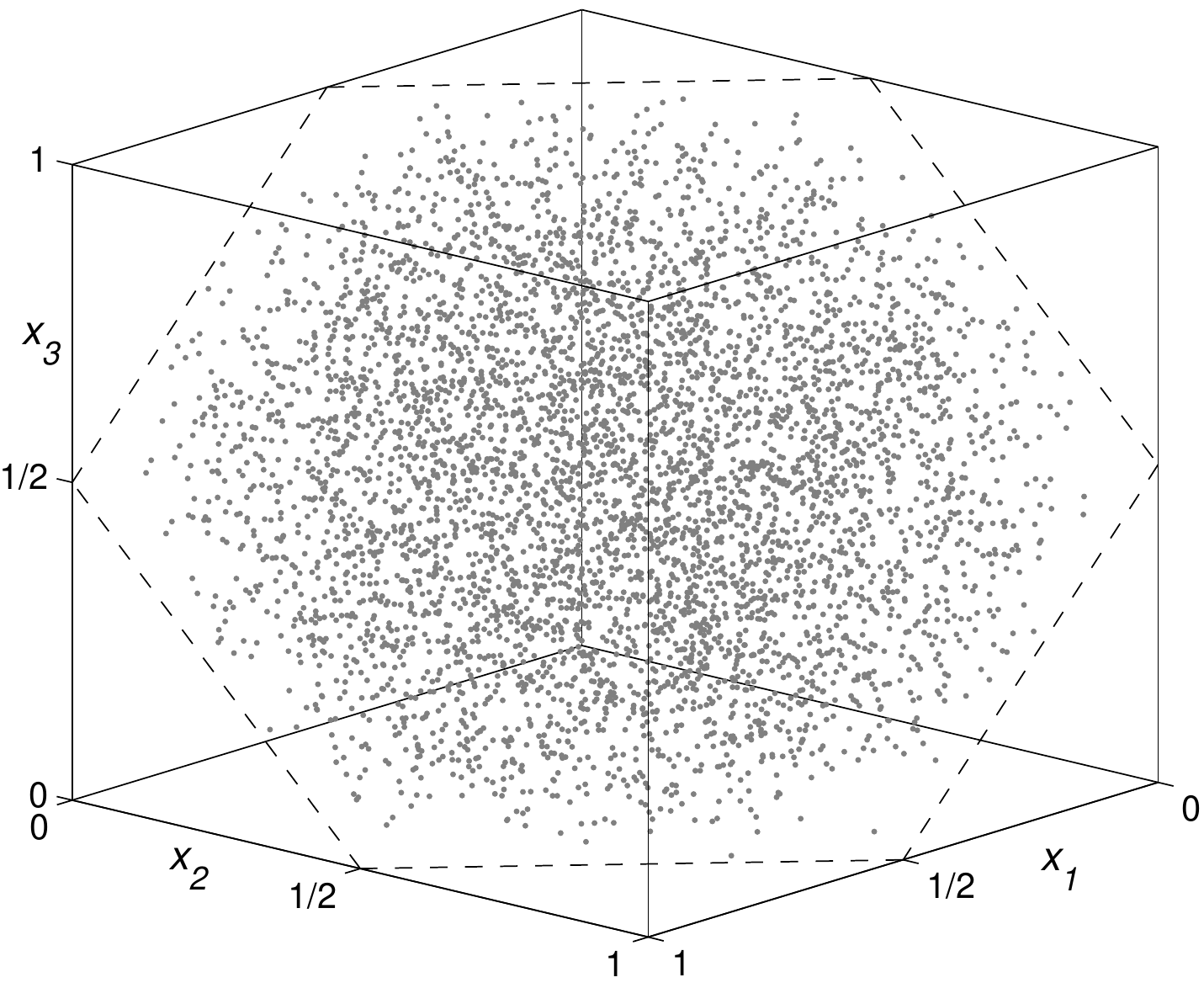} \\
\small  $b=0.5$ & \small $b=1.25$ &\small  $b=2$ \\
\multicolumn{3}{c}{\small Marginal distribution ($1,5$ million draws)}\\ 
\includegraphics[width=0.3\linewidth]{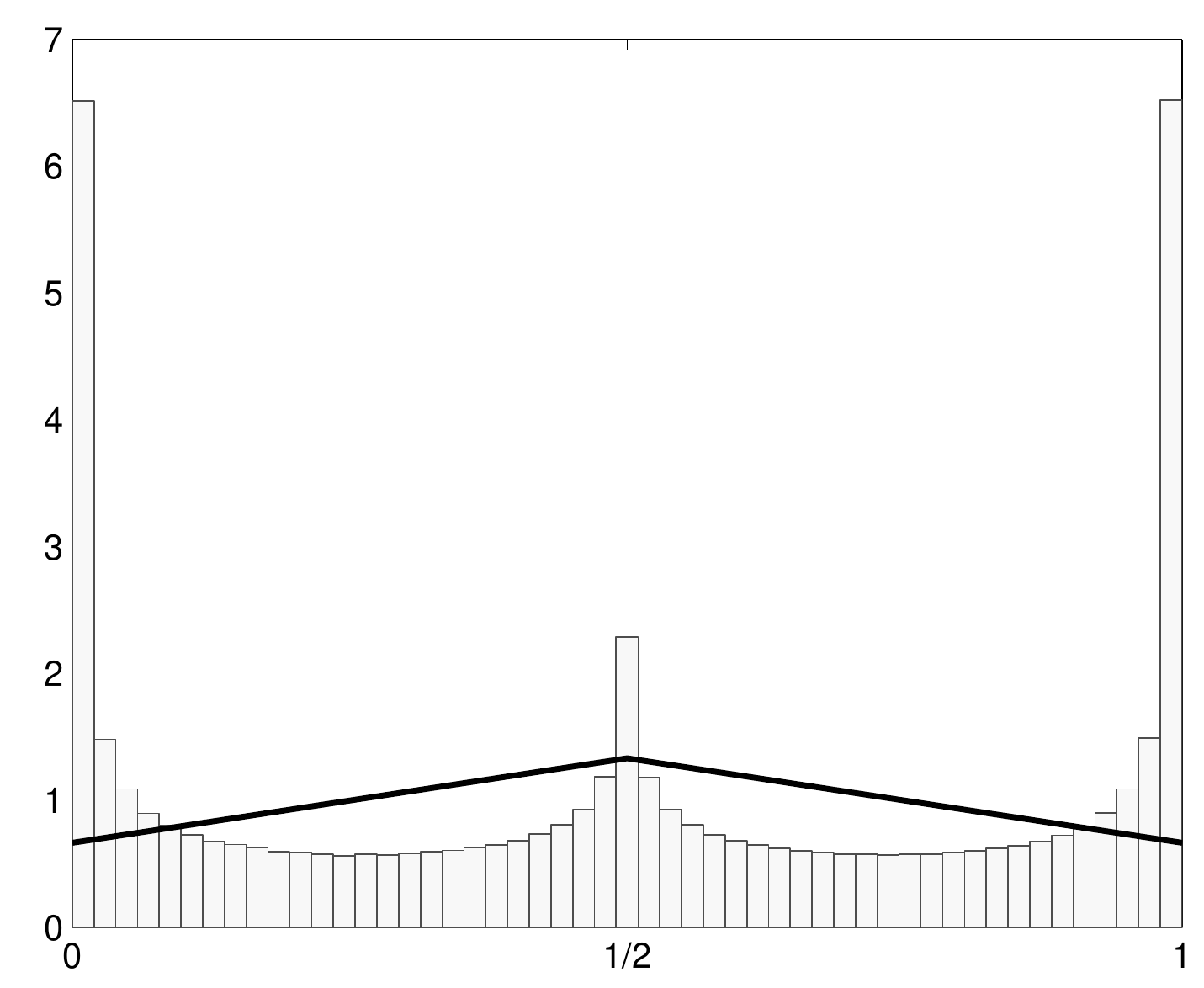} & 
\includegraphics[width=0.3\linewidth]{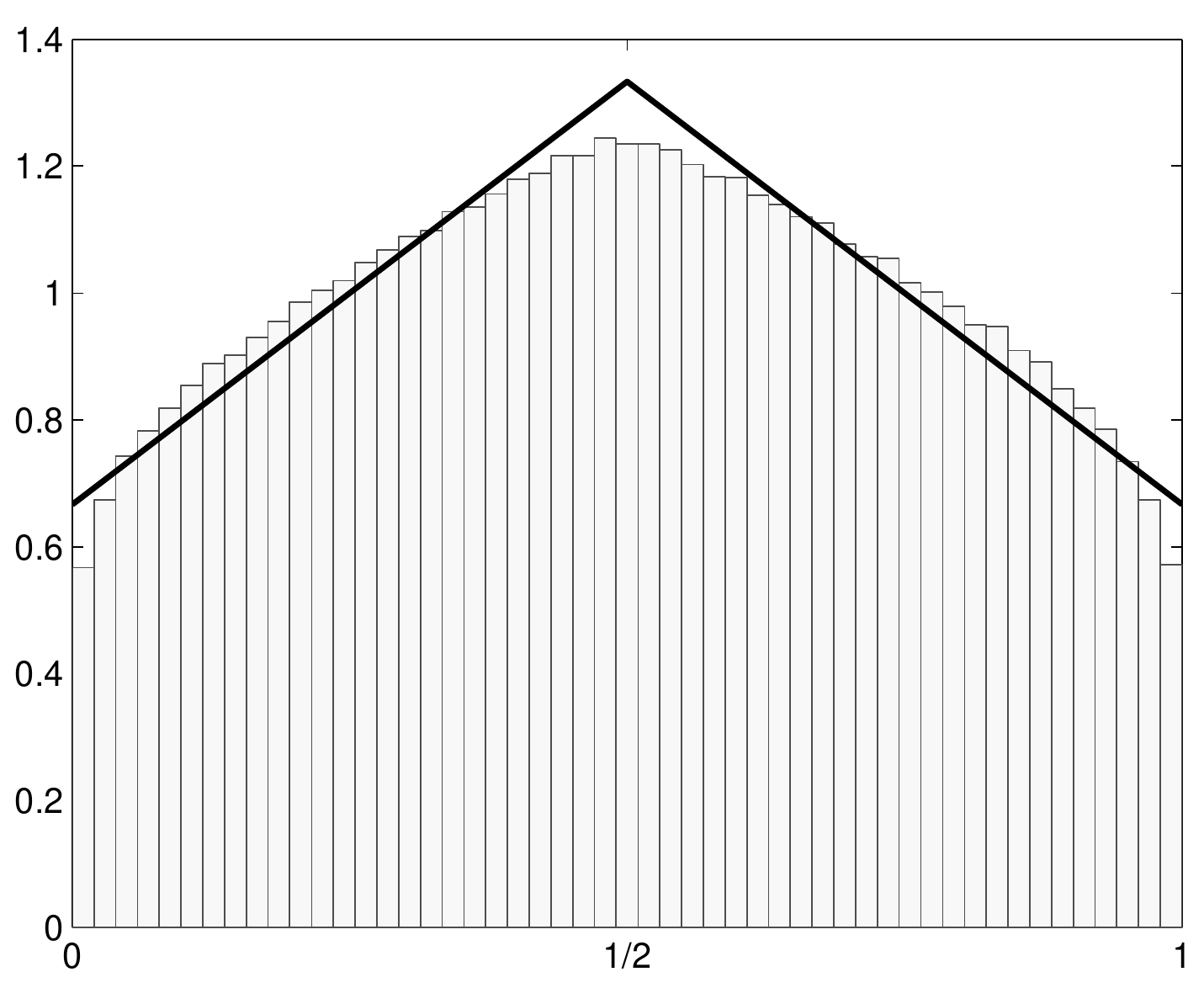} & 
\includegraphics[width=0.3\linewidth]{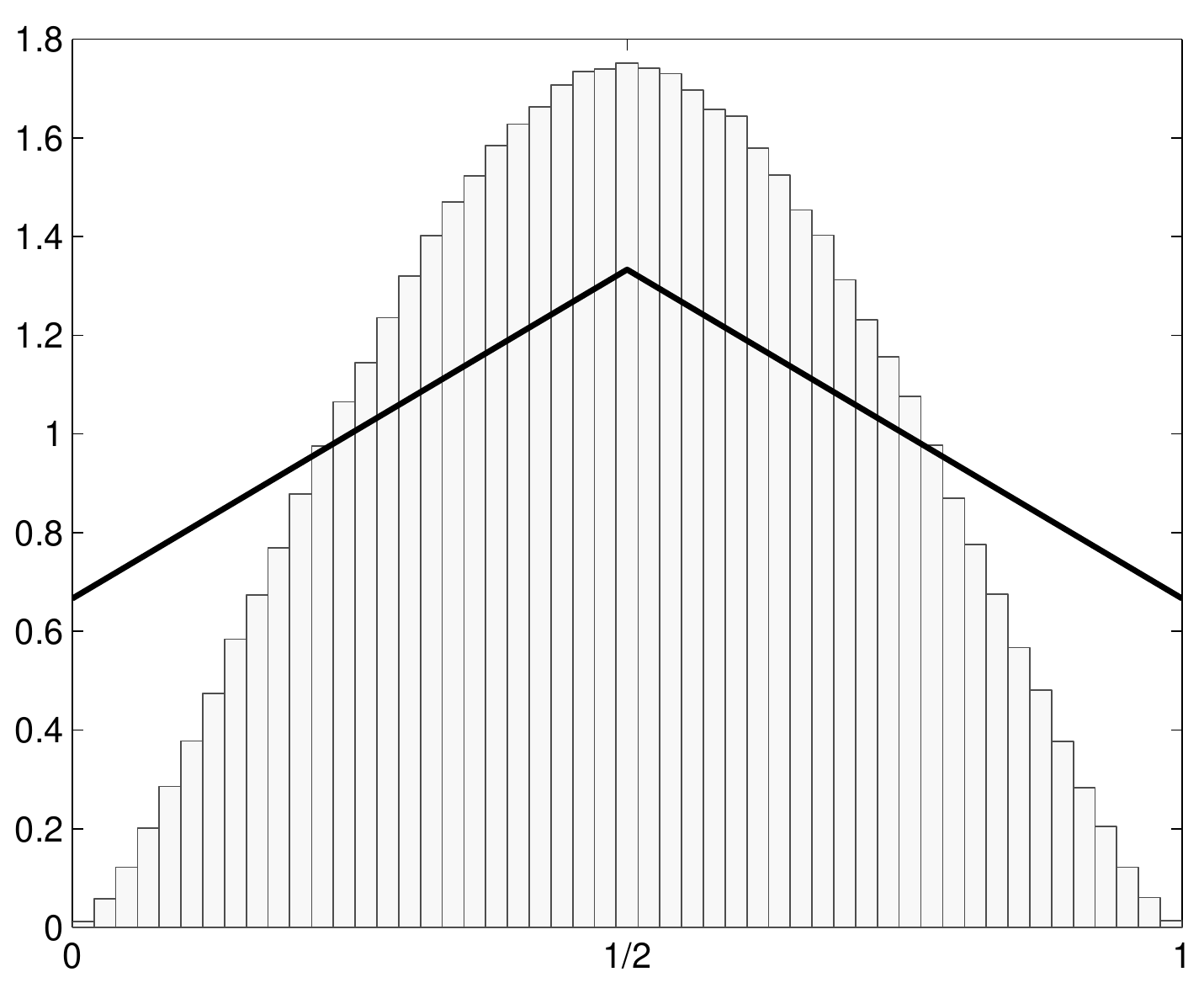} \\
\small  $b=0.5$ &\small  $b=1.25$ &\small  $b=2$ \\
\end{tabular}
\caption{An illustration of the indirect maximum entropy bandwidth}
\label{Fig:Gaussian.n3}
\end{figure}

The \emph{indirect maximum entropy} (iMaxEnt)  bandwidth can be defined as the bandwidth which results in the distribution of $V_{i}$'s being close to the uniform distribution over $\Pi_{n}$, \eqref{Eq:Marginal.Finite.n}. 
To illustrate the idea, let $n=3$ and $f$ and $k$ be the standard normal densities. 
The top panel of Figure \ref{Fig:Gaussian.n3} shows $5,000$ random draws of $V(b)=\left(V_{1}(b),V_{2}(b),V_{3}(b)\right)\transp$ with $b=0.5,\,1.25$, and $2$. The corresponding marginal density of $V_{i}(b)$ (histogram) together with $l_{3}$ (solid line) is shown in the bottom panel. 
The distribution of $V(b)$ with $b=1.25$ is visually uniform over $\Pi_{3}$ with the marginal being close to $l_{3}$, whereas the left and right panels correspond to bandwidths which are too small and too big, respectively.

To operationalise the definition of the iMaxEnt bandwidth the remainder of this section considers three criteria of closeness
chosen chiefly for their simplicity including simple calculating forms. Small sample performance is evaluated in a simulation study reported in Section \ref{Sec:Monte.Carlo}. 

\subsubsection{Minimum Cram\'{e}r-von Mises estimators}\label{S.Sec:CvM.iMaxEntbw}
Let $F$ and $G$ be two distribution functions with common support. The weighted Cram\'{e}r--von Mises (CvM) discrepancy \citep{smirnov1937} is defined as 
\begin{equation}
\label{Eq:WCvM.def}
\omega_{\psi}^{2}(G,F) = \int_{-\infty}^{\infty}\left[G(z)-F(z)\right]^{2}\psi\left(F(z)\right)dF(z)
 = \int_{0}^{1}\left[G\left(F^{-1}(t)\right)-t\right]^{2}\psi(t)dt,
\end{equation}
where $\psi$ is a weight function which is typically taken to be non-negative and smooth. 
Special cases of \eqref{Eq:WCvM.def} include the classical CvM criterion, $\omega^{2}$, when $\psi(t)=1$, $t\in[0,1]$, and the Anderson-Darling (AD) statistic, $\omega_{AD}^{2}$, when $\psi(t)=t^{-1}(1-t)^{-1}$, $t\in(0,1)$  \citep{anderson1952}. 

The min-CvM iMaxEnt bandwidth is defined as 
\begin{equation}
\label{Eq:iMaxEnt.BW.WCvM}
\hat{b}_{CvM} = \argmin_{b>0}\;\omega_{\psi}^{2}(G_{n},L_{n}),
\end{equation}
where $G_{n}=G_{n}(v;b)$ is the EDF of $V_{i}(b)$, $i=1,\ldots,n$. 
The CvM criterion $\omega^{2}_{\alpha;\epsilon}=\omega_{\psi}^{2}$ with the weight function $\psi(t)=t^{\alpha-1}(1-t)^{\alpha-1}\I{\epsilon\leq t\leq 1-\epsilon}$, where $\alpha\geq0$, and $0\leq\epsilon\ll1/2$ is a (small) trimming constant, is attractive as it has a simple calculating form given in Appendix \ref{App:CvM.comp} and includes both $\omega^{2}=\omega^{2}_{1;0}$ and $\omega_{AD}^{2}=\omega^{2}_{0;0}$ as special cases. 
In practice, $L_{n}$ can be estimated on a grid of points in $[0,1]$ as described in Appendix \ref{Sec:Permutohedron}, and evaluation of $u_{i}=L_{n}\left(V_{i}(b)\right)$ can be done by interpolation (linear or otherwise). Then $\omega_{\alpha;\epsilon}^{2}(G_{n},L_{n})$ can be easily calculated from the order statistics $u_{(1)}\leq\cdots\leq u_{(n)}$. 

An interesting connection exists between the above procedure and the \citet{sarda1993} cross-validation criterion (Appendix 
\ref{S.Sec:KCDFE.CV}). In particular, the unweighted version of the cross-validation criterion is numerically equivalent, up to an additive constant, to the CvM discrepancy between $G_{n}$ and the uniform d.f. on $[0,1]$, $R$, viz. $\CV_{1}(b) = \omega^{2}(G_{n},R)+1/(6n^{2})$. However, while asymptotically the minimiser of $\omega^{2}(G,R)$ is well defined, the integrated variance of $G_{n}$, $\IVar G_{n}$, adds an extra term making $\E\CV_{1}$ an increasing function of $b$. That is to say, the $\CV_{1}$ criterion cannot work.

\begin{figure}[htbp]\centering
\includegraphics[width=\linewidth]{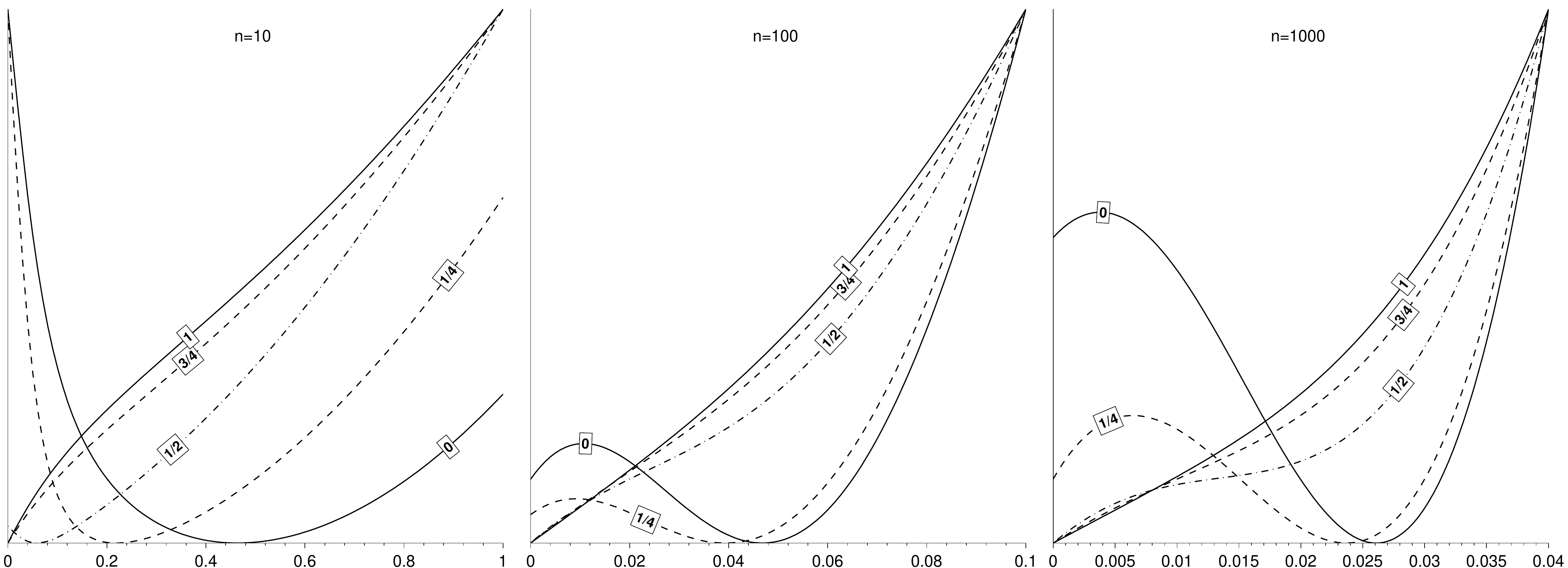}
\vspace*{-0.5\baselineskip}
\begin{flushleft}
\footnotesize Horizontal axes--bandwidth; vertical axes (not to scale)--CvM criteria, $\omega^{2}_{\alpha;\epsilon}$, for $\alpha=0,1/4,1/2,3/4$, and $1$, and $\epsilon=0.001$. 
\end{flushleft}
\vspace*{-\baselineskip}
\caption{CvM criteria for the strongly skewed density}
\label{Fig:EXiMaxEntBwCvMbetaD3M1K}
\end{figure}

The $\bigO{bn^{-1}}$ term from $\IVar G_{n}$ causing $CV_{1}$ to break down is still present in the expansion of  $\E\omega^{2}(G_{n},L_{n})$, of course, but it does not dominate asymptotically as in $\E CV_{1}$. In particular, if Conjecture \ref{Conj:Limit.Distr}(b) in Section \ref{Sec:Asy.conj} holds, then by the same method as in the proof of Lemma \ref{Lemma:KCDFE:Sarda.CV.CvM}, the leading terms in an expansion of $\omega^{2}(G,L)$ are $a_{1}b^{4}-a_{2}b^{2}n^{-1/2}+a_{3}n^{-1}$, where $a_{1}=\mu_{2}^{2}(k)\zeta_{1}(F)/4$, as in Lemma \ref{Lemma:KCDFE:Sarda.CV.CvM}, and $a_{2},a_{3}$ are some positive constants depending on $\gamma_{m}$'s in Conjecture \ref{Conj:Limit.Distr}(b). 
Thus $\E\omega^{2}(G_{n},L_{n})$ generally has a well-defined minimum away from zero, and the minimiser is of order $n^{-1/4}$ asymptotically. However, in finite samples, there is typically a second (local or global) minimum at $b=0$, even for the normal density. In some particularly unfavourable cases, for example for the strongly skewed distribution (\#3 in Figure \ref{Fig:sMWNMdens1-6}), only $\omega^{2}_{\alpha;\epsilon}$ with $\alpha$ very close to zero appears to perform well provided care is taken to ensure the correct minimum is chosen; see Figure \ref{Fig:EXiMaxEntBwCvMbetaD3M1K} for an illustration. 

\subsubsection{Moment-based estimating equations}
Moment-based estimators provide a simple, appealing alternative alleviating some of the problems with the min-CvM procedure.
Let $m_{j,n}$, $j=2,3,\ldots$ be the central moments of the distribution $l_{n}(u)$ defined in \eqref{Eq:Marginal.Finite.n}. 
The $r$-moment iMaxEnt bandwidth, $\hat{b}_{MEE}$, can be defined as the near-solution to the $(r-1)$ estimating equations
\begin{equation}
\label{Eq:iMaxEnt.BW.GEE}
n^{-1}\sum_{i=1}^{n}(V_{i}(\hat{b})-1/2)^{j}-m_{j,n} = 0,\quad j=2,3,\ldots,r.
\end{equation}
(Since $V_{i}$'s sum to $n/2$ for all $b$, the first moment contains no identifying information.) 
$\hat{b}$ in \eqref{Eq:iMaxEnt.BW.GEE} can be estimated using Generalised Empirical Likelihood (GEL) methods \citep{newey2004}. 

\begin{figure}[htbp]\centering
\includegraphics[width=0.7\linewidth]{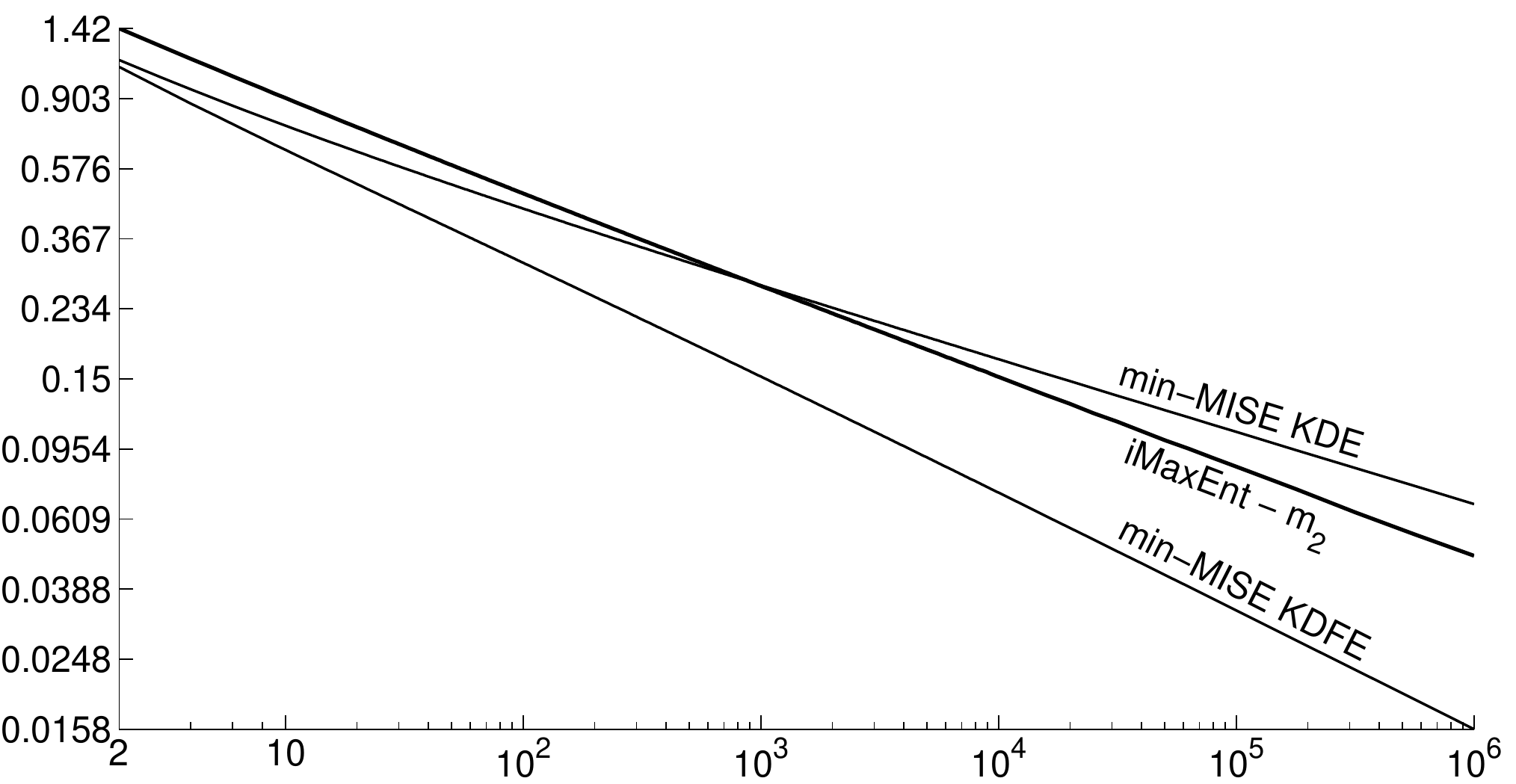}
\vspace*{-0.5\baselineskip}
\begin{flushleft}
\footnotesize Horizontal axis--$n$, log scale; vertical axis--bandwidth, $b$, log scale.
\end{flushleft}
\vspace*{-\baselineskip}
\caption{$2^{nd}$ moment-based iMaxEnt bandwidth}
\label{Fig:exGaussVarBw}
\end{figure}

The simplest moment-based bandwidth sets the sample variance of $V_{i}$'s equal to $m_{2,n}$. For example when $f$ and $k$ are both standard normal densities, 
\begin{equation*}
\Var V_{1}(b)  = \frac{n-2}{n-1}\frac{1}{\pi}\arctan\sqrt{\frac{3+b^{2}}{1+b^{2}}} + \frac{1}{n-1}\frac{1}{\pi}\arctan\sqrt{\frac{4+b^{2}}{b^{2}}} -\frac{1}{4}.
\end{equation*}
The bandwidth $b$ solving $\Var V_{1}(b)=m_{2,n}$ is shown in Figure \ref{Fig:exGaussVarBw}  (iMaxEnt--$m_{2}$). Bandwidths minimising the exact MISE of $\hat{f}$ (min-MISE KDE) and $\widehat{F}$ (min-MISE KDFE) are shown for comparison. 

When the true density is close to normal the $2^{nd}$ moment-based iMaxEnt bandwidth can be expected to perform well in the sense of the resultant distribution of $V_{i}$ being close to $l_{n}$ in other metrics as well. This is confirmed by the simulation study (Figure \ref{Fig:SIMiMaxEntBwKDE.1bw} in Section \ref{Sec:Monte.Carlo}). 
In general, inclusion of the $3^{rd}$ and $4^{th}$ moments is advisable to capture skewness and kurtosis, respectively. 
However, estimators based on much higher moments are likely to become increasingly sensitive to outliers. 

\subsubsection{Neyman smooth test}
A useful alternative to the moment-based estimators can be obtained by inverting the \citet{neyman1937} smooth test statistic. 
Let $\rho_{k}(v)$, $k=0,1,2,\ldots$ be the orthonormal shifted Legendre polynomials\footnote{\label{Fn:Legendre.polynomials}
$\rho_{k}(v)$ can be defined by the Rodrigues formula 
$\rho_{k}(v) = \frac{\sqrt{2k+1}}{k!}\frac{d^{k}}{dv^{k}}(v^{2}-v)^{k}$
or the explicit expression 
$\rho_{k}(v) = (-1)^{k}\sqrt{2k+1}\sum_{j=0}^{k}\binom{k}{j}\binom{k+j}{j}(-v)^{j}$.
Properties of the polynomials $\rho_{k}(v)$ can be obtained from those of the 
`usual' Legendre polynomials $P_{k}(v)$, orthogonal to the uniform density on the interval $[-1,1]$, using the relationship 
 $\rho_{k}(v)=\sqrt{2k+1}P_{k}(2v-1)$; see e.g. \citet[Ch.18]{Olver2010}. }, i.e. orthonormal polynomials with respect to the uniform density on $[0,1]$. 
The Neyman smooth test statistic based on the first $r$ polynomials is defined as 
\begin{equation}\label{Eq:Neyman.Crit}
S_{r}(b) = \sum_{j=1}^{r}\left[\frac{1}{\sqrt{n}}\sum_{i=1}^{n}\rho_{j}\left(u_{i}(b)\right)\right]^{2} 
= n\bar{\rho}_{r}\transp\bar{\rho}_{r}, 
\end{equation}
where $\bar{\rho}_{r} = n^{-1}\sum_{i=1}^{n}\left(\rho_{1}(u_{i}(b)),\ldots,\rho_{r}(u_{i}(b))\right)\transp$ and $u_{i}(b)=L_{n}(V_{i}(b))$ as before. The estimator of the iMaxEnt bandwidth based on the first $r\geq 2$ polynomials is defined as a minimiser of \eqref{Eq:Neyman.Crit}, viz. $\hat{b}_{NS}=\argmin_{b>0}S_{r}(b)$. 

Quite generally, $\hat{b}_{MEE}$ and $\hat{b}_{NS}$ with the same $r$ can be expected to be similar
 as, loosely speaking, $\hat{b}_{NS}$ can be viewed as a restricted version of $\hat{b}_{MEE}$. 
$\hat{b}_{NS}$ is computationally faster and can be more stable with high $r$ since it does not involve computation of the covariance matrix of MEE \eqref{Eq:iMaxEnt.BW.GEE}. The choice of $r$ can be based on the same considerations as above; see also discussion in \citet{bera2002b} and references therein.

\subsection{Remarks on asymptotic behaviour}\label{Sec:Asy.conj}
To obtain asymptotic expansions for the proposed bandwidth estimator the knowledge of the limiting behaviour of  $l_{n}(u)$ in required. While there is no formal proof, intuition and simulation evidence (see Supplement) suggest that the following conjecture is true. 

\begin{conjecture}
\label{Conj:Limit.Distr}
As $n\to\infty$
\begin{enumerate}[(a)]
\item $l_{n}(u)$ approaches the uniform density, $l_{n}(u)\to 1$ for all $u\in(0,1)$, except at the end points where $l_{n}(0) = l_{n}(1) = \left(1-1/n\right)^{n-2} \to e^{-1}$, and 
\item the moments of a random variable with density $l_{n}(u)$ approach the moments of the uniform distribution at a rate $n^{-1/2}$ from below, i.e. $\int_{0}^{1}u^{m}l_{n}(u)du = \frac{1}{m+1}-\gamma_{m}n^{-1/2}+\littleO{n^{-1/2}}$ for $m\geq2$ and some finite constants $\gamma_{m}>0$.
\hfill$\square$
\end{enumerate}
\end{conjecture}

If Assumptions \ref{Ass:DGP}(a,b), \ref{Ass:kernel}(a,b), and \ref{Ass:bandwidth} hold, and Conjecture \ref{Conj:Limit.Distr}(b) is true, then in view of Lemma \ref{Lemma:KCDFE.PIT.Moments}, the bandwidth setting $G=L_{n}$ will be of order $n^{-1/4}$ as $n\to\infty$. Asymptotically, the iMaxEnt bandwidth will be in-between the KDE and KDFE MISE-minimising bandwidths, which are of order $n^{-1/5}$ and $n^{-1/3}$, respectively. 
This is true at least for the simple examples such as shown in Figure \ref{Fig:exGaussVarBw}. 

\section{Simulation study}\label{Sec:Monte.Carlo}
\begin{figure}[htbp]\centering
\begin{tabular}{ccc}
\#1: Gaussian & \#2: Skewed unimodal & \#3: Strongly skewed  \\
  \includegraphics[width=0.3\linewidth]{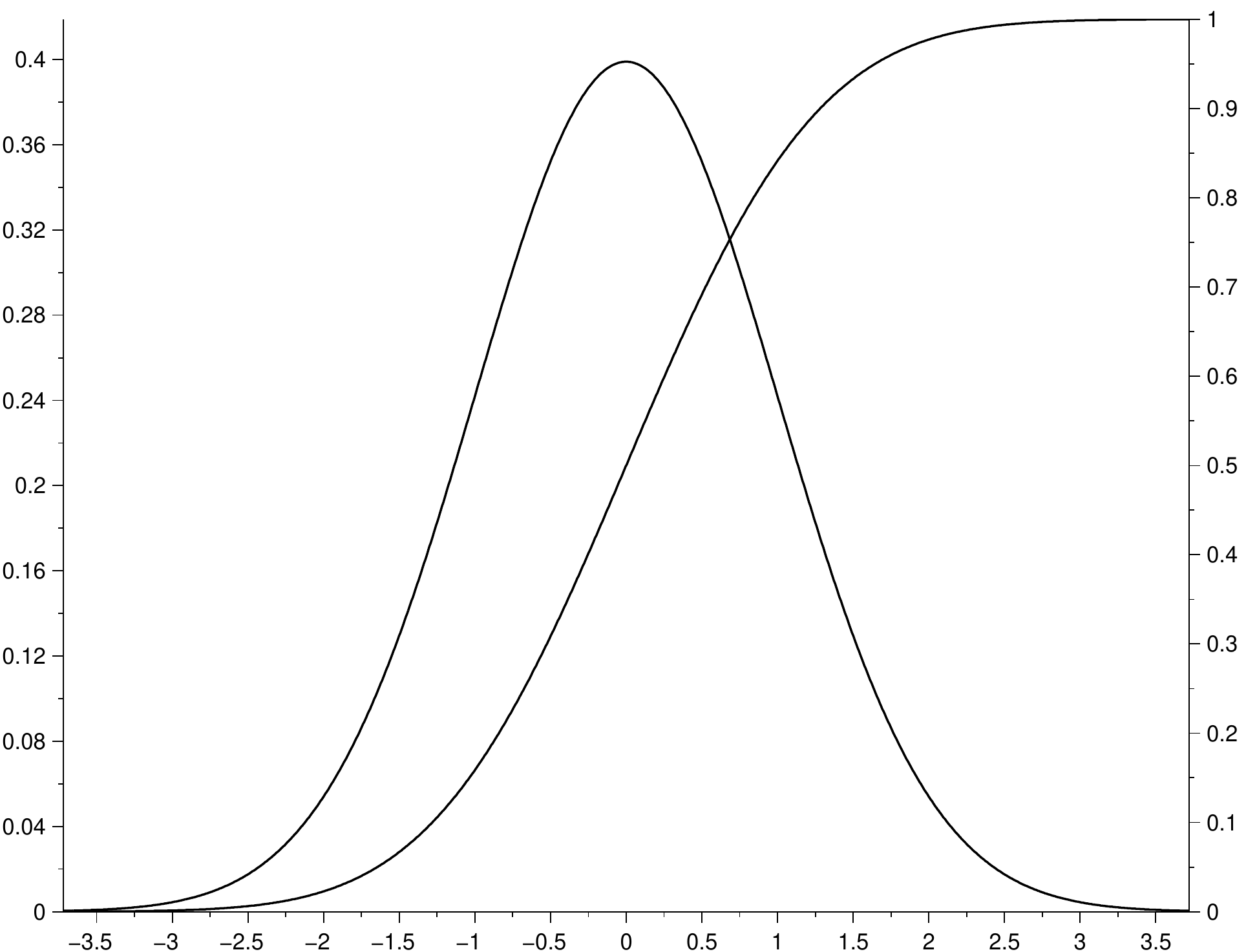} & 
  \includegraphics[width=0.3\linewidth]{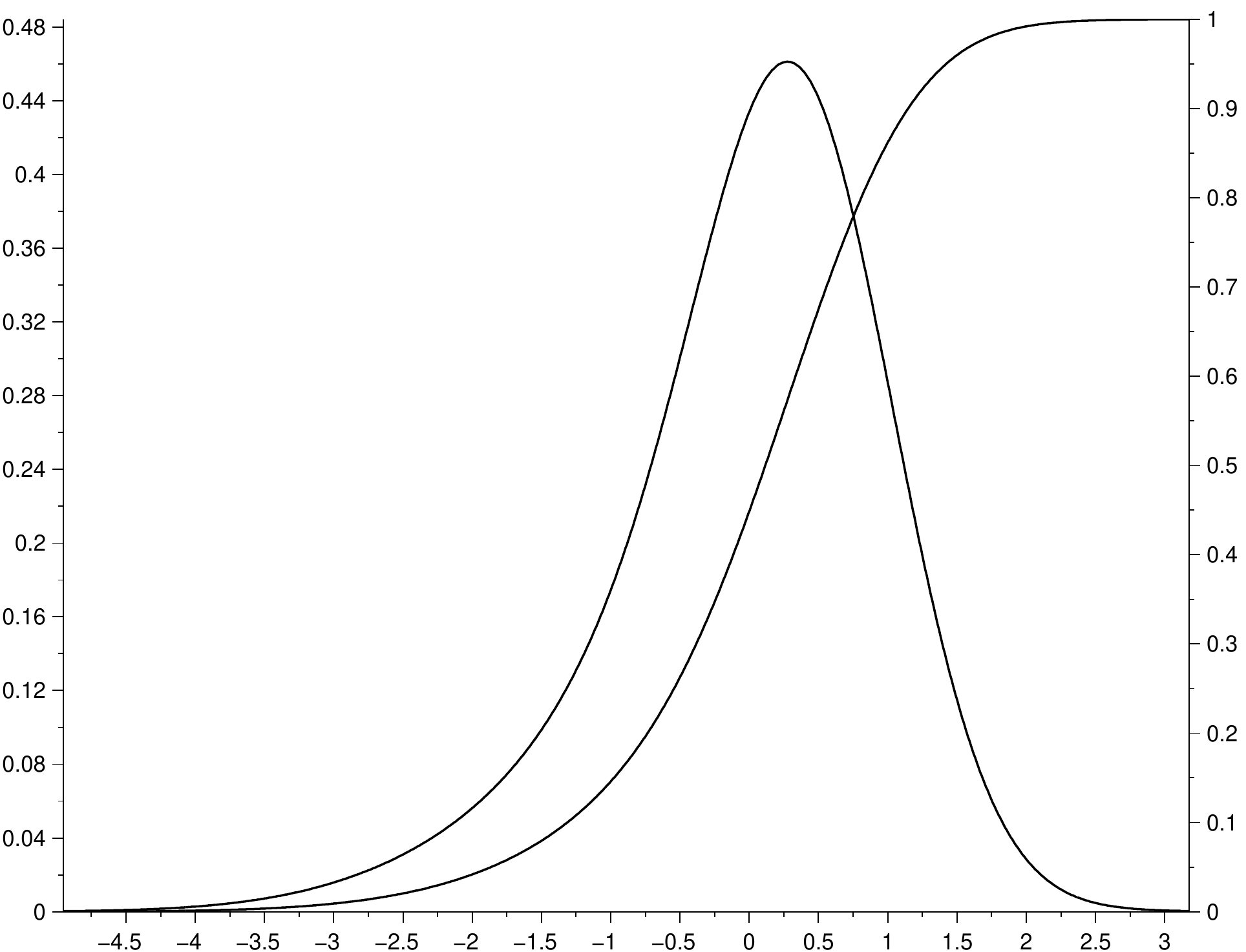} & 
  \includegraphics[width=0.3\linewidth]{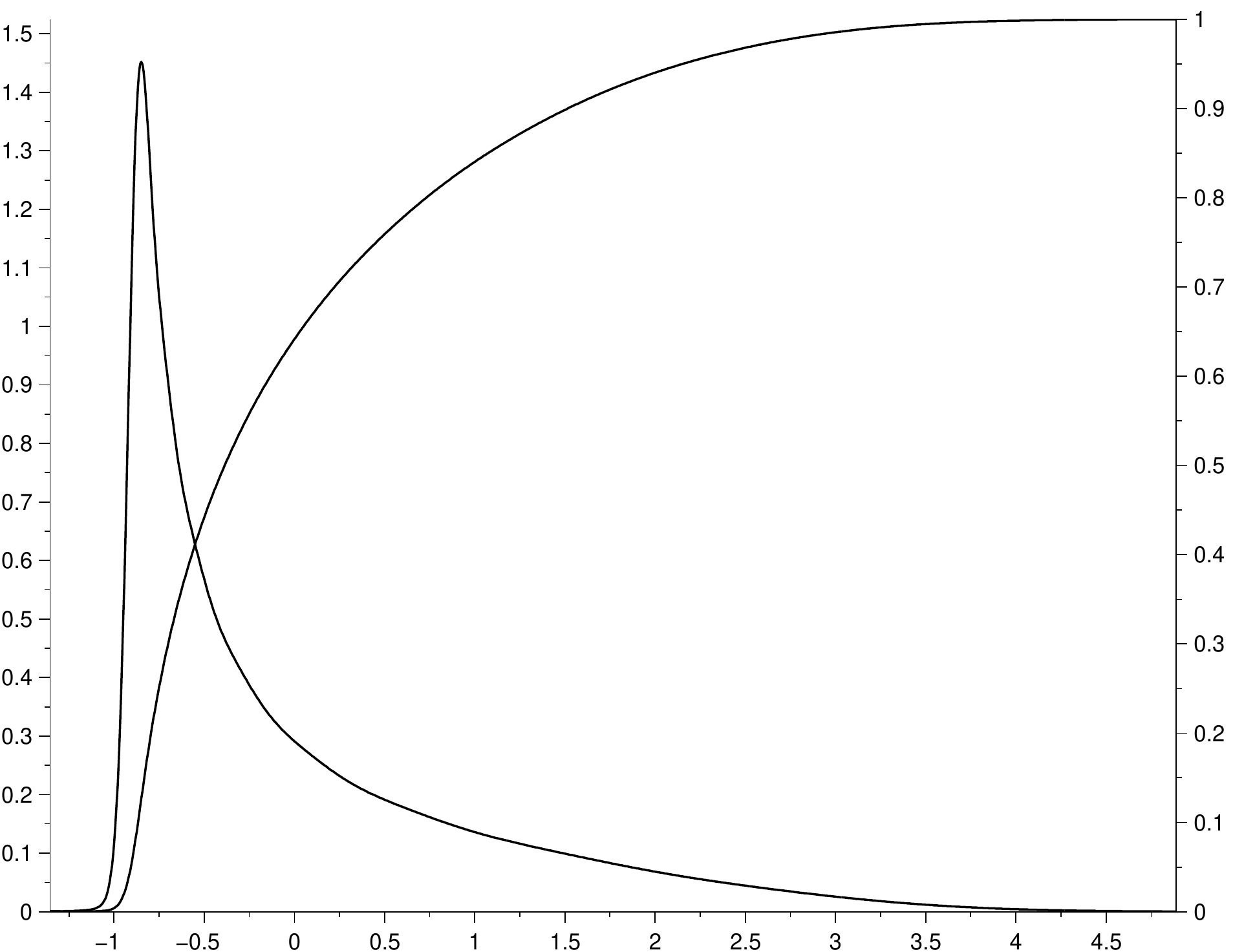} \\ 
\#4: Kurtotic unimodal & \#5: Outlier & \#6: Bimodal  \\
  \includegraphics[width=0.3\linewidth]{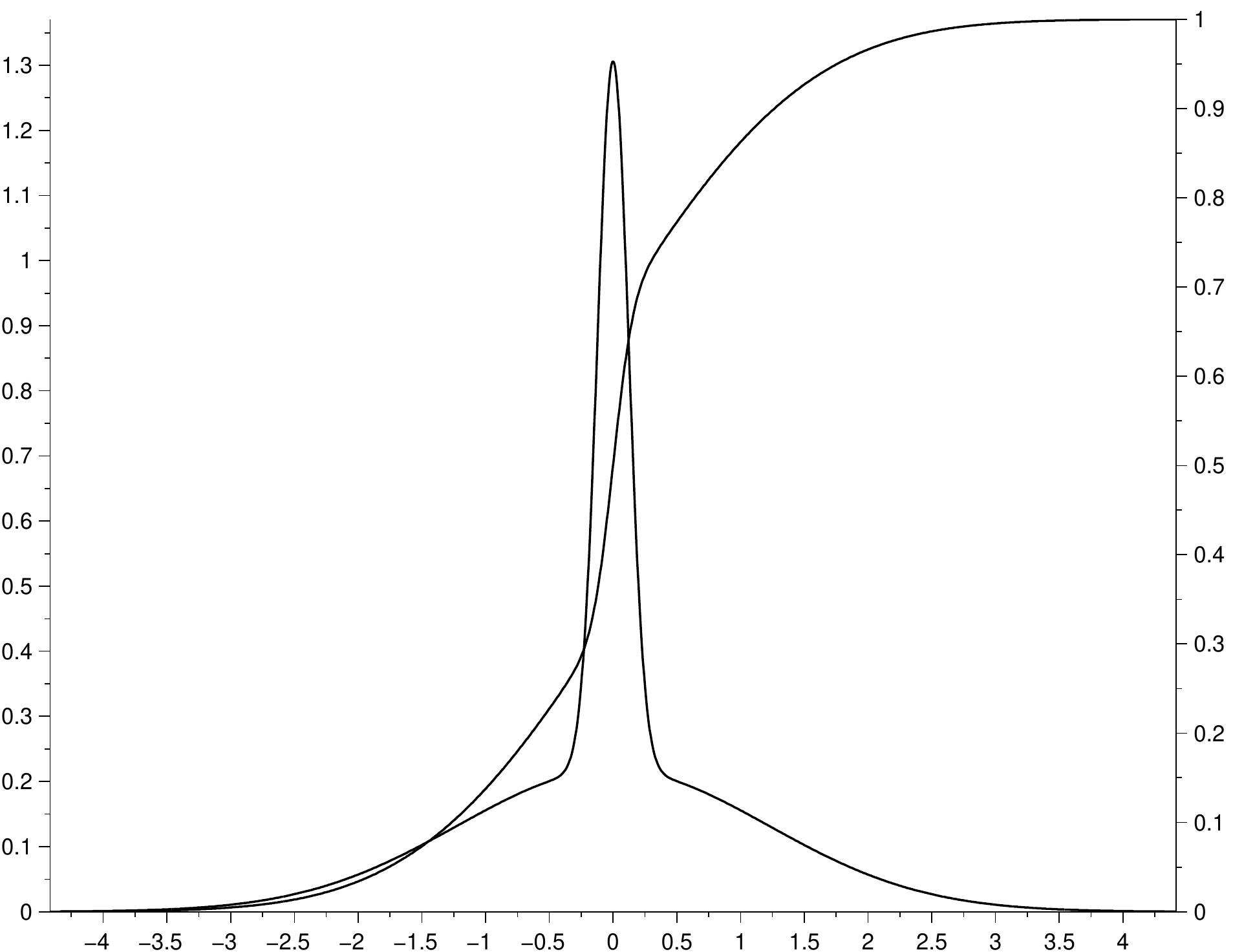} & 
  \includegraphics[width=0.3\linewidth]{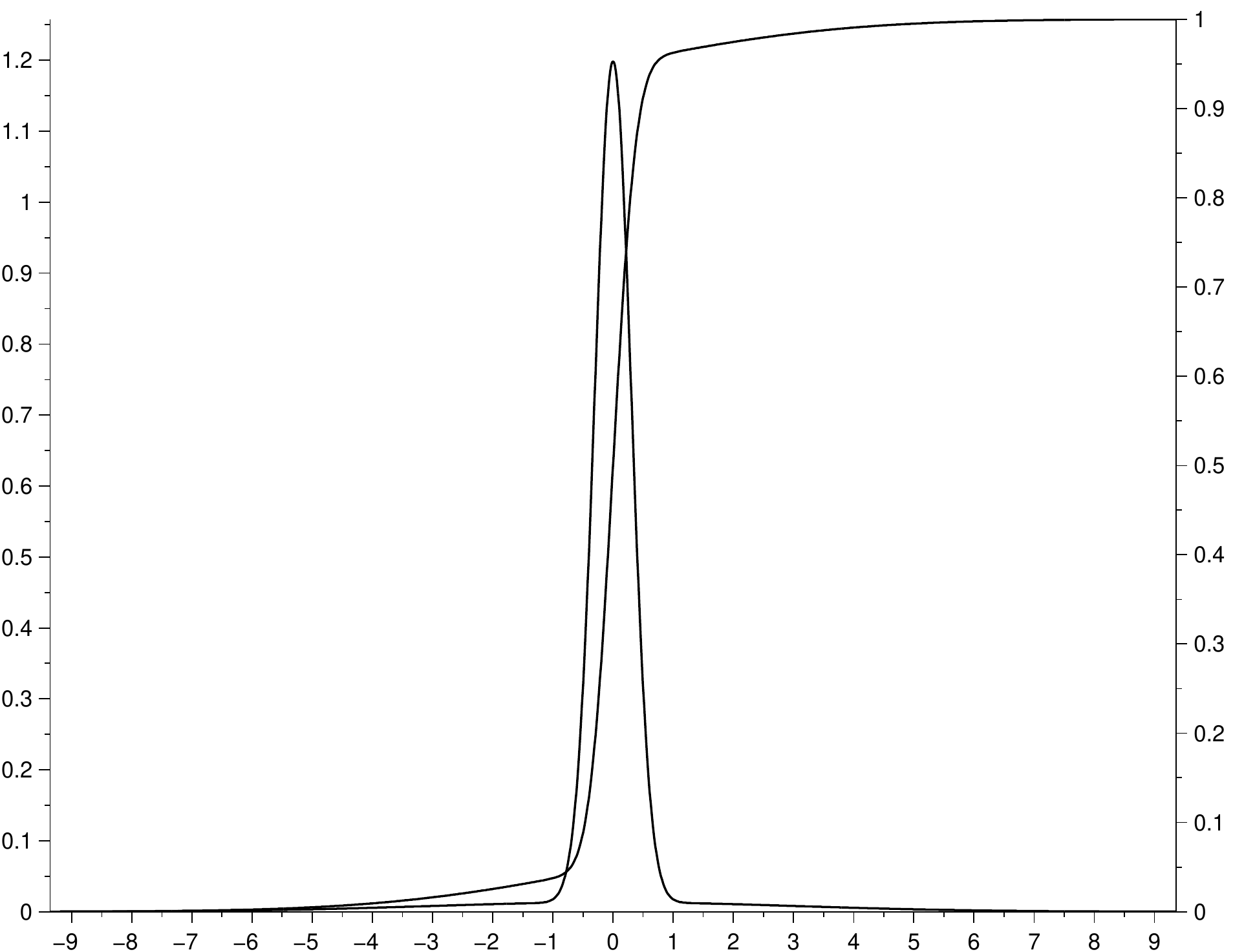} & 
  \includegraphics[width=0.3\linewidth]{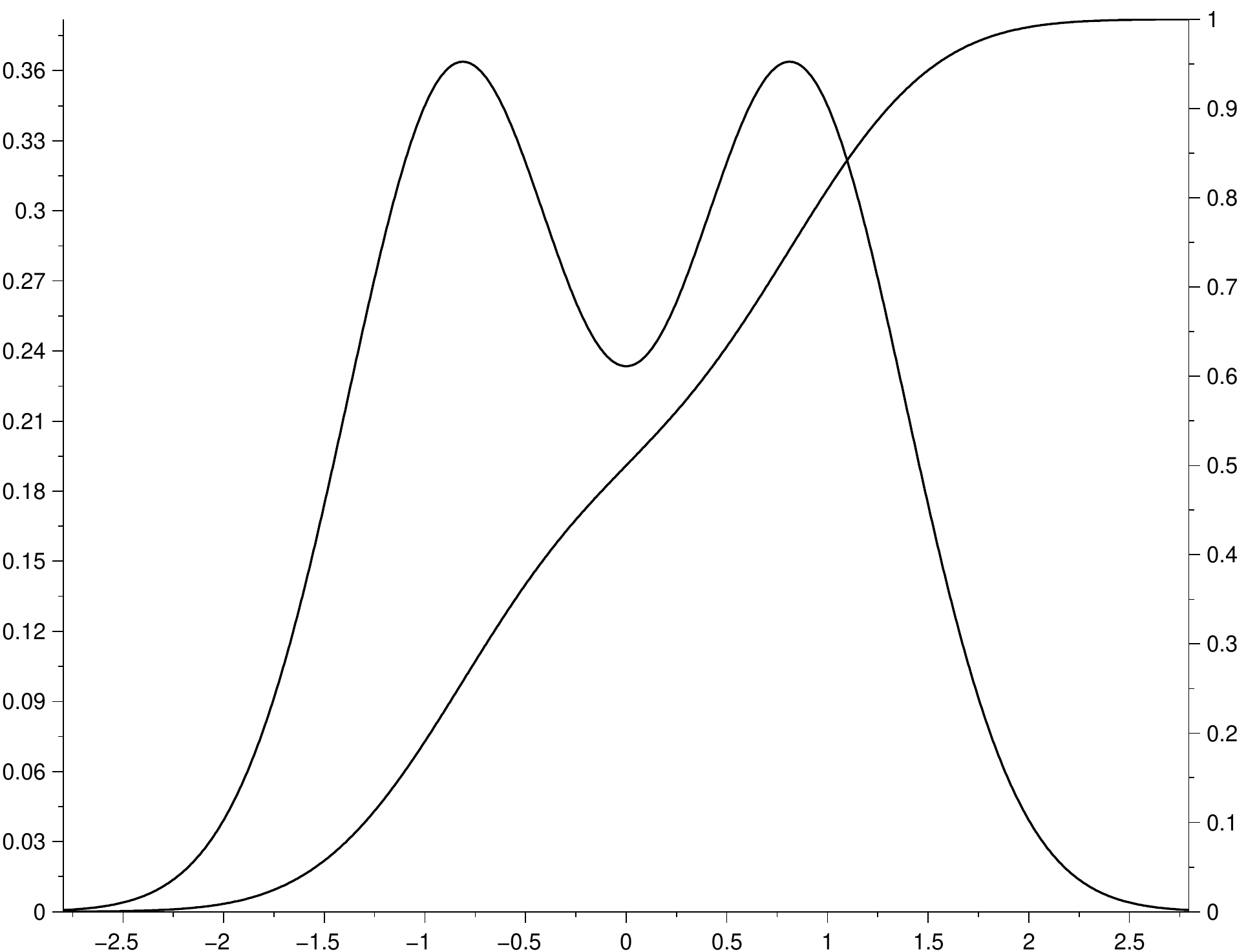} \\[1ex]
\multicolumn{3}{l}{Density on the left vertical axes, d.f. on the right.}
\end{tabular}
\caption{Selected normal mixture distributions}
\label{Fig:sMWNMdens1-6}
\end{figure}

\begin{figure}[htbp]\centering
\begin{tabular}{@{}>{\footnotesize}c@{}}
\#1: Gaussian\tabularnewline
\includegraphics[width=\linewidth]{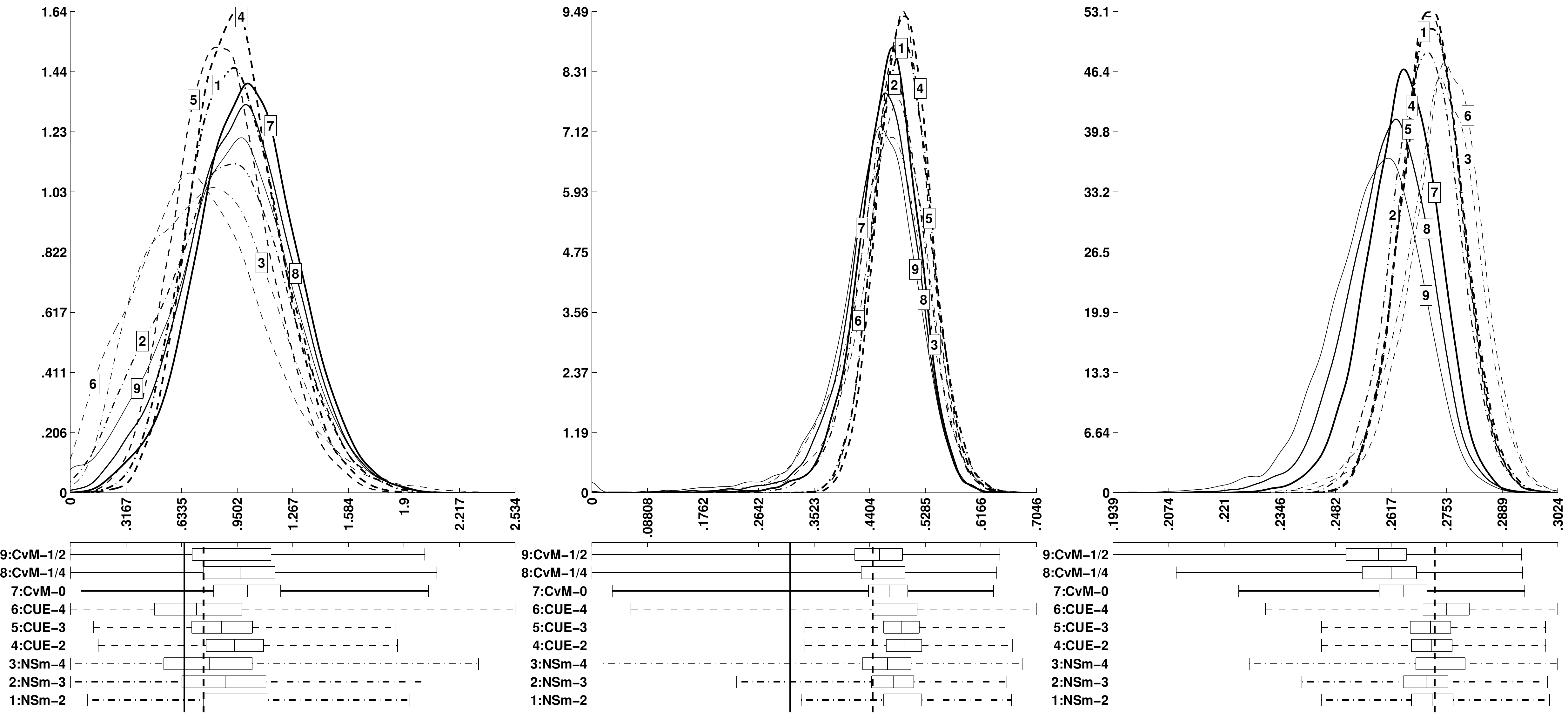} \tabularnewline
\#2: Skewed unimodal\tabularnewline
\includegraphics[width=\linewidth]{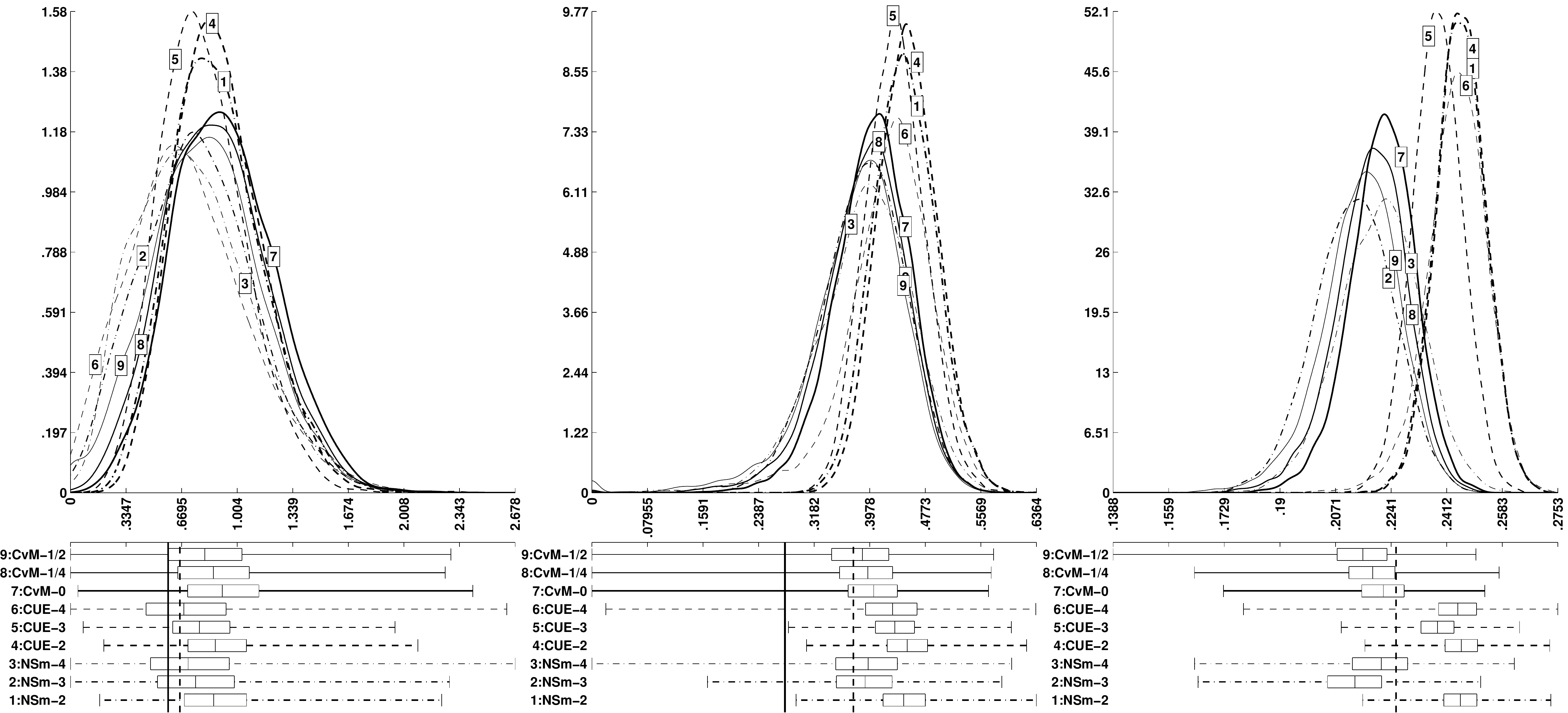} \tabularnewline
\#3: Strongly skewed\tabularnewline
\includegraphics[width=\linewidth]{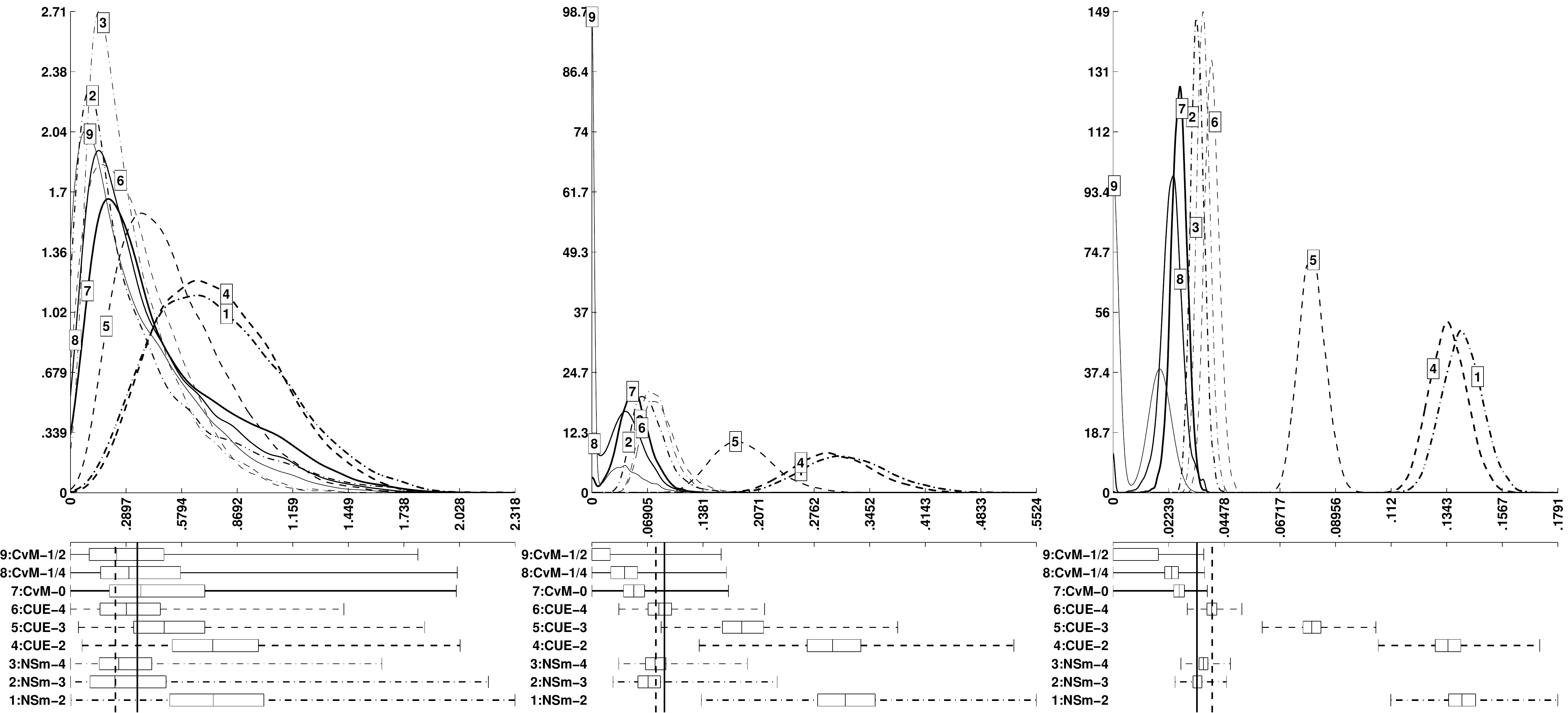}
\end{tabular}
\caption{Performance of selected iMaxEnt bandwidth estimators}
\label{Fig:SIMiMaxEntBwKDE.1bw}
\end{figure}

This section investigates performance of selected iMaxEnt bandwidth estimators in small and medium samples via simulation using the first six normal mixture distributions defined in Table 1 of \citet{marron1992} as examples (Figure \ref{Fig:sMWNMdens1-6}). The densities have been scaled to have zero mean and unit variance. 
Among the advantages of working with normal mixtures is the availability of exact MISE expressions for the class of Gaussian-based kernels given in \citet{marron1992} for KDE and \citet{oryshchenko2016} for KDFE. The values of the exact MISE-minimising bandwidths for KDE and KDFE are shown in Figure \ref{Fig:SimLOOPITGKmargDens.1} (`D' and `C' bandwidths, respectively) and drawn in Figure \ref{Fig:SIMiMaxEntBwKDE.1bw} as vertical dashed and solid lines, respectively. These provide reference points for the iMaxEnt bandwidth, expected to be in-between the two asymptotically.

For each mixture distribution, Figure \ref{Fig:SIMiMaxEntBwKDE.1bw} shows density estimates and boxplots (minimum, $1^{st}$, $2^{nd}$, and $3^{rd}$ quartiles, and maximum) of the nine iMaxEnt bandwidth estimators:
\begin{itemize}
\item[-] The min-CvM estimators with weight function $\psi(t)=t^{\alpha-1}(1-t)^{\alpha-1}\I{\epsilon\leq t\leq 1-\epsilon}$ for $\alpha=0,\,1/2$, and $1/4$ (CvM-0 or AD, CvM-1/2, and CvM-1/4, respectively), and $\epsilon=0.001$.
\item[-] The MEE estimators based on the $2^{nd}$ (CUE-2), $2^{nd}$ and $3^{rd}$ (CUE-3), and $2^{nd}$, $3^{rd}$, and $4^{th}$ moments (CUE-4). These are the continuously updating estimators (CUE), which is a special case of GEL. Performance of other members of the GEL class of estimators, including Empirical Likelihood, was much worse and is therefore not shown.
\item[-] The minimum Neyman smooth test estimators based on the first 2, 3, and 4 polynomials (NSm-2, -3, and -4, respectively). 
\end{itemize} 
Figure \ref{Fig:SimLOOPITGKmargDens.1} shows the $l_{n}$ density (thick solid line) and estimated densities of $V_{i}(b)$ for five selected values of the bandwidth: the KDE and KDFE min-MISE bandwidths (`D' and `C') and the median values of CvM-0 (AD), NSm-2, and NSm-4 iMaxEnt bandwidths. 
In both Figure \ref{Fig:SIMiMaxEntBwKDE.1bw} and \ref{Fig:SimLOOPITGKmargDens.1}  the kernel is Gaussian, and the sample sizes are (left to right) 10, 100, and 1000. Results are based on 10 and 100 thousand replications, respectively. 

As expected, all bandwidth estimators produce similar results for the normal density. This is also largely true for the moderately skewed density, although the importance of the third moment is noticeable. Somewhat surprisingly, the estimators produce not too dissimilar results for the outlier density in moderate samples, with the importance of the fourth moment now being pronounced, and the larger right tail in small samples reflecting the presence of outliers. 

The situation is strikingly different for the densities with significant departures from normality. As was noted in Section \ref{S.Sec:CvM.iMaxEntbw} (Figure \ref{Fig:EXiMaxEntBwCvMbetaD3M1K}), the min-CvM bandwidth for the strongly skewed density is likely to be zero for $\alpha$ much larger than zero, which is indeed the case. The NSm-3, NSm-4, and CUE-4, but not CUE-3, yield similar results, followed by the min-AD bandwidth; all delivering the distribution of $V_{i}$ reasonably close to $l_{n}$ (Figure \ref{Fig:SimLOOPITGKmargDens.1}). The NSm-2 and CUE-2 behave differently, of course, since they only use the information in the second moment. The difference in behaviour of NSm-3 and CUE-3 is somewhat puzzling. 
Similar picture emerges for the bimodal density (\#6) with the estimators using the first four moments producing the most promising results, followed closely by AD. 

None of the moment-based (or Neyman smooth test) bandwidth estimators can be recommended for the kurtotic unimodal density (\#4): the NSm-4 and CUE-4 estimators deliver a bandwidth which is too small, whereas the estimates using the $2^{nd}$ and $3^{rd}$ moments are too big. The min-CvM bandwidths appear to be the best in this case. 

In general, the min-CvM bandwidth with $\alpha=0$ (min-AD) can be recommended for densities that exhibit significant departures from normality, although it is advisable to examine the profile of the criterion over the relevant range of $b$ to ensure the correct minimum is selected. For regularly shaped densities, the moment-based estimators may be preferred. An estimate of the resultant density or d.f. of $V_{i}$ can be used as a further diagnostic tool to ascertain the adequacy of the chosen bandwidth by either a visual examination or comparison of the distances from $l_{n}$ or $L_{n}$. 

\renewcommand{\arraystretch}{0.5} 

\renewcommand{\thefigure}{\arabic{figure} (Cont.)}
\addtocounter{figure}{-1}
\begin{figure}[htbp]\centering
\begin{tabular}{@{}>{\footnotesize}c@{}}
\#4: Kurtotic unimodal\tabularnewline
\includegraphics[width=\linewidth]{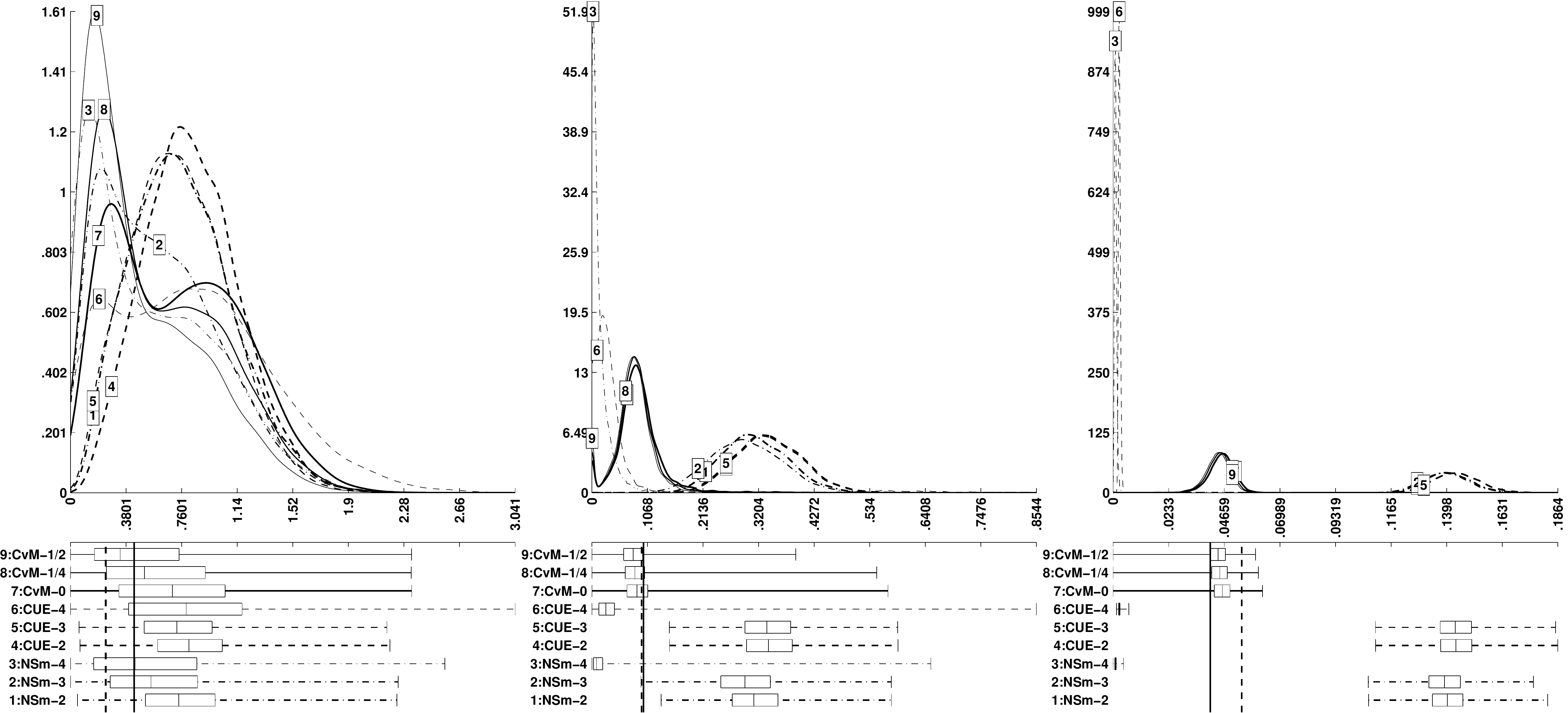} \tabularnewline
\#5: Outlier\tabularnewline
\includegraphics[width=\linewidth]{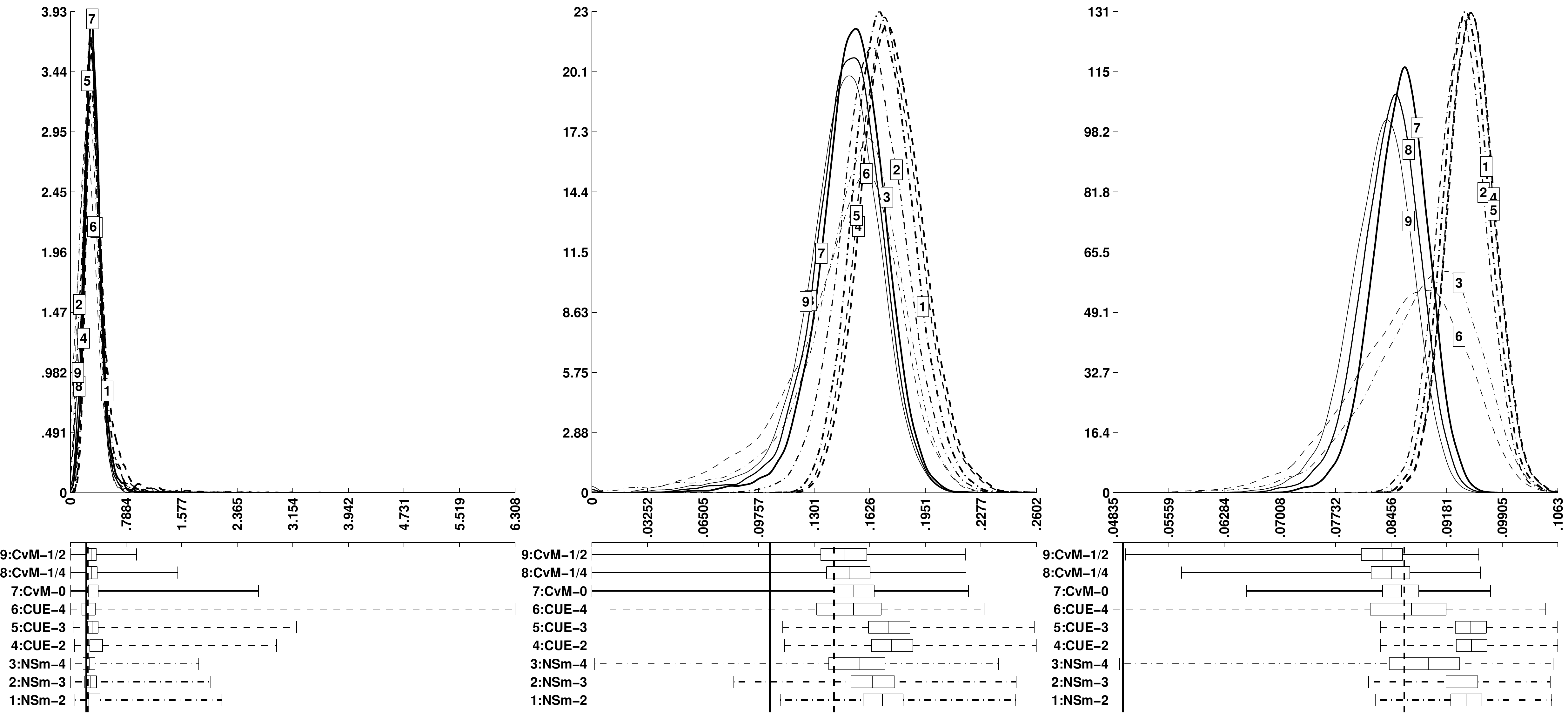} \tabularnewline
\#6: Bimodal\tabularnewline
\includegraphics[width=\linewidth]{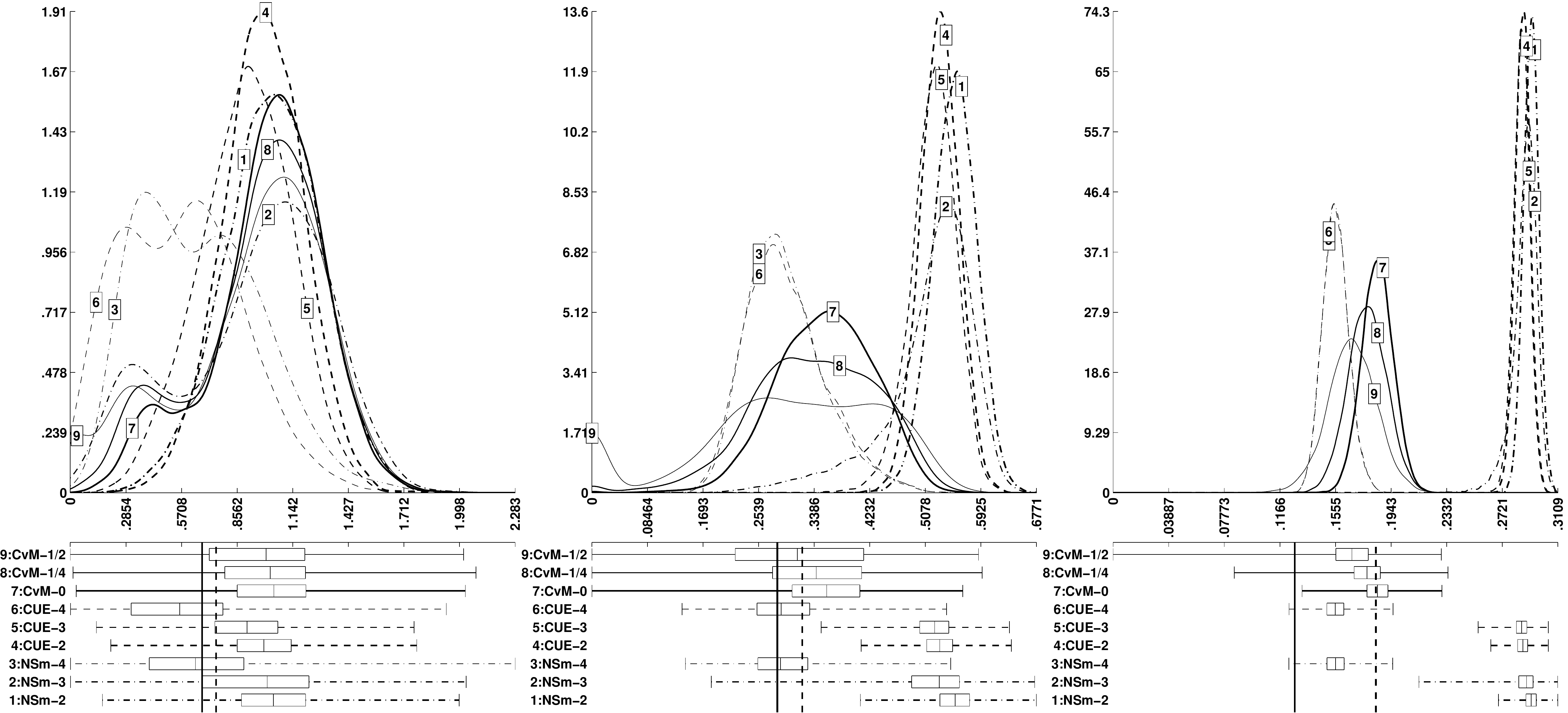} 
\end{tabular}
\caption{Performance of selected iMaxEnt bandwidth estimators}
\end{figure}
\renewcommand{\thefigure}{\arabic{figure}}

\begin{figure}[htbp]\centering
\begin{tabularx}{\linewidth}{*{3}{@{}>{\centering\arraybackslash}X@{}}}
\multicolumn{3}{@{}c@{}}{\#1: Gaussian}\tabularnewline
\multicolumn{3}{@{}c@{}}{\includegraphics[width=\linewidth]{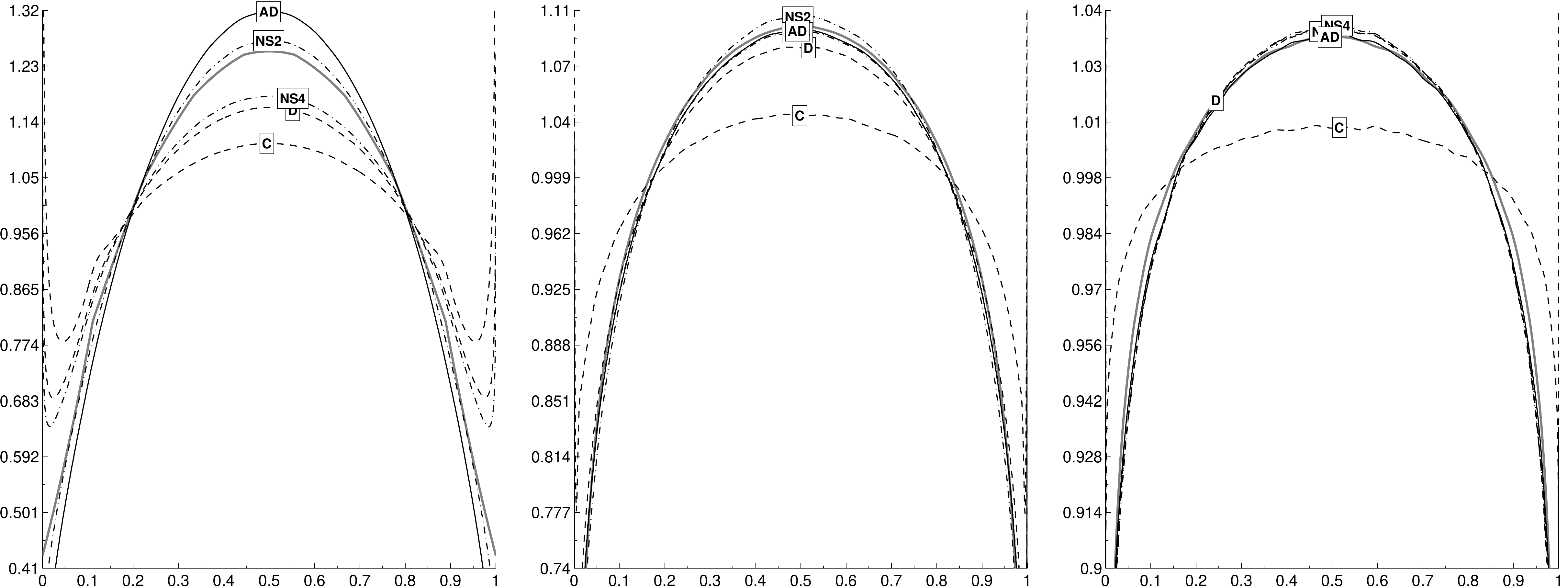}}\tabularnewline
\begin{tabular}{*{2}{>{\scriptsize}l}}
 Bandwidths: &NS2: 0.9374 \tabularnewline
C: 0.6495 &NS4: 0.7915 \tabularnewline
D: 0.7585 &AD:~  1.0091 \tabularnewline
\end{tabular}&
\begin{tabular}{*{2}{>{\scriptsize}l}}
 Bandwidths: &NS2: 0.4931 \tabularnewline
C: 0.3147 &NS4: 0.4687 \tabularnewline
D: 0.4455 &AD:~  0.4712 \tabularnewline
\end{tabular}&
\begin{tabular}{*{2}{>{\scriptsize}l}}
 Bandwidths: &NS2: 0.2717 \tabularnewline
C: 0.1517 &NS4: 0.2740 \tabularnewline
D: 0.2723 &AD:~  0.2648 \tabularnewline
\end{tabular}\tabularnewline[5mm]
\multicolumn{3}{@{}c@{}}{\#2: Skewed unimodal}\tabularnewline
\multicolumn{3}{@{}c@{}}{\includegraphics[width=\linewidth]{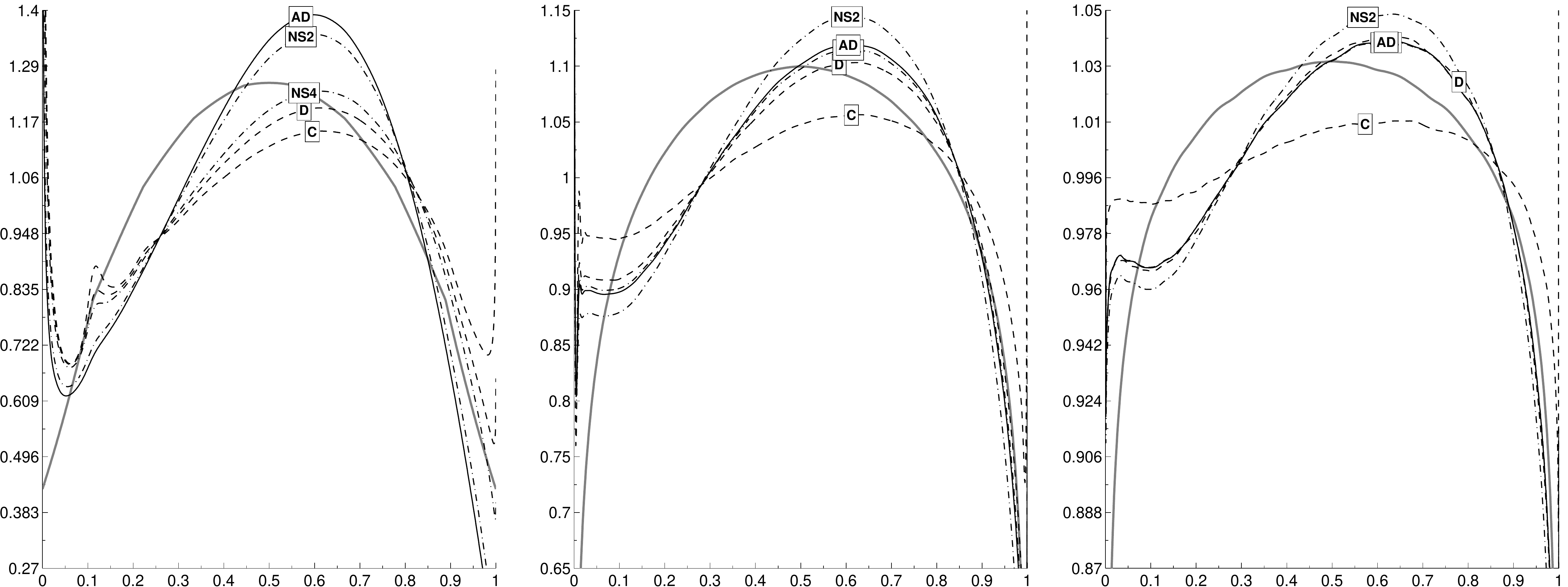}}\tabularnewline
\begin{tabular}{*{2}{>{\scriptsize}l}}
 Bandwidths: &NS2: 0.8636 \tabularnewline
C: 0.5896 &NS4: 0.7095 \tabularnewline
D: 0.6602 &AD:~  0.9145 \tabularnewline
\end{tabular}&
\begin{tabular}{*{2}{>{\scriptsize}l}}
 Bandwidths: &NS2: 0.4465 \tabularnewline
C: 0.2764 &NS4: 0.3955 \tabularnewline
D: 0.3743 &AD:~  0.4034 \tabularnewline
\end{tabular}&
\begin{tabular}{*{2}{>{\scriptsize}l}}
 Bandwidths: &NS2: 0.2454 \tabularnewline
C: 0.1317 &NS4: 0.2211 \tabularnewline
D: 0.2257 &AD:~  0.2218 \tabularnewline
\end{tabular}\tabularnewline[5mm]
\multicolumn{3}{@{}c@{}}{\#3: Strongly skewed}\tabularnewline
\multicolumn{3}{@{}c@{}}{\includegraphics[width=\linewidth]{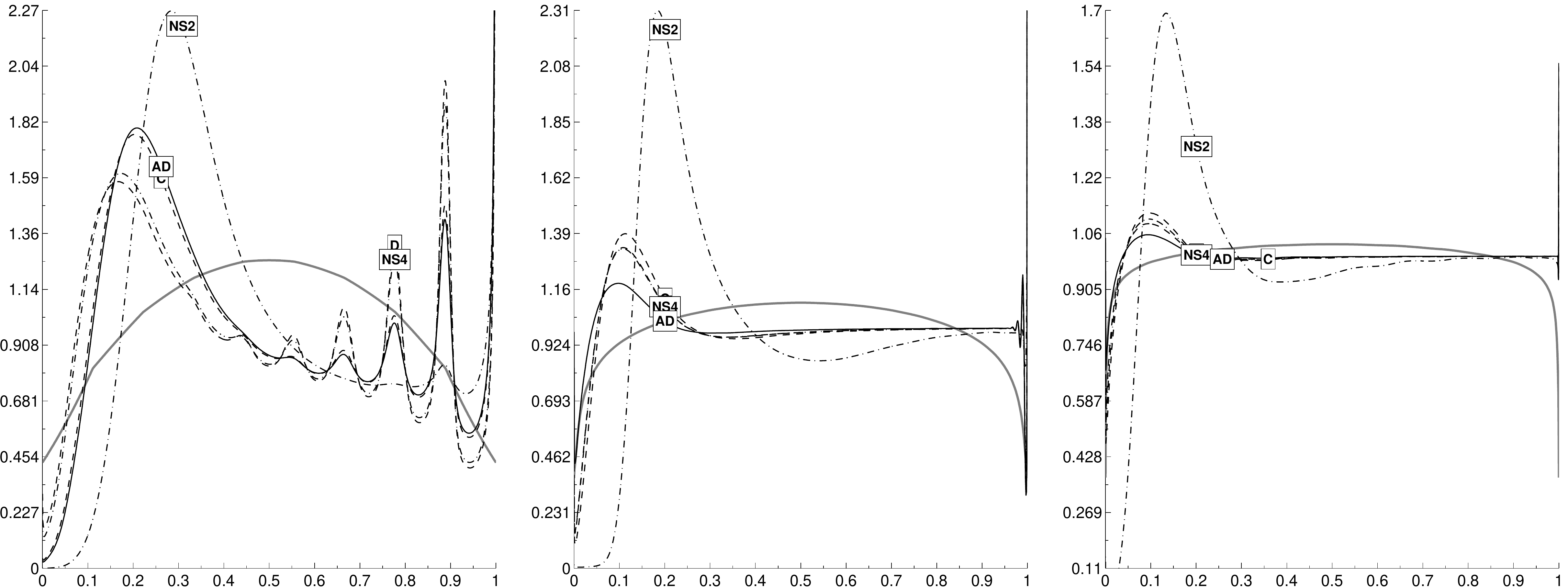}}\tabularnewline
\begin{tabular}{*{2}{>{\scriptsize}l}}
 Bandwidths: &NS2: 0.7446 \tabularnewline
C: 0.3496 &NS4: 0.2528 \tabularnewline
D: 0.2355 &AD:~  0.3673 \tabularnewline
\end{tabular}&
\begin{tabular}{*{2}{>{\scriptsize}l}}
 Bandwidths: &NS2: 0.3149 \tabularnewline
C: 0.0904 &NS4: 0.0787 \tabularnewline
D: 0.0796 &AD:~  0.0523 \tabularnewline
\end{tabular}&
\begin{tabular}{*{2}{>{\scriptsize}l}}
 Bandwidths: &NS2: 0.1405 \tabularnewline
C: 0.0338 &NS4: 0.0363 \tabularnewline
D: 0.0399 &AD:~  0.0266 \tabularnewline
\end{tabular}\tabularnewline[5mm]
\end{tabularx}
\caption{Marginal density of PITs for selected bandwidth values}
\label{Fig:SimLOOPITGKmargDens.1}
\end{figure}

\renewcommand{\thefigure}{\arabic{figure} (Cont.)}
\addtocounter{figure}{-1}
\begin{figure}[htbp]\centering
\begin{tabularx}{\linewidth}{*{3}{@{}>{\centering\arraybackslash}X@{}}}
\multicolumn{3}{@{}c@{}}{\#4: Kurtotic unimodal}\tabularnewline
\multicolumn{3}{@{}c@{}}{\includegraphics[width=\linewidth]{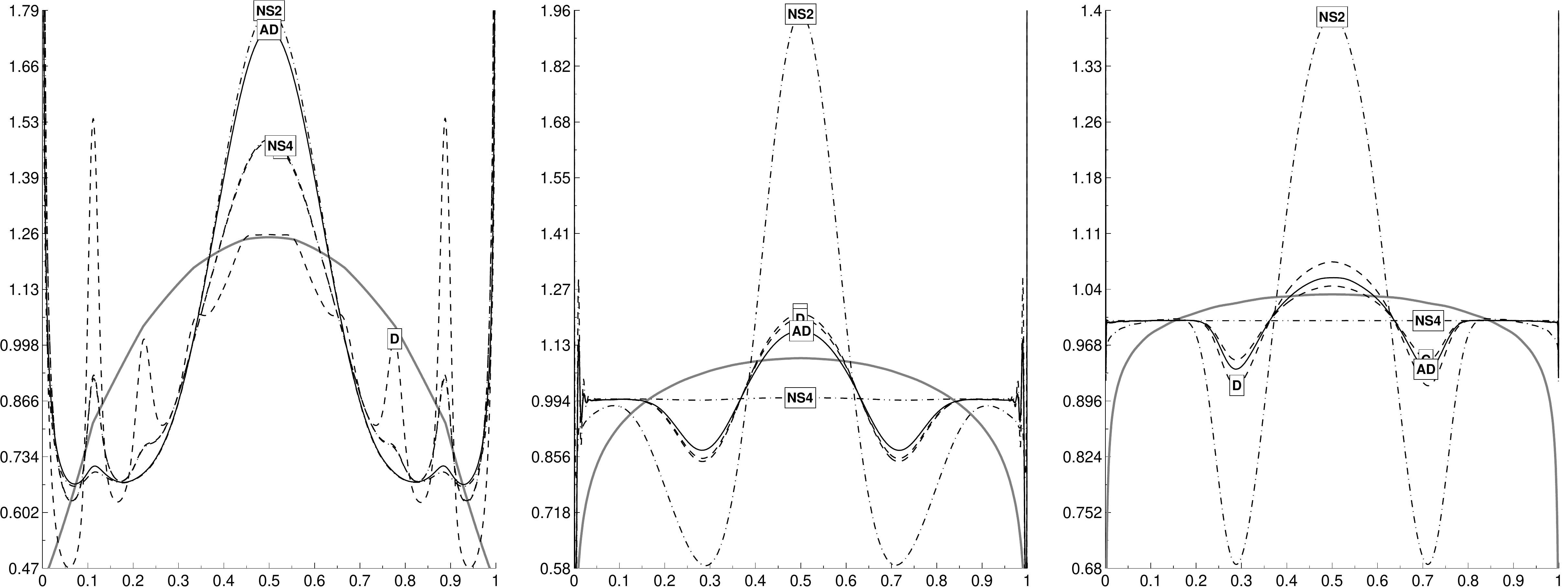}}\tabularnewline
\begin{tabular}{*{2}{>{\scriptsize}l}}
 Bandwidths: &NS2: 0.7414 \tabularnewline
C: 0.4362 &NS4: 0.4418 \tabularnewline
D: 0.2419 &AD:~  0.6980 \tabularnewline
\end{tabular}&
\begin{tabular}{*{2}{>{\scriptsize}l}}
 Bandwidths: &NS2: 0.3113 \tabularnewline
C: 0.0993 &NS4: 0.0091 \tabularnewline
D: 0.0958 &AD:~  0.0868 \tabularnewline
\end{tabular}&
\begin{tabular}{*{2}{>{\scriptsize}l}}
 Bandwidths: &NS2: 0.1400 \tabularnewline
C: 0.0407 &NS4: 0.0010 \tabularnewline
D: 0.0539 &AD:~  0.0458 \tabularnewline
\end{tabular}\tabularnewline[5mm]
\multicolumn{3}{@{}c@{}}{\#5: Outlier}\tabularnewline
\multicolumn{3}{@{}c@{}}{\includegraphics[width=\linewidth]{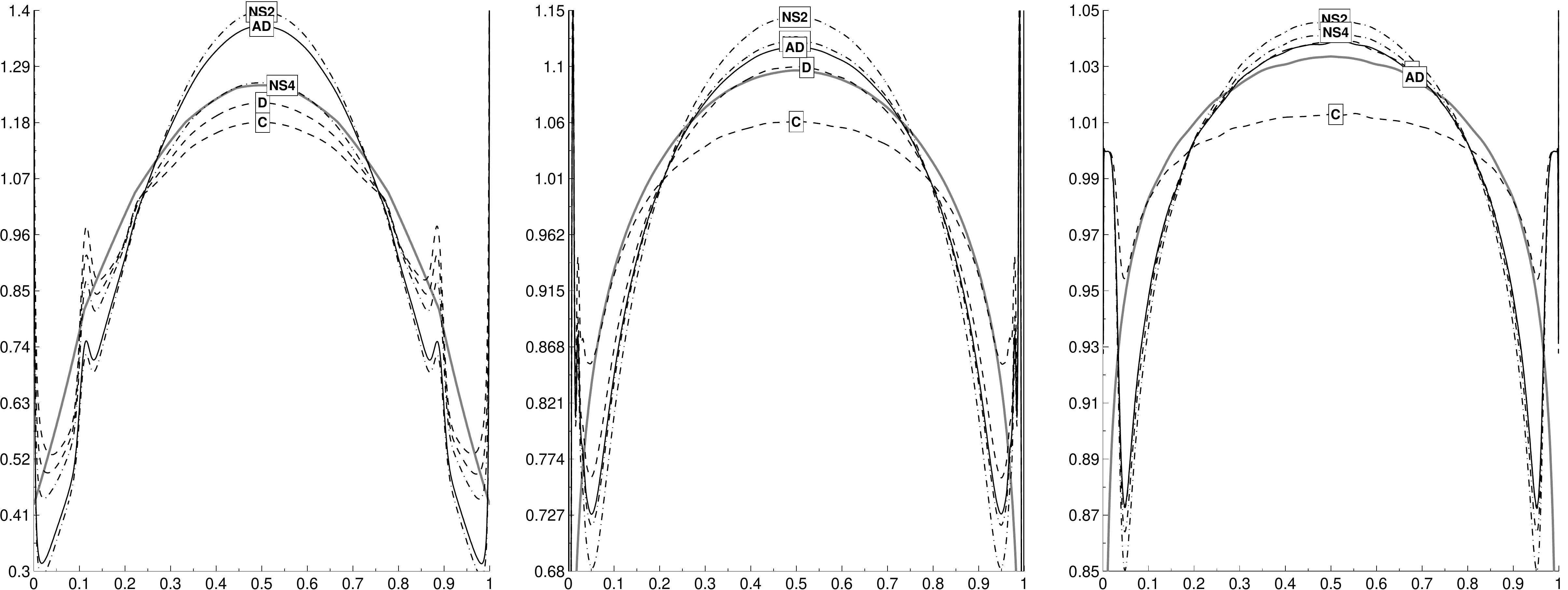}}\tabularnewline
\begin{tabular}{*{2}{>{\scriptsize}l}}
 Bandwidths: &NS2: 0.3331 \tabularnewline
C: 0.2250 &NS4: 0.2664 \tabularnewline
D: 0.2457 &AD:~  0.3202 \tabularnewline
\end{tabular}&
\begin{tabular}{*{2}{>{\scriptsize}l}}
 Bandwidths: &NS2: 0.1700 \tabularnewline
C: 0.1042 &NS4: 0.1569 \tabularnewline
D: 0.1418 &AD:~  0.1534 \tabularnewline
\end{tabular}&
\begin{tabular}{*{2}{>{\scriptsize}l}}
 Bandwidths: &NS2: 0.0944 \tabularnewline
C: 0.0496 &NS4: 0.0894 \tabularnewline
D: 0.0863 &AD:~  0.0859 \tabularnewline
\end{tabular}\tabularnewline[5mm]
\multicolumn{3}{@{}c@{}}{\#6: Bimodal}\tabularnewline
\multicolumn{3}{@{}c@{}}{\includegraphics[width=\linewidth]{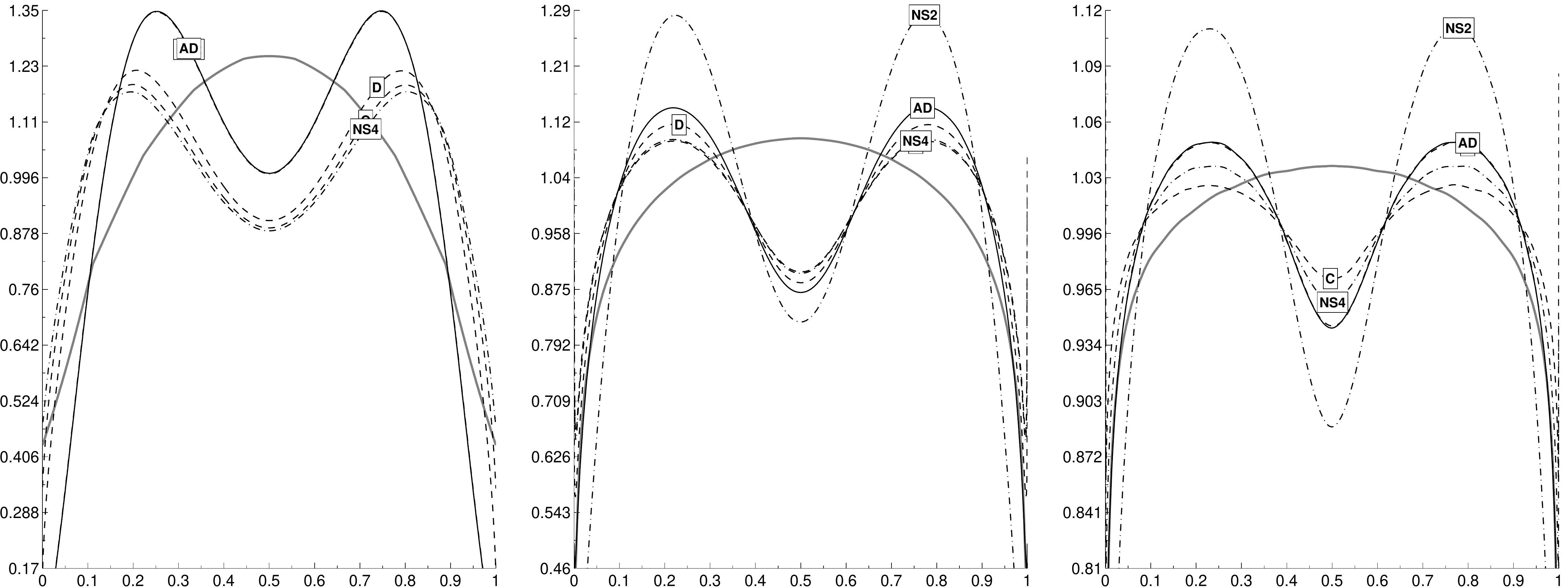}}\tabularnewline
\begin{tabular}{*{2}{>{\scriptsize}l}}
 Bandwidths: &NS2: 1.0421 \tabularnewline
C: 0.6762 &NS4: 0.6422 \tabularnewline
D: 0.7477 &AD:~  1.0449 \tabularnewline
\end{tabular}&
\begin{tabular}{*{2}{>{\scriptsize}l}}
 Bandwidths: &NS2: 0.5532 \tabularnewline
C: 0.2825 &NS4: 0.2875 \tabularnewline
D: 0.3207 &AD:~  0.3578 \tabularnewline
\end{tabular}&
\begin{tabular}{*{2}{>{\scriptsize}l}}
 Bandwidths: &NS2: 0.2923 \tabularnewline
C: 0.1270 &NS4: 0.1554 \tabularnewline
D: 0.1838 &AD:~  0.1848 \tabularnewline
\end{tabular}\tabularnewline[5mm]
\end{tabularx}
\caption{Marginal density of PITs for selected bandwidth values}
\end{figure}
\renewcommand{\thefigure}{\arabic{figure}}
\renewcommand{\arraystretch}{1} 

\section{Concluding remarks}\label{Sec:Conclusions}
This paper proposes a framework for choosing a single bandwidth for kernel estimators of both density and distribution functions motivated by the role of probability integral transforms in classical parametric modelling and their relationship to entropy maximisation. 
The leave-one-out estimates of a d.f. at sample values, $V_{i}$'s, are the analogues of PITs in parametric models. If the maximum entropy distribution were achieved, the $V_{i}$'s would be jointly uniformly distributed over the rescaled regular permutohedron $\Pi_{n}$. In general this cannot be achieved in finite sample, but the bandwidth can be chosen so that the distribution of $V_{i}$'s is close to the MaxEnt distribution. 
The proposed estimation procedures for the iMaxEnt bandwidth are fully operationalisable for finite $n$. 
Several estimators are examined in a simulation study, and the bandwidth minimising the Anderson-Darling version of the Cram\'{e}r-von Mises discrepancy is found to perform reliably and---in a loose sense---robustly  for a variety of distribution shapes.

Validity of Conjecture \ref{Conj:Limit.Distr} remains an open question that further research could usefully address. 
If it holds, the resultant iMaxEnt bandwidth will be found to asymptotically occupy the in-between position relative to the KDE and KDFE MISE-minimising bandwidths, reinforcing the use of the iMaxEnt procedure as a unifying framework for kernel estimation of density and distribution functions. However, the absence of the proof is not a hindrance for the practical implementation of the proposed method in finite samples.

\section*{Acknowledgements}
The author would like to thank Johan Koskinesn, Bent Nielsen, and Alexei Onatski for helpful comments. An earlier version of the paper was presented at the conferences and workshops in Cambridge, Bristol, 
Oxford, and Manchester. The comments of the participants are gratefully acknowledged. 
Any errors are the responsibility of the author.

\section*{Appendices}
\begin{appendix}
\numberwithin{equation}{section}
\numberwithin{lemma}{section}
\numberwithin{conjecture}{section}
\footnotesize

\section{Uniform distribution over a regular permutohedron}
\label{Sec:Permutohedron}
\subsection{Exact calculation of the marginal density}
The  expression for the marginal density given in Lemma \ref{Lemma:Unif.Pn.Marg} is exact but requires calculation of volume of certain permutohedra. 
The latter can be computed exactly using the following result due to  \citet[Theorem 3.1.]{postnikov2009}. 
Suppose that $x_{1}>\cdots>x_{n}$, and let $\lambda_{1},\ldots,\lambda_{n}\in\R$ be fixed distinct numbers. Then the volume of the permutohedron $P_{n} = P_{n}(x_{1},\ldots,x_{n})$ is equal to 
\begin{equation}
\label{Eq:Vol:Postnikov.Thm.3.1}
\Vol[n-1]{P_{n}} = \frac{1}{(n-1)!}\sum_{\sigma\in S_{n}}\frac{\left(\lambda_{\sigma(1)}x_{1}+\cdots+\lambda_{\sigma(n)}x_{n}\right)^{n-1}}
{(\lambda_{\sigma(1)}-\lambda_{\sigma(2)})(\lambda_{\sigma(2)}-\lambda_{\sigma(3)})\cdots(\lambda_{\sigma(n-1)}-\lambda_{\sigma(n)})},
\end{equation}
where $S_{n}$ is the symmetric group. 
(See also Theorem 3.2 and section 16 in \citet{postnikov2009} for alternative expressions for the volume of permutohedra.)
The $\lambda_{i}$'s in \eqref{Eq:Vol:Postnikov.Thm.3.1} cancel out and the resultant expression for $\Vol[n-1]{P_{n}}$ is a homogeneous polynomial of degree $n-1$ in $x_{1},\ldots,x_{n}$. 
For a given $n$, the marginal density \eqref{Eq:Marginal.Finite.n} is then a piecewise $(n-2)$-degree polynomial on intervals $[(j-1)/(n-1),\;j/(n-1)]$, $j=1,\ldots,n-1$. For example, $l_{2}(u)=1$ if $u\in[0,1]$, 
\begin{equation*}
 l_{3}(u) = \begin{cases}
 \frac{2}{3}\left(2u + 1\right) & \;\text{if}\;u\in[0;1/2]\\
 \frac{2}{3}\left(3 - 2u\right) & \;\text{if}\;u\in[1/2;1]\\
\end{cases}
\end{equation*}
\begin{equation*}
 l_{4}(u) = \begin{cases}
 \frac{3}{32}\left(9u^2 + 18u + 6\right) & \;\text{if}\;u\in[0;1/3]\\
 \frac{3}{32}\left(- 18u^2 + 18u + 9\right) & \;\text{if}\;u\in[1/3;2/3]\\
 \frac{3}{32}\left(9u^2 - 36u + 33\right) & \;\text{if}\;u\in[2/3;1]\\
\end{cases}
\end{equation*}
and so on. 

\subsection{Approximation of the marginal density by simulation}
As $n$ increases, it quickly becomes impractical to use expression \eqref{Eq:Vol:Postnikov.Thm.3.1} as it involves summation over $n!$ permutations. One can then resort to numerical approximation of $l_{n}$ based on random draws from a uniform distribution over $\Pi_{n}$ which can be obtained using a simple rejection method. Specifically, draw random variates $u_{1},\ldots,u_{n-1}$ from a uniform distribution over a cube $I^{n-1}$, set $u_{n}=n/2-(u_{1}+\cdots+u_{n-1})$, and reject if $u=(u_{1},\ldots,u_{n})\transp\notin I^{n}$. The accepted draws are distributed uniformly over $I^{n}\cap H_{\ones{n},n/2}$, the central section of $I^{n}$. The draws are then rejected\footnote{To reduce the number of queries to the membership oracle, one can also reject $u$ if $\norm{u-\ones{n}/2}>\sqrt{\frac{n(n+1)}{12(n-1)}}$, which is the circumradius of $\Pi_{n}$.}
 if $u\notin\Pi_{n}$ by calling a membership oracle. 

The following result due to \citet{rado1952} requires only sorting to verify membership of $u$ in $\Pi_{n}$. This can be done only in $\bigO{n\log n}$ operations.
For any $x\in\R^{n}$ let $x_{[1]}\geq\cdots\geq x_{[n]}$ denote the components of $x$ in decreasing order (reverse order statistics). Then for $x,y\in\R^{n}$, $x$ is said to be majorized by $y$ ($y$ majorizes $x$), denoted as $x\prec y$ (or $y\succ x$), if
\begin{equation*}
\sum_{i=1}^{k}x_{[i]}\leq\sum_{i=1}^{k}y_{[i]},\quad k=1,\ldots,n,
\end{equation*}
where there is equality with $k=n$. 
Furthermore, $x\prec y$ if and only if $x\in P_{n}(y)$. 
That is, to verify if $u\in\Pi_{n}$, one has only to verify if $u\in H_{\ones{n},n/2}$ is majorized by $(1,(n-2)/(n-1),(n-3)/(n-1),\ldots,1/(n-1),0)$. 
The overall acceptance rate of the above method is $\bigO{n^{-1}}$. 

The density $l_{n}$ and the corresponding d.f., moments, or any other functionals of $l_{n}$ can be approximated from the simulated data by the standard techniques. 

\section{Sarda (1993) cross-validation criterion}\label{S.Sec:KCDFE.CV}
\citet{sarda1993} proposed choosing the bandwidth for KDFE by minimising the following cross-validation (CV) criterion: 
\begin{equation}\label{Eq:CV.crit.Sarda-1993}
\CV_{\psi}(b) = n^{-1}\sum_{i=1}^{n}\left[V_{i}(b)-F_{n}(x_{i})\right]^{2}\psi(x_{i}),
\end{equation}
where $\psi$ is a non-negative weight function. $\CV_{\psi}(b)$ is designed to approximate the weighted MISE
\begin{equation*}
\MISE_{\psi} \widehat{F}(\cdot;b) = \E \int_{-\infty}^{\infty}\left[\widehat{F}(x;b)-F(x)\right]^{2}\psi(x)dF(x).
\end{equation*}

Summing in \eqref{Eq:CV.crit.Sarda-1993} over $i$ in the order of $x_{(i)}$ (see footnote \ref{Fn:KCDFE.PIT.Ordering}), it can be seen that the CV criterion with $\psi(x)\equiv1$ is computationally equivalent (up to an additive constant) to the classical CvM discrepancy between the EDF of $V$, $G_{n}$, and the uniform d.f. on $[0,1]$, $R$, viz. $\CV_{1}(b) = \omega^{2}(G_{n},R)+1/(6n^{2})$. 
\begin{lemma}\label{Lemma:KCDFE:Sarda.CV.CvM}
If Assumptions \ref{Ass:DGP}(a,b), \ref{Ass:kernel}(a,b), and \ref{Ass:bandwidth} hold, then as $n\to\infty$ and $b\to0$,
\begin{equation*}\begin{split}
\omega^{2}(G,R) 
& = \frac{1}{4}\mu_{2}^{2}(k)\zeta_{1}(F)b^{4} - \frac{1}{2}\mu_{2}(k)R_{3}(f)b^{2}n^{-1} +\frac{1}{12}n^{-2}
+\frac{1}{2}\mu_{2}(k)\psi_{2,1}(K)\zeta_{2}(F)b^{3}n^{-1} \\&\quad
 -\psi_{2,1}(K)R_{2}(f)bn^{-2}  + \bigO{\max(b^{4}n^{-1},b^{6},b^{2}n^{-2},n^{-3})},
\end{split}\end{equation*}
where $R_{j}(f) = \int f(x)^{j}dx$, and  $\zeta_{1}(F)$ and $\zeta_{1}(F)$ are constants depending on $F$ defined in \eqref{Eq:CvM.Zeta.Def}. 
\hfill$\square$
\end{lemma}

In view of Lemma \ref{Lemma:KCDFE.PIT.Moments}, one expects the bandwidth obtained by minimising a discrepancy between $G$ and the uniform distribution to be of order $n^{-1/2}$, which is indeed the same as the bandwidth minimising the two leading terms in $\omega^{2}(G,R)$. 
However, $\E\CV_{1}(b) = \omega^{2}(G,R)+\IVar G_{n}+1/(6n^{2})$, where $\IVar G_{n}=\int_{0}^{1}\E\left[G_{n}(t)-G(t)\right]^{2}dt$ is the integrated variance of $G_{n}$. This adds an extra term of order $b/n$, entering $\E\CV_{1}$ with a positive sign. This term dominates asymptotically and therefore $\E\CV_{1}$ is an increasing function of $b$, minimized by $b=0$. 
This provides an alternative confirmation of the failure of $\CV_{1}$ first pointed out in \citet{altman1995}.
In can be shown that the offending term is due to the correlation between $\I{V_{i}\leq t}$ and $\I{V_{j}\leq t}$, $i\neq j$.

\section{Proofs}\label{App:Proofs}

\noindent\textbf{Proof of Lemma \ref{Lemma:KCDFE.PIT.Support}}. 
By Assumption \ref{Ass:kernel}(a), $K_{b}(0)=1/2$ and $K_{b}(-z)=1-K_{b}(z)$. Collecting the $\binom{n}{2}$ values of $\kappa_{ij}=K_{b}(X_{i}-X_{j})$, $i,j=1,\ldots,n$, $j>i$, in an $n(n-1)/2$-vector $\kappa = \left(\kappa_{12},\ldots,\kappa_{1n},\kappa_{23},\ldots,\kappa_{2n},\ldots,\kappa_{(n-1)n}\right)\transp$, the leave-one-out PITs \eqref{Eq:KCDFE.PIT.def}, $V=(V_{1},\ldots,V_{n})\transp$, can be written as 
$(n-1)V +1= p(\kappa) = A(2\kappa-1) + (n+1)/2$, where $A=[a_{1},\ldots,a_{n(n-1)/2}]$ is an $n\times n(n-1)/2$ matrix with $a_{l}=(e_{i}-e_{j})/2$, $i=1,\ldots,n-1$, $j=i+1,\ldots,n$, where $e_{i},\ldots,e_{n}$ are the standard basis vectors in $\R^{n}$. 

Remark now that $2\kappa-1$ lies in the $n(n-1)/2$-cube $C^{n(n-1)/2}=[-1,1]^{n(n-1)/2}$, and the image of $C^{n(n-1)/2}$ under the affine projection $p(\kappa)$ is the permutohedron $P_{n}(1,2,\ldots,n)$; see \citet[Sec.7.3]{ziegler1995}. Therefore, the support of $V$ is  $\Pi_{n}$, as required.\hfill$\blacksquare$


\vspace*{0.5\baselineskip}

\noindent\textbf{Proof of Lemma \ref{Lemma:KCDFE.PIT.Moments}}.
If Assumptions \ref{Ass:DGP}(a,b) and \ref{Ass:kernel}(a,b) hold, then by an expansion around $x$ as $b\to0$, for $r=1,2,\ldots$,
\begin{equation*}\begin{split}
\E_{X_{1}}K_{b}^{r}(x-X_{1})  
& = \int_{-\infty}^{\infty}rK^{r-1}(z)k(z)F(x-bz)dz \\
& =  F(x) -\psi_{r,1}(K)f(x)b + \frac{1}{2}\psi_{r,2}(K)f^{(1)}(x)b^{2} - \frac{1}{6}\psi_{r,3}(K)f^{(2)}(x)b^{3} + \bigO{b^{4}},
\end{split}\end{equation*}
where for $j=0,1,2,\ldots$, $\psi_{r,j}(K) = \int_{-\infty}^{\infty}z^{j}rK^{r-1}(z)k(z)dz$; 
$\psi_{r,0}=1$ and $\psi_{1,j} = \mu_{j}(k)$, which is zero if $j$ is odd by symmetry of $k$. 
Then $\E_{X_{1}}K_{b}(x-X_{1})   =  F(x) + \frac{1}{2}\mu_{2}(k)f^{(1)}(x)b^{2}  + \bigO{b^{4}}$, and $\E_{X_{1}}K_{b}^{2}(x-X_{1})  
 =  F(x) -\psi_{2,1}(K)f(x)b + \bigO{b^{2}}$. 
Thus, for $r\geq2$,
\begin{equation*}\begin{split}
\E V_{1}^{r} & = \int_{-\infty}^{\infty}\left[\frac{1}{(n-1)^r}\sum_{j_{1}=2}^{n}\cdots\sum_{j_{r}=2}^{n}\E_{X_{j_{1}},\ldots,X_{j_{r}}}\prod_{i=1}^{r}K_{b}(x-X_{j_{i}})\right]f(x)dx \\
& =   \left[1-\binom{r}{2}n^{-1}\right]\int_{-\infty}^{\infty}\left[\E_{X_{1}}K_{b}(x-X_{1})\right]^{r}f(x)dx
\\&\qquad  + n^{-1}\binom{r}{2}\int_{-\infty}^{\infty}\left[\E_{X_{1}}K_{b}(x-X_{1})\right]^{r-2}\left[\E_{X_{1}}K_{b}^{2}(x-X_{1})\right]f(x)dx
		+\bigO{n^{-2}}  \\
& = \frac{1}{r+1} - \frac{1}{2}\mu_{2}(k)\xi_{2,r}(F)b^{2}   +\frac{r-1}{2(r+1)}n^{-1}  -\psi_{2,1}(K)\xi_{1,r}(F)bn^{-1} 
  + \bigO{\max(b^{2}n^{-1},h^{4},n^{-2})},
\end{split}\end{equation*}
where $\xi_{2,r}(F) = - \int_{-\infty}^{\infty}rF^{r-1}(x)f^{(1)}(x)f(x)dx$ and $\xi_{1,r}(F) = \int_{-\infty}^{\infty}\frac{r(r-1)}{2}F^{r-2}(x)f^{2}(x)dx$. 
Furthermore, integration by parts yields $\xi_{2,r}(F) = \frac{r(r-1)}{2}\int_{-\infty}^{\infty}F^{r-2}(x)f^{3}(x)dx > 0$. 
Finally, by H\"{o}lder's inequality with exponents $1$ and $\infty$, $\xi_{2,r}(F)\leq \frac{r}{2}\sup_{x}f^{2}(x)$, and $\xi_{1,r}(F)\leq \frac{r}{2}\sup_{x}f(x)$.
\hfill$\blacksquare$


\vspace*{0.5\baselineskip}

\noindent\textbf{Proof of Lemma \ref{Lemma:Unif.Pn.Marg}}. 
Since $\Vol[n-1]{\Pi_{n}}=n^{n-2}/(n-1)^{n-1}$, for $u\in[0,1]$, $l_{n}(u)=(n-1)^{n-1}\Vol[n-2]{\Pi_{n}\cap H_{e_{1},u}}/n^{n-2}$. Recognising that the sections $\Pi_{n}\cap H_{e_{1},u}$ are themselves permutohedra and applying Theorem 1 in \citet{gaiha1977}, we obtain for $u\in\left[\frac{j-1}{n-1},\frac{j}{n-1}\right]$, $j=1,\ldots,n-1$, 
\begin{equation*}
\Pi_{n}\cap H_{e_{1},u} = (n-1)^{-1}P_{n-1}(n-1,n-2,\ldots,j+1,2j-1-(n-1)u,j-2,\ldots,0),
\end{equation*}
which gives \eqref{Eq:Marginal.Finite.n} as required. 
Noting that the sections $\Pi_{n}\cap H_{e_{1},0}$ and $\Pi_{n}\cap H_{e_{1},1}$ are translated copies of $\Pi_{n-1}$, one immediately obtains that the marginal density  at $0$ and $1$ is $l_{n}(0) = l_{n}(1) = \left(1-1/n\right)^{n-2} \xrightarrow{n\to\infty} e^{-1}$.
\hfill$\blacksquare$


\vspace*{0.5\baselineskip}

\noindent\textbf{Proof of Lemma \ref{Lemma:KCDFE:Sarda.CV.CvM}}. 
Let $\rho_{r}(v)$, $r=0,1,2,\ldots$, be the shifted Legendre polynomials  orthonormal with respect to the uniform density on the unit interval, viz. $\int_{0}^{1}\rho_{r}(v)\rho_{s}(v)dv=\delta_{rs}$, where $\delta_{rs}$ is the Kronecker delta (see footnote \ref{Fn:Legendre.polynomials}). Let $\rho_{r}^{(s)}(v) = \frac{d^{s}}{dv^{s}}\rho_{r}(v)$ for $s=0,1,2,\ldots$ denote the derivatives, and $\rho_{r}^{(-1)}(v) = \int_{0}^{v}\rho_{r}(t)dt$ be the first antiderivative of $\rho_{r}$.
Then $q(v)$ can be expanded into a generalised Fourier series as $q(v) = 1+\sum_{r=1}^{\infty}c_{r}\rho_{r}(v)$, where the coefficients are given by 
\begin{equation*}
c_{r} = \int_{0}^{1}\rho_{r}(v)g(v)dv 
= (-1)^{r}\sqrt{2r+1}\sum_{i=0}^{r}\binom{r}{i}\binom{r+i}{i}(-1)^{i} \E V_{1}^{i}.
\end{equation*}
Specifically, $c_{0}=1$, $c_{1}=0$, and for $r\geq2$, $c_{r}=\sqrt{2r+1}\tilde{c}_{r}$, with
\begin{equation*}
\tilde{c}_{r} = -\frac{1}{2}\mu_{2}(k)\tilde{\xi}_{2,r}(F)b^{2} + \I{r\;\text{even}}n^{-1}
-\psi_{2,1}(K)\tilde{\xi}_{1,r}(F)bn^{-1} + \bigO{\max(b^{2}n^{-1},b^{4},n^{-2})}, 
\end{equation*}
where $\tilde{\xi}_{2,r}(F) = \frac{1}{2}\int_{-\infty}^{\infty}\frac{\rho_{r}^{(2)}\left(F(x)\right)}{\sqrt{2r+1}}f^{3}(x)dx = \int_{-\infty}^{\infty}\frac{\rho_{r}\left(F(x)\right)}{\sqrt{2r+1}}f^{(2)}(x)dx$ and 
$\tilde{\xi}_{1,r}(F) = \frac{1}{2}\int_{-\infty}^{\infty}\frac{\rho_{r}^{(2)}\left(F(x)\right)}{\sqrt{2r+1}}f^{2}(x)dx$.
Note that $\tilde{\xi}_{2,2}(F) = 6R_{3}(f)$ and $\tilde{\xi}_{1,2}(F) = 6R_{2}(f)$.

Expanding $G(v) = v + \sum_{r=2}^{\infty}c_{r}\rho_{r}^{-1}(v)$ and using repeated integration by parts to find that for $n,m\geq2$, 
\begin{equation*}
\sqrt{2n+1}\sqrt{2m+1}\int_{0}^{1}\rho_{n}^{(-1)}(v)\rho_{m}^{(-1)}(v) dv = 
\begin{cases}
\frac{2n+1}{2(2n-1)(2n+3)} & \text{ if } n=m, \\
-\frac{1}{4(2n-1)} & \text{ if } \abs{n-m}=2, \\
0 & \text{ otherwise},
\end{cases}
\end{equation*}
yields
\begin{equation*}\begin{split}
\omega^{2}(G,R) 
& = \sum_{r=2}^{\infty}\sum_{s=2}^{\infty}\tilde{c}_{r}\tilde{c}_{s}\sqrt{2r+1}\sqrt{2s+1}\int_{0}^{1}\rho_{r}^{-1}(v)\rho_{s}^{-1}(v)dv 
= \sum_{r=2}^{\infty}\left[\frac{(2r+1)\tilde{c}_{r}^{2}}{2(2r-1)(2r+3)} - \frac{\tilde{c}_{r}\tilde{c}_{r+2}}{2(2r+3)}\right] \\
& = \frac{1}{4}\mu_{2}^{2}(k)\zeta_{1}(F)b^{4} - \frac{1}{2}\mu_{2}(k)\theta_{1}b^{2}n^{-1} +\theta_{2}n^{-2}
+\frac{1}{2}\mu_{2}(k)\psi_{2,1}(K)\zeta_{2}(F)b^{3}n^{-1} -\psi_{2,1}(K)\theta_{3}bn^{-2} \\
&\quad + \bigO{\max(b^{4}n^{-1},b^{6},b^{2}n^{-2},n^{-3})},
\end{split}\end{equation*}
where
\begin{equation}\label{Eq:CvM.Zeta.Def}\begin{split}
\zeta_{1}(F) & =\sum_{r=2}^{\infty}\left[\frac{(2r+1)}{2(2r-1)(2r+3)}\tilde{\xi}_{2,r}^{2}(F) -\frac{1}{2(2r+3)}\tilde{\xi}_{2,r}(F)\tilde{\xi}_{2,r+2}(F)\right], \\
\zeta_{2}(F) & =\sum_{r=2}^{\infty}\left[\frac{2(2r+1)}{2(2r-1)(2r+3)}\tilde{\xi}_{1,r}(F)\tilde{\xi}_{2,r}(F)
-\frac{1}{2(2r+3)}\left(\tilde{\xi}_{1,r+2}(F)\tilde{\xi}_{2,r}(F)+ \tilde{\xi}_{2,r+2}(F)\tilde{\xi}_{1,r}(F)\right)\right],
\end{split}\end{equation}
\begin{equation*}
\theta_{1}=\sum_{r=2}^{\infty}\left[\frac{\tilde{\xi}_{2,r}(F)}{2(2r-1)} -\frac{\tilde{\xi}_{2,r+2}(F)}{2(2r+3)}\right]\I{r\;\text{even}}
=\frac{1}{6}\tilde{\xi}_{2,2}(F)+\sum_{s=2}^{\infty}\frac{\tilde{\xi}_{2,2s}(F)}{2(4s-1)} -\sum_{s=1}^{\infty}\frac{\tilde{\xi}_{2,2(s+1)}(F)}{2(4(s+1)-1)}
= R_{3}(f),
\end{equation*}
\begin{equation*}
\theta_{2} = \sum_{r=2}^{\infty}\frac{\I{r\;\text{even}}}{(2r-1)(2r+3)}
= \sum_{s=1}^{\infty}\frac{1}{(4(s-1)+3)(4s+3)} = \frac{1}{12},
\end{equation*}
\begin{equation*}
\theta_{3}=\sum_{r=2}^{\infty}\left[\frac{\tilde{\xi}_{1,r}(F)}{2(2r-1)} -\frac{\tilde{\xi}_{1,r+2}(F)}{2(2r+3)}\right]\I{r\;\text{even}}
=\frac{1}{6}\tilde{\xi}_{1,2}(F) + \sum_{s=2}^{\infty}\frac{\tilde{\xi}_{1,2s}(F)}{2(4s-1)} -\sum_{s=1}^{\infty}\frac{\tilde{\xi}_{1,2(s+1)}(F)}{2(4(s+1)-1)} = R_{2}(f),
\end{equation*}
as required \hfill$\blacksquare$

\section[~Beta-weighted Cram\'{e}r--von Mises criteria]{Beta-weighted Cram\'{e}r--von Mises criteria}\label{App:CvM.comp}
Let $\omega_{\alpha,\beta;\epsilon}^{2}$ denote the CvM discrepancy \eqref{Eq:WCvM.def} with $\psi(t)=t^{\alpha-1}(1-t)^{\beta-1}\I{\epsilon\leq t\leq 1-\epsilon}$, $\alpha,\beta\geq0$, $0\leq\epsilon<1/2$, and let $B(x;\alpha,\beta)=\int_{0}^{x}t^{\alpha-1}(1-t)^{\beta-1}dt$ and $B(\alpha,\beta)=B(1;\alpha,\beta)$ denote the incomplete and complete beta functions, respectively. 
Let $U_{i}=F(Z_{i})$ and $u_{(1)}\leq\cdots\leq u_{(n)}$ be the respective order statistics with $u_{(0)}\equiv0$ and $u_{(n+1)}\equiv1$. Then in view of the second equality in \eqref{Eq:WCvM.def} and because  $G_{n}\circ F^{-1}$ is the EDF of $U$, 
unless $\alpha=\beta=0$, 
\begin{equation}\label{S:Eq:CvM.beta.comp.form}
\omega_{\alpha,\beta;\epsilon}^{2} 
= \sum_{j=0}^{n}\int_{u_{(j)}}^{u_{(j+1)}}\left[\frac{j}{n}-t\right]^{2}\psi(t)dt
= \frac{1}{n}\sum_{j=j_{min}}^{j_{max}}\left[2B(u_{(j)};\alpha+1,\beta)  -\frac{2j-1}{n}B(u_{(j)};\alpha,\beta)\right] 
+ A_{\alpha,\beta;\epsilon},
\end{equation}
where $j_{min}$ is the smallest index $j=1,\ldots,n$ such that $u_{(j_{min})}\geq\epsilon$, $j_{max}$ is the greatest index such that $u_{(j_{max})}\leq1-\epsilon$, $u_{(j_{min}-1)}\equiv\epsilon$,  $u_{(j_{max}+1)}\equiv1-\epsilon$, and 
\begin{equation*}\begin{split}
A_{\alpha,\beta;\epsilon}  &= \int_{0}^{1-\epsilon} \left[t -\frac{j_{max}}{n}\right]^{2}t^{\alpha-1}(1-t)^{\beta-1}dt 
 -\int_{0}^{\epsilon}\left[t-\frac{j_{min}-1}{n}\right]^{2} t^{\alpha-1}(1-t)^{\beta-1}dt. \\
 & =  B(1-\epsilon;\alpha+2,\beta) - B(\epsilon;\alpha+2,\beta)
  \\ & \qquad
 -\frac{2j_{max}}{n}B(1-\epsilon;\alpha+1,\beta)  +\frac{2(j_{min}-1)}{n}B(\epsilon;\alpha+1,\beta)
 + \frac{j_{max}^{2}}{n^{2}}B(1-\epsilon;\alpha,\beta)  -\frac{(j_{min}-1)^{2}}{n^{2}}B(\epsilon;\alpha,\beta).
\end{split}\end{equation*}
When there is no trimming  ($\epsilon=0$), $j_{min}=1$, $j_{max}=n$, and $A_{\alpha,\beta;\epsilon}=B(\alpha,\beta+2)$.
Another special case of interest is the trimmed classical CvM when $\alpha=\beta=1$, viz.
\begin{equation*}
\omega_{1,1;\epsilon}^{2} = \frac{1}{n}\sum_{j=j_{min}}^{j_{max}}\left[u_{(j)}-\frac{2j-1}{2n}\right]^{2}
+\frac{j_{max}-j_{min}+1}{12n^{3}} 
+\frac{1}{3}\left(1-\epsilon-\frac{j_{max}}{n}\right)^{3} - \frac{1}{3}\left(\epsilon-\frac{j_{min}-1}{n}\right)^{3}.
\end{equation*}
(When $\epsilon=0$ the second term becomes $1/(12n^{2})$ and the last two terms are zero). 

For the case $\alpha=\beta=0$ (Anderson-Darling criterion) and $\epsilon>0$  we have
\begin{equation*}\begin{split}
  \omega_{0,0;\epsilon}^{2}
  & = \frac{1}{n}\sum_{j=j_{min}}^{j_{max}}\left[ -2\ln(1-u_{(j)}) -\frac{2j-1}{n}\left(\ln(u_{(j)}) -\ln(1-u_{(j)})\right)\right]
  \\ & \qquad  
  -1+2\epsilon + \left[\frac{j_{max}^{2}}{n^{2}} + \left(\frac{j_{min}-1}{n}-1\right)^{2}\right]\ln(1-\epsilon)
  - \left[\left(\frac{j_{max}}{n}-1\right)^{2}+ \frac{(j_{min}-1)^{2}}{n^{2}}\right]\ln(\epsilon).
\end{split}\end{equation*}
With $\epsilon=0$,   
$\omega_{AD}^{2}=\omega_{0,0;0}^{2}
   = -\frac{1}{n^{2}}\sum_{j=1}^{n}(2j-1)\left[\ln(u_{(j)})+\ln(1-u_{(n-j+1)})\right] -1$.


\end{appendix}

\addcontentsline{toc}{section}{References}

\clearpage
\addcontentsline{toc}{section}{Supplement}
\renewcommand{\thepage}{S.\arabic{page}}
\setcounter{page}{1}
\renewcommand{\thesection}{S.\arabic{section}}
\setcounter{section}{0}
\numberwithin{equation}{section}
\numberwithin{table}{section}
\numberwithin{figure}{section}

\title{Supplement to `Indirect Maximum Entropy Bandwidth'}
\author{Vitaliy Oryshchenko\textsuperscript{$\ast$}  \\ 
University of Manchester }
\maketitle

\section[2nd---10th even central moments of a random variable with density ln(u) for n=2,...,1,000,000]{$2^{\text{nd}}$---$10^{\text{th}}$ even central moments of a random variable with density $l_{n}(u)$ for $n=2,\ldots,1,000,000$} \label{OS:Subsec:Moments.Tab}
Tables~\ref{Tab:PnUnifCmomT1}--\ref{Tab:PnUnifCmomT4} list values of the first five even central moments of a random variable with density $l_{n}(u)$ for $n$ between $2$ and $1,000,000$. The values for $n=2,\ldots,11$ are exact. 
For $12\leq n\leq100,000$, the values are obtained by simulation from $\ceil{1e5\times n}/1e4$ random draws from a uniform distribution on $\Pi_{n}$ and exploiting the fact that all marginals are identical. For $n=177,828,\; 316,228,\; 562,341$, and $1,000,000$ the number of random draws is $6000,\; 4000,\; 900$, and $500$, respectively.

The values in the tables are displayed as $m_{2}\times1e5$, $m_{4}\times1e6$, $m_{6}\times1e6$, $m_{8}\times1e7$, and $m_{10}\times1e8$, rounded to a nearest integer. For example, for $n=2$, $m_{2}\approx0.08333$, $m_{4}\approx0.012500$,  $m_{6}\approx0.002232$,  $m_{8}\approx0.0004340$, and $m_{10}\approx0.00008878$. 

\begin{figure}[htbp]\centering
\begin{tabular}{c}
(a) $2^{\text{nd}}$ central moment \\
\includegraphics[width=0.925\linewidth]{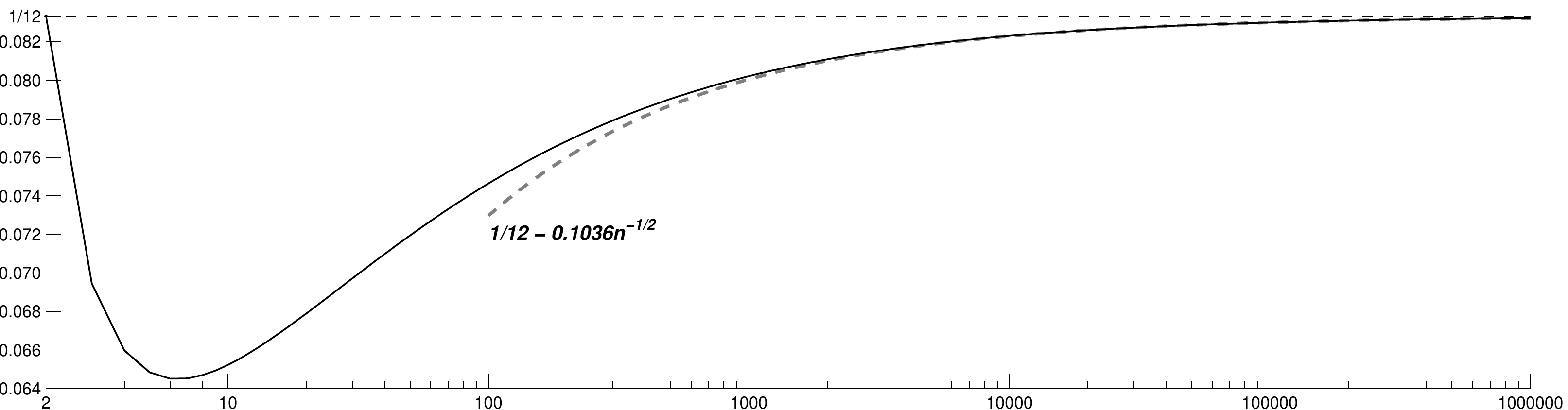} \\
(b) $4^{\text{th}}$ central moment \\
\includegraphics[width=0.925\linewidth]{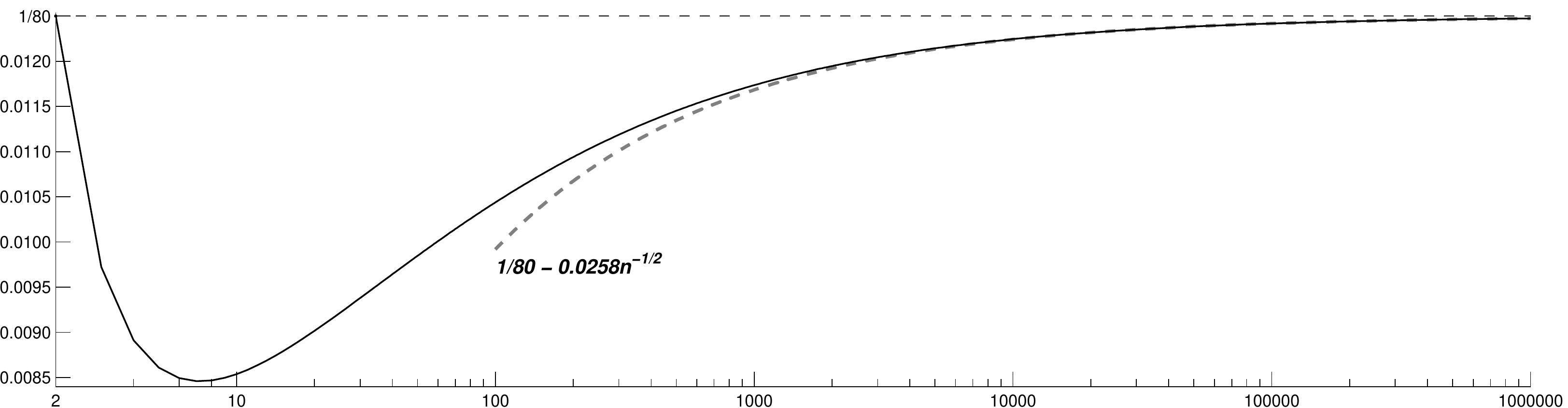} \\
(c) $6^{\text{th}}$ central moment \\
\includegraphics[width=0.925\linewidth]{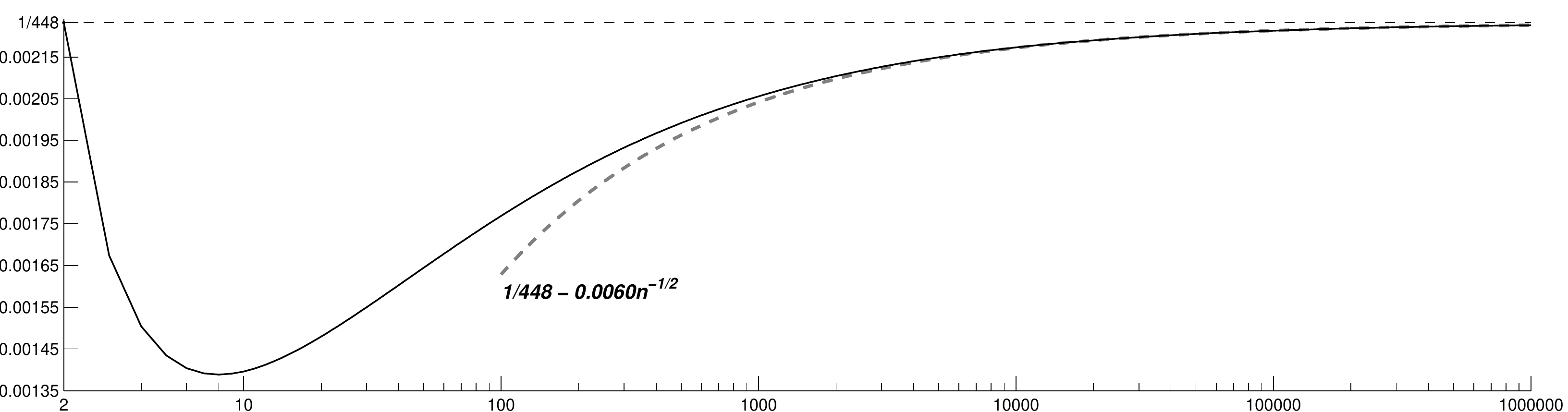} \\
(d) $8^{\text{th}}$ central moment \\
\includegraphics[width=0.925\linewidth]{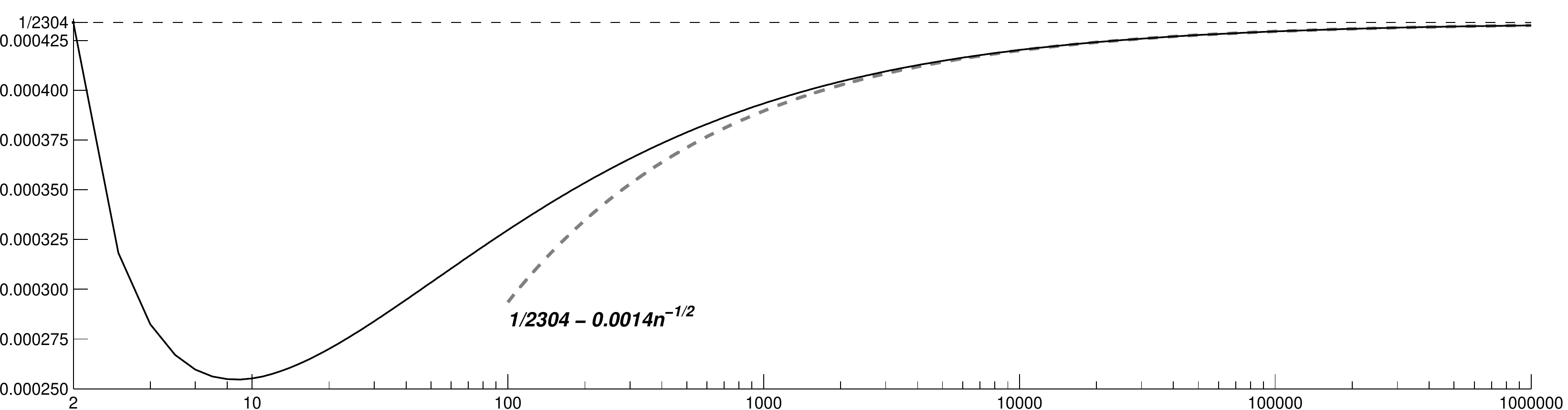} \\
(e) $10^{\text{th}}$ central moment \\
\includegraphics[width=0.925\linewidth]{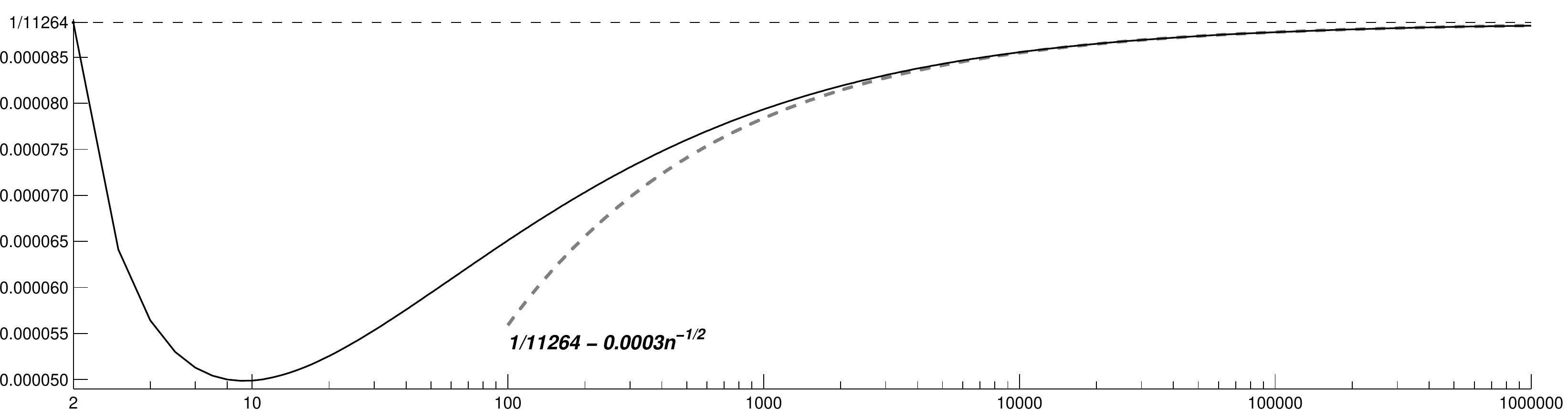} \\
Horizontal axes: sample size, $n$, log-scale. 
\end{tabular}
\begin{flushleft}\footnotesize
Black line---central moments of $U_{i}$; dashed grey line---$a_{j,n}=m_{j,0}-c_{j}n^{-1/2}$, where $m_{j,0}$ are the $j^{th}$ central moments of the uniform distribution, and constants $c_{j}$ are determined numerically.
\end{flushleft}
\vspace*{-\baselineskip}
\caption{Even central moments of $U_{i}$ and asymptotic fit}
\label{Fig:MargUnifDistr:Even.Central.Moments.FIT.2}
\end{figure}

Figure \ref{Fig:MargUnifDistr:Even.Central.Moments.FIT.2} plots the moments against $n$ (on a log-scale). In each case the following regression was estimated on subsamples $n\geq10$, $n\geq100$,  $n\geq1,000$,  $n\geq10,000$:
\begin{equation}
\label{SO:Eq:Moments.REG}
\log_{10}\left(\hat{m}_{2k}-m_{2k,0}\right) = \alpha + \beta\log_{10}(n), \quad k=1,\ldots,5,
\end{equation}
where $\hat{m}_{2k}$ are the tabulated moments of a random variable with density $l_{n}(u)$ and $m_{2k,0}$ are the central moments of a uniform random variable on $[0,1]$. The estimates are given in Table \ref{SO:Tab:Moments.REG}. The constants shown in Figure \ref{Fig:MargUnifDistr:Even.Central.Moments.FIT.2} are determined by restricting $\beta=-1/2$ in \eqref{SO:Eq:Moments.REG} and using the subsample $n\geq10,000$. 

\begin{table}[!h]\centering
\caption{Regression fit for the marginal even central moments of a random variable distributed uniformly on the permutohedron $\Pi_{n}$}
\label{SO:Tab:Moments.REG}
\begin{tabular}{l|rrrr||rrrr}\hline\hline
 & \multicolumn{4}{c||}{Estimates of $\beta$ in \eqref{SO:Eq:Moments.REG}} & \multicolumn{4}{c}{Estimates of $10^{\alpha}$ in \eqref{SO:Eq:Moments.REG}}\\\hline
$2k$& $n\geq10$&  $n\geq100$&  $n\geq1,000$&  $n\geq10,000$  & $n\geq10$&  $n\geq100$&  $n\geq1,000$&  $n\geq10,000$  \\\hline
2  & -0.4512  & -0.4748  & -0.4896  & -0.4944  &  0.0677  &  0.0814  &  0.0927  &  0.0974 \\
4  & -0.4388  & -0.4677  & -0.4868  & -0.4942  &  0.0151  &  0.0189  &  0.0224  &  0.0242 \\
6  & -0.4288  & -0.4619  & -0.4844  & -0.4935  &  0.0032  &  0.0042  &  0.0051  &  0.0056 \\
8  & -0.4204  & -0.4569  & -0.4823  & -0.4927  &  0.0007  &  0.0009  &  0.0012  &  0.0013 \\
10 & -0.4131  & -0.4525  & -0.4804  & -0.4921  &  0.0002  &  0.0002  &  0.0003  &  0.0003 \\
\hline\hline
\end{tabular}
\end{table}

These finite sample results motivate the {\it conjecture} that as $n\to\infty$, the marginal moments of a random variable distributed uniformly on the permutohedron $\Pi_{n}$ approach the respective moments of the uniform random variable on $[0,1]$ {\it from below} at a rate $n^{-1/2}$.

\begin{table}[htbp]\centering
\caption{Central moments of $l_{n}$, $n=2,\ldots,99$}
\label{Tab:PnUnifCmomT1}
\begin{tabular}{r|rrrrr||r|rrrrr}\hline\hline
$n$&$m_{2}$&$m_{4}$&$m_{6}$&$m_{8}$&$m_{10}$ &$n$&$m_{2}$&$m_{4}$&$m_{6}$&$m_{8}$&$m_{10}$\\\hline
2 &8333 &12500 &2232 &4340 &8878 &51 &7202 &9864 &1648 &3041 &5957 \\
3 &6944 &9722 &1674 &3183 &6412 &52 &7211 &9881 &1652 &3048 &5973 \\
4 &6597 &8912 &1504 &2824 &5642 &53 &7219 &9899 &1655 &3056 &5989 \\
5 &6483 &8609 &1434 &2670 &5300 &54 &7226 &9915 &1659 &3063 &6004 \\
6 &6451 &8493 &1404 &2596 &5129 &55 &7234 &9932 &1662 &3070 &6020 \\
7 &6452 &8460 &1391 &2562 &5042 &56 &7242 &9948 &1665 &3077 &6034 \\
8 &6469 &8468 &1388 &2548 &5000 &57 &7249 &9963 &1669 &3084 &6049 \\
9 &6494 &8497 &1391 &2546 &4986 &58 &7256 &9979 &1672 &3091 &6064 \\
10 &6522 &8537 &1396 &2552 &4988 &59 &7263 &9993 &1675 &3097 &6077 \\
11 &6552 &8584 &1403 &2562 &5001 &60 &7270 &10009 &1678 &3104 &6092 \\
12 &6582 &8634 &1411 &2575 &5022 &61 &7276 &10023 &1681 &3111 &6106 \\
13 &6611 &8684 &1419 &2589 &5046 &62 &7283 &10037 &1684 &3117 &6119 \\
14 &6640 &8735 &1428 &2604 &5073 &63 &7289 &10051 &1687 &3123 &6132 \\
15 &6667 &8785 &1437 &2620 &5102 &64 &7296 &10065 &1690 &3129 &6144 \\
16 &6694 &8834 &1446 &2636 &5132 &65 &7302 &10079 &1693 &3135 &6158 \\
17 &6719 &8882 &1454 &2652 &5162 &66 &7308 &10093 &1696 &3141 &6171 \\
18 &6744 &8928 &1463 &2669 &5194 &67 &7314 &10106 &1698 &3147 &6184 \\
19 &6767 &8973 &1471 &2685 &5225 &68 &7320 &10118 &1701 &3153 &6196 \\
20 &6790 &9016 &1480 &2700 &5255 &69 &7326 &10130 &1704 &3158 &6207 \\
21 &6811 &9058 &1487 &2716 &5285 &70 &7331 &10143 &1706 &3164 &6220 \\
22 &6832 &9099 &1495 &2731 &5315 &71 &7337 &10155 &1709 &3169 &6231 \\
23 &6852 &9138 &1503 &2745 &5345 &72 &7342 &10167 &1711 &3174 &6242 \\
24 &6871 &9176 &1510 &2760 &5373 &73 &7348 &10179 &1714 &3180 &6254 \\
25 &6889 &9213 &1517 &2774 &5402 &74 &7353 &10191 &1716 &3185 &6265 \\
26 &6907 &9248 &1524 &2788 &5430 &75 &7358 &10202 &1719 &3190 &6276 \\
27 &6924 &9282 &1531 &2801 &5456 &76 &7363 &10213 &1721 &3195 &6287 \\
28 &6940 &9315 &1537 &2814 &5483 &77 &7368 &10224 &1723 &3200 &6298 \\
29 &6956 &9347 &1544 &2827 &5509 &78 &7373 &10234 &1726 &3205 &6308 \\
30 &6971 &9379 &1550 &2839 &5535 &79 &7378 &10245 &1728 &3210 &6319 \\
31 &6986 &9408 &1556 &2851 &5559 &80 &7383 &10256 &1730 &3214 &6329 \\
32 &7000 &9437 &1562 &2863 &5583 &81 &7387 &10266 &1732 &3219 &6339 \\
33 &7014 &9466 &1567 &2874 &5607 &82 &7392 &10276 &1734 &3223 &6349 \\
34 &7027 &9493 &1573 &2885 &5630 &83 &7396 &10287 &1737 &3228 &6359 \\
35 &7040 &9520 &1578 &2896 &5652 &84 &7401 &10297 &1739 &3233 &6369 \\
36 &7052 &9546 &1583 &2907 &5675 &85 &7406 &10307 &1741 &3237 &6379 \\
37 &7064 &9571 &1588 &2917 &5697 &86 &7410 &10316 &1743 &3242 &6388 \\
38 &7076 &9596 &1593 &2927 &5718 &87 &7414 &10325 &1745 &3246 &6398 \\
39 &7087 &9620 &1598 &2937 &5738 &88 &7418 &10335 &1747 &3250 &6407 \\
40 &7099 &9643 &1603 &2947 &5759 &89 &7422 &10344 &1749 &3255 &6416 \\
41 &7109 &9665 &1607 &2956 &5778 &90 &7426 &10353 &1751 &3259 &6425 \\
42 &7119 &9687 &1612 &2966 &5798 &91 &7430 &10362 &1753 &3263 &6434 \\
43 &7130 &9709 &1616 &2975 &5817 &92 &7435 &10371 &1755 &3267 &6443 \\
44 &7140 &9731 &1621 &2984 &5836 &93 &7438 &10380 &1756 &3271 &6452 \\
45 &7150 &9751 &1625 &2993 &5854 &94 &7442 &10388 &1758 &3275 &6460 \\
46 &7159 &9771 &1629 &3001 &5873 &95 &7446 &10397 &1760 &3279 &6469 \\
47 &7168 &9791 &1633 &3010 &5890 &96 &7450 &10405 &1762 &3283 &6478 \\
48 &7177 &9810 &1637 &3018 &5908 &97 &7453 &10414 &1764 &3286 &6486 \\
49 &7186 &9828 &1641 &3026 &5925 &98 &7457 &10422 &1766 &3290 &6495 \\
50 &7195 &9847 &1645 &3033 &5941 &99 &7460 &10429 &1767 &3294 &6502 \\
\hline\hline
\end{tabular}
\end{table}

\begin{table}[htbp]\centering
\caption{Central moments of $l_{n}$, $n=100,\ldots,1,000$}
\label{Tab:PnUnifCmomT2}
\begin{tabular}{r|rrrrr||r|rrrrr}\hline\hline
$n$&$m_{2}$&$m_{4}$&$m_{6}$&$m_{8}$&$m_{10}$ &$n$&$m_{2}$&$m_{4}$&$m_{6}$&$m_{8}$&$m_{10}$\\\hline
100 &7465 &10438 &1769 &3298 &6510 &320 &7807 &11223 &1940 &3674 &7345 \\
102 &7472 &10454 &1772 &3305 &6527 &327 &7812 &11235 &1943 &3680 &7358 \\
105 &7482 &10476 &1777 &3315 &6549 &335 &7817 &11248 &1946 &3686 &7373 \\
107 &7488 &10491 &1780 &3322 &6564 &343 &7823 &11261 &1949 &3693 &7388 \\
110 &7498 &10513 &1785 &3332 &6586 &351 &7828 &11273 &1951 &3699 &7402 \\
112 &7504 &10527 &1788 &3339 &6600 &359 &7833 &11286 &1954 &3705 &7415 \\
115 &7513 &10547 &1792 &3348 &6621 &368 &7839 &11298 &1957 &3711 &7430 \\
118 &7521 &10567 &1796 &3357 &6641 &376 &7844 &11310 &1960 &3717 &7443 \\
120 &7527 &10579 &1799 &3363 &6654 &385 &7849 &11322 &1962 &3723 &7456 \\
123 &7536 &10598 &1803 &3372 &6674 &394 &7854 &11334 &1965 &3729 &7470 \\
126 &7544 &10617 &1807 &3381 &6693 &404 &7859 &11347 &1968 &3736 &7485 \\
129 &7551 &10634 &1811 &3389 &6710 &413 &7864 &11358 &1970 &3741 &7497 \\
132 &7559 &10651 &1815 &3397 &6729 &423 &7869 &11370 &1973 &3747 &7511 \\
135 &7566 &10667 &1818 &3405 &6745 &433 &7874 &11382 &1976 &3753 &7524 \\
138 &7573 &10684 &1822 &3412 &6762 &443 &7879 &11393 &1978 &3759 &7537 \\
142 &7582 &10704 &1826 &3422 &6784 &453 &7883 &11404 &1981 &3764 &7549 \\
145 &7589 &10719 &1829 &3429 &6799 &464 &7888 &11416 &1983 &3770 &7563 \\
148 &7595 &10734 &1833 &3436 &6815 &475 &7893 &11427 &1986 &3776 &7576 \\
152 &7604 &10753 &1837 &3445 &6834 &486 &7898 &11439 &1988 &3782 &7589 \\
156 &7612 &10771 &1841 &3454 &6855 &498 &7903 &11450 &1991 &3787 &7602 \\
159 &7618 &10785 &1844 &3460 &6868 &509 &7907 &11460 &1993 &3792 &7614 \\
163 &7625 &10802 &1848 &3469 &6887 &521 &7911 &11470 &1996 &3798 &7626 \\
167 &7632 &10819 &1851 &3477 &6904 &534 &7916 &11482 &1998 &3804 &7639 \\
171 &7640 &10835 &1855 &3484 &6922 &546 &7920 &11491 &2000 &3808 &7650 \\
175 &7647 &10852 &1858 &3492 &6939 &559 &7925 &11503 &2003 &3814 &7663 \\
179 &7653 &10866 &1862 &3499 &6955 &572 &7929 &11513 &2005 &3819 &7675 \\
183 &7660 &10881 &1865 &3507 &6971 &586 &7934 &11524 &2008 &3825 &7687 \\
187 &7666 &10896 &1868 &3514 &6987 &599 &7938 &11533 &2010 &3830 &7699 \\
192 &7673 &10913 &1872 &3522 &7005 &614 &7942 &11543 &2012 &3835 &7709 \\
196 &7679 &10927 &1875 &3528 &7020 &628 &7946 &11554 &2014 &3840 &7721 \\
201 &7686 &10942 &1878 &3536 &7037 &643 &7950 &11564 &2017 &3845 &7733 \\
206 &7693 &10959 &1882 &3544 &7055 &658 &7955 &11573 &2019 &3850 &7744 \\
210 &7699 &10971 &1885 &3550 &7068 &673 &7958 &11583 &2021 &3855 &7756 \\
215 &7705 &10987 &1888 &3558 &7085 &689 &7963 &11593 &2023 &3860 &7767 \\
221 &7713 &11004 &1892 &3566 &7104 &705 &7967 &11602 &2025 &3865 &7778 \\
226 &7719 &11018 &1895 &3573 &7119 &722 &7970 &11611 &2027 &3870 &7789 \\
231 &7725 &11032 &1898 &3580 &7134 &739 &7974 &11621 &2030 &3875 &7800 \\
236 &7731 &11045 &1901 &3586 &7148 &756 &7978 &11630 &2032 &3879 &7811 \\
242 &7737 &11060 &1904 &3594 &7165 &774 &7982 &11639 &2034 &3884 &7822 \\
248 &7744 &11075 &1908 &3601 &7181 &792 &7986 &11649 &2036 &3889 &7833 \\
254 &7750 &11090 &1911 &3608 &7197 &811 &7990 &11658 &2038 &3893 &7843 \\
260 &7756 &11104 &1914 &3615 &7213 &830 &7993 &11667 &2040 &3898 &7853 \\
266 &7761 &11117 &1917 &3621 &7227 &850 &7997 &11676 &2042 &3903 &7865 \\
272 &7767 &11130 &1920 &3628 &7242 &870 &8001 &11684 &2044 &3907 &7875 \\
278 &7773 &11143 &1922 &3634 &7256 &890 &8004 &11693 &2046 &3911 &7884 \\
285 &7779 &11158 &1926 &3641 &7272 &911 &8008 &11702 &2048 &3916 &7895 \\
292 &7785 &11172 &1929 &3648 &7288 &933 &8012 &11710 &2050 &3920 &7905 \\
298 &7790 &11183 &1931 &3654 &7301 &955 &8015 &11719 &2052 &3925 &7915 \\
305 &7795 &11196 &1934 &3660 &7315 &977 &8019 &11727 &2054 &3929 &7925 \\
313 &7802 &11211 &1938 &3668 &7332 &1000 &8022 &11735 &2055 &3933 &7934 \\
\hline\hline
\end{tabular}
\end{table}

\begin{table}[htbp]\centering
\caption{Central moments of $l_{n}$, $n=1,000,\ldots,10,000$}
\label{Tab:PnUnifCmomT3}
\begin{tabular}{r|rrrrr||r|rrrrr}\hline\hline
$n$&$m_{2}$&$m_{4}$&$m_{6}$&$m_{8}$&$m_{10}$ &$n$&$m_{2}$&$m_{4}$&$m_{6}$&$m_{8}$&$m_{10}$\\\hline
1000 &8022 &11735 &2055 &3933 &7934 &3199 &8155 &12058 &2129 &4102 &8324 \\
1024 &8025 &11744 &2057 &3938 &7945 &3275 &8157 &12062 &2130 &4105 &8329 \\
1048 &8029 &11751 &2059 &3942 &7954 &3352 &8158 &12067 &2132 &4107 &8335 \\
1072 &8032 &11759 &2061 &3946 &7963 &3430 &8161 &12072 &2133 &4110 &8341 \\
1097 &8035 &11767 &2063 &3950 &7973 &3511 &8163 &12077 &2134 &4113 &8347 \\
1123 &8038 &11775 &2065 &3954 &7982 &3594 &8164 &12082 &2135 &4115 &8353 \\
1150 &8041 &11783 &2066 &3958 &7991 &3678 &8166 &12086 &2136 &4117 &8358 \\
1177 &8045 &11791 &2068 &3962 &8001 &3765 &8168 &12091 &2137 &4120 &8364 \\
1205 &8048 &11798 &2070 &3966 &8010 &3854 &8170 &12096 &2138 &4122 &8370 \\
1233 &8051 &11806 &2072 &3970 &8019 &3944 &8172 &12101 &2139 &4125 &8376 \\
1262 &8054 &11813 &2073 &3974 &8028 &4037 &8174 &12105 &2140 &4127 &8381 \\
1292 &8057 &11821 &2075 &3977 &8036 &4132 &8175 &12109 &2141 &4129 &8387 \\
1322 &8060 &11828 &2077 &3982 &8045 &4229 &8177 &12113 &2142 &4132 &8391 \\
1353 &8063 &11835 &2078 &3985 &8054 &4329 &8179 &12117 &2143 &4134 &8396 \\
1385 &8066 &11843 &2080 &3989 &8063 &4431 &8181 &12122 &2144 &4136 &8402 \\
1417 &8069 &11849 &2082 &3993 &8071 &4535 &8183 &12126 &2145 &4139 &8408 \\
1451 &8072 &11857 &2083 &3997 &8080 &4642 &8184 &12130 &2146 &4141 &8412 \\
1485 &8075 &11864 &2085 &4000 &8088 &4751 &8186 &12134 &2147 &4143 &8417 \\
1520 &8078 &11871 &2086 &4004 &8096 &4863 &8188 &12139 &2148 &4145 &8423 \\
1556 &8081 &11877 &2088 &4007 &8105 &4977 &8189 &12143 &2149 &4147 &8428 \\
1592 &8083 &11884 &2089 &4011 &8112 &5094 &8191 &12147 &2150 &4149 &8433 \\
1630 &8086 &11891 &2091 &4014 &8120 &5214 &8192 &12150 &2151 &4152 &8438 \\
1668 &8089 &11898 &2093 &4018 &8129 &5337 &8194 &12154 &2152 &4153 &8442 \\
1707 &8092 &11904 &2094 &4021 &8136 &5462 &8195 &12158 &2152 &4155 &8447 \\
1748 &8095 &11911 &2096 &4025 &8145 &5591 &8197 &12162 &2153 &4157 &8452 \\
1789 &8097 &11917 &2097 &4028 &8153 &5722 &8199 &12166 &2154 &4160 &8457 \\
1831 &8099 &11923 &2098 &4031 &8159 &5857 &8200 &12170 &2155 &4162 &8461 \\
1874 &8102 &11930 &2100 &4035 &8167 &5995 &8202 &12174 &2156 &4164 &8466 \\
1918 &8105 &11936 &2101 &4038 &8174 &6136 &8203 &12177 &2157 &4166 &8470 \\
1963 &8107 &11942 &2103 &4041 &8183 &6280 &8204 &12180 &2158 &4167 &8474 \\
2009 &8110 &11948 &2104 &4044 &8189 &6428 &8206 &12184 &2159 &4169 &8479 \\
2057 &8112 &11955 &2106 &4048 &8198 &6579 &8207 &12187 &2159 &4171 &8484 \\
2105 &8115 &11960 &2107 &4051 &8205 &6734 &8209 &12191 &2160 &4173 &8488 \\
2154 &8117 &11966 &2108 &4054 &8211 &6893 &8210 &12195 &2161 &4175 &8492 \\
2205 &8120 &11972 &2110 &4057 &8219 &7055 &8212 &12198 &2162 &4177 &8496 \\
2257 &8122 &11978 &2111 &4060 &8225 &7221 &8213 &12201 &2163 &4179 &8501 \\
2310 &8124 &11983 &2112 &4063 &8233 &7391 &8214 &12205 &2163 &4181 &8505 \\
2364 &8126 &11989 &2113 &4066 &8239 &7565 &8216 &12208 &2164 &4182 &8509 \\
2420 &8129 &11995 &2115 &4069 &8246 &7743 &8217 &12211 &2165 &4184 &8513 \\
2477 &8131 &12000 &2116 &4072 &8253 &7925 &8219 &12215 &2166 &4186 &8517 \\
2535 &8133 &12005 &2117 &4075 &8260 &8111 &8220 &12218 &2166 &4188 &8521 \\
2595 &8135 &12011 &2118 &4077 &8265 &8302 &8221 &12221 &2167 &4189 &8525 \\
2656 &8138 &12016 &2120 &4080 &8273 &8498 &8222 &12224 &2168 &4191 &8529 \\
2719 &8140 &12022 &2121 &4083 &8279 &8697 &8223 &12226 &2168 &4192 &8532 \\
2783 &8142 &12028 &2122 &4086 &8287 &8902 &8225 &12230 &2169 &4194 &8536 \\
2848 &8144 &12032 &2124 &4089 &8292 &9112 &8226 &12233 &2170 &4196 &8540 \\
2915 &8146 &12038 &2125 &4092 &8299 &9326 &8227 &12236 &2171 &4197 &8544 \\
2984 &8149 &12043 &2126 &4094 &8305 &9545 &8228 &12239 &2171 &4199 &8547 \\
3054 &8151 &12048 &2127 &4097 &8311 &9770 &8230 &12243 &2172 &4201 &8552 \\
3126 &8153 &12053 &2128 &4100 &8318 &10000 &8231 &12246 &2173 &4203 &8556 \\
\hline\hline
\end{tabular}
\end{table}

\begin{table}[htbp]\centering
\caption{Central moments of $l_{n}$, $n=12,589,\ldots,1,000,000$}
\label{Tab:PnUnifCmomT4}
\begin{tabular}{l|rrrrr||l|rrrrr}\hline\hline
$n$&$m_{2}$&$m_{4}$&$m_{6}$&$m_{8}$&$m_{10}$ &$n$&$m_{2}$&$m_{4}$&$m_{6}$&$m_{8}$&$m_{10}$\\\hline
12589 &8242 &12273 &2179 &4217 &8589 &63096 &8292 &12397 &2208 &4284 &8746 \\
15849 &8252 &12297 &2185 &4230 &8620 &79433 &8296 &12408 &2211 &4290 &8761 \\
19953 &8260 &12318 &2190 &4241 &8647 &100000 &8300 &12418 &2213 &4296 &8773 \\
25119 &8268 &12338 &2194 &4252 &8672 &177828 &8308 &12438 &2218 &4307 &8799 \\
31623 &8275 &12355 &2198 &4261 &8693 &316228 &8315 &12454 &2221 &4315 &8819 \\
39811 &8281 &12371 &2202 &4270 &8713 &562341 &8319 &12465 &2224 &4321 &8834 \\
50119 &8287 &12385 &2205 &4278 &8731 &1000000 &8323 &12474 &2226 &4326 &8844 \\
\hline\hline
\end{tabular}
\end{table}

\renewcommand{\arraystretch}{0.5} 
\renewcommand{\arraystretch}{1} 

\section[~Additional Monte Carlo results]{Additional Monte Carlo results}

Tables \ref{MC:Tab:SIMiMaxEntBwD1M10KSumm}--\ref{MC:Tab:SIMiMaxEntBwD6M10KSumm} list the $0$, $0.025$, $0.25$, $0.5$, $0.75$, $0.975$, and $1^{st}$ quantiles (denoted as min, $2.5^{th}$-p., $1^{st}$ Q., median, $3^{rd}$ Q., $97.5^{th}$-p., and max, respectively), as well as the sample average and standard deviation of the simulated distribution of selected bandwidth estimators for sample sizes $10$, $100$, and $1,000$. The estimators are: minimum Neyman Smooth Test (min-Neyman Sm.) estimators using the first $k=2$, $3$, or $4$ polynomials; the moment-based estimators (MEE) based on the first $m=2$, $3$, and $4$ moments; and the minimum weighted Cram\'{e}r-von Mises estimators with the weight function $\psi(t)=t^{\alpha-1}(1-t)^{\alpha-1}\I{\epsilon\leq t\leq 1-\epsilon}$ for $\alpha=0$ (the Anderson-Darling criterion), $1/4$, and $1/2$, and $\epsilon= 0.001$. 

\begin{table}[htbp]\centering
\caption{Performance of iMaxEnt bandwidth estimators: Density \#1 (Gaussian)}
\label{MC:Tab:SIMiMaxEntBwD1M10KSumm}
\begin{tabular}{l|*{9}{r}}\hline\hline
 & \multicolumn{3}{c}{min-Neyman Sm.} & \multicolumn{3}{c}{MEE (CUE)} & \multicolumn{3}{c}{min-CvM}\\\cline{2-10}
&$k=2$ &$k=3$ &$k=4$ &$m=2$ &$m=3$ &$m=4$ &$\alpha=0$ &$\alpha=1/4$ &$\alpha=1/2$ \\ \hline\hline
\multicolumn{10}{l}{$n=10$; min-MISE bandwidths: 0.6495 (KCDFE), 0.7585 (KDE).}\\[1mm]\hline
min&0.0969 &0.0000 &0.0000 &0.1591 &0.1329 &0.0000 &0.0609 &0.0001 &0.0000 \\
$2.5^{th}$-p.&0.4605 &0.2117 &0.2155 &0.4988 &0.4074 &0.1090 &0.4247 &0.3368 &0.1985 \\
$1^{st}$ Q.&0.7580 &0.6359 &0.5319 &0.7734 &0.6914 &0.4785 &0.8161 &0.7582 &0.6949 \\
median&0.9374 &0.8828 &0.7915 &0.9350 &0.8601 &0.7193 &1.0091 &0.9674 &0.9262 \\
$3^{rd}$ Q.&1.1260 &1.1140 &1.0366 &1.1009 &1.0369 &0.9777 &1.1982 &1.1657 &1.1433 \\
$97.5^{th}$-p.&1.5279 &1.5389 &1.4991 &1.4542 &1.4050 &1.5845 &1.5854 &1.5696 &1.5631 \\
max&1.9337 &2.0041 &2.3252 &1.8648 &1.8544 &2.5339 &2.0393 &2.0875 &2.0204 \\
avg. &0.9510 &0.8745 &0.8005 &0.9454 &0.8712 &0.7456 &1.0090 &0.9623 &0.9105 \\
st.dev. &0.2726 &0.3460 &0.3456 &0.2449 &0.2556 &0.3765 &0.2891 &0.3083 &0.3432 \\\hline
\multicolumn{10}{l}{$n=100$; min-MISE bandwidths: 0.3147 (KCDFE), 0.4455 (KDE).}\\[1mm]\hline
min&0.3319 &0.2294 &0.0180 &0.3380 &0.3379 &0.0620 &0.0325 &0.0000 &0.0000 \\
$2.5^{th}$-p.&0.4084 &0.3678 &0.3226 &0.4165 &0.4108 &0.3354 &0.3600 &0.3272 &0.2734 \\
$1^{st}$ Q.&0.4628 &0.4436 &0.4291 &0.4668 &0.4626 &0.4442 &0.4386 &0.4268 &0.4168 \\
median&0.4931 &0.4776 &0.4687 &0.4949 &0.4911 &0.4805 &0.4712 &0.4629 &0.4562 \\
$3^{rd}$ Q.&0.5231 &0.5104 &0.5060 &0.5231 &0.5191 &0.5154 &0.5008 &0.4961 &0.4922 \\
$97.5^{th}$-p.&0.5832 &0.5734 &0.5735 &0.5771 &0.5738 &0.5819 &0.5576 &0.5556 &0.5565 \\
max&0.6654 &0.6577 &0.6822 &0.6666 &0.6627 &0.7046 &0.6366 &0.6416 &0.6468 \\
avg. &0.4936 &0.4759 &0.4635 &0.4955 &0.4911 &0.4756 &0.4677 &0.4579 &0.4475 \\
st.dev. &0.0442 &0.0514 &0.0642 &0.0411 &0.0417 &0.0620 &0.0512 &0.0595 &0.0734 \\\hline
\multicolumn{10}{l}{$n=1000$; min-MISE bandwidths: 0.1517 (KCDFE), 0.2723 (KDE).}\\[1mm]\hline
min&0.2447 &0.2400 &0.2271 &0.2448 &0.2447 &0.2311 &0.2246 &0.2092 &0.1939 \\
$2.5^{th}$-p.&0.2566 &0.2541 &0.2522 &0.2569 &0.2567 &0.2554 &0.2456 &0.2391 &0.2329 \\
$1^{st}$ Q.&0.2667 &0.2647 &0.2678 &0.2667 &0.2664 &0.2695 &0.2588 &0.2546 &0.2507 \\
median&0.2717 &0.2703 &0.2740 &0.2716 &0.2714 &0.2753 &0.2648 &0.2617 &0.2586 \\
$3^{rd}$ Q.&0.2768 &0.2756 &0.2798 &0.2766 &0.2763 &0.2808 &0.2705 &0.2680 &0.2655 \\
$97.5^{th}$-p.&0.2867 &0.2857 &0.2903 &0.2861 &0.2859 &0.2909 &0.2804 &0.2790 &0.2771 \\
max&0.3004 &0.3000 &0.3023 &0.2995 &0.2994 &0.3024 &0.2944 &0.2938 &0.2936 \\
avg. &0.2717 &0.2702 &0.2734 &0.2716 &0.2713 &0.2748 &0.2644 &0.2609 &0.2576 \\
st.dev. &0.0076 &0.0080 &0.0095 &0.0074 &0.0074 &0.0089 &0.0089 &0.0102 &0.0115 \\\hline
\hline
\end{tabular}
\end{table}

\begin{table}[htbp]\centering
\caption{Performance of iMaxEnt bandwidth estimators: Density \#2 (Skewed unimodal)}
\label{MC:Tab:SIMiMaxEntBwD2M10KSumm}
\begin{tabular}{l|*{9}{r}}\hline\hline
 & \multicolumn{3}{c}{min-Neyman Sm.} & \multicolumn{3}{c}{MEE (CUE)} & \multicolumn{3}{c}{min-CvM}\\\cline{2-10}
&$k=2$ &$k=3$ &$k=4$ &$m=2$ &$m=3$ &$m=4$ &$\alpha=0$ &$\alpha=1/4$ &$\alpha=1/2$ \\ \hline\hline
\multicolumn{10}{l}{$n=10$; min-MISE bandwidths: 0.5896 (KCDFE), 0.6602 (KDE).}\\[1mm]\hline
min&0.1763 &0.0000 &0.0000 &0.2014 &0.0772 &0.0000 &0.0455 &0.0000 &0.0000 \\
$2.5^{th}$-p.&0.4025 &0.1658 &0.1935 &0.4388 &0.3590 &0.1052 &0.3667 &0.2980 &0.1737 \\
$1^{st}$ Q.&0.6860 &0.5250 &0.4822 &0.7083 &0.6171 &0.4557 &0.7062 &0.6472 &0.5849 \\
median&0.8636 &0.7536 &0.7095 &0.8726 &0.7772 &0.6834 &0.9145 &0.8610 &0.8100 \\
$3^{rd}$ Q.&1.0614 &0.9872 &0.9573 &1.0604 &0.9610 &0.9375 &1.1354 &1.0772 &1.0333 \\
$97.5^{th}$-p.&1.5016 &1.4683 &1.4505 &1.4538 &1.3567 &1.5366 &1.5682 &1.5241 &1.4975 \\
max&2.2357 &2.2825 &2.6780 &2.0919 &1.9552 &2.6282 &2.4228 &2.2569 &2.2912 \\
avg. &0.8872 &0.7650 &0.7373 &0.8942 &0.8004 &0.7158 &0.9301 &0.8725 &0.8136 \\
st.dev. &0.2820 &0.3373 &0.3384 &0.2606 &0.2550 &0.3670 &0.3085 &0.3141 &0.3356 \\\hline
\multicolumn{10}{l}{$n=100$; min-MISE bandwidths: 0.2764 (KCDFE), 0.3743 (KDE).}\\[1mm]\hline
min&0.2923 &0.1658 &0.0000 &0.3073 &0.2814 &0.0200 &0.0000 &0.0000 &0.0000 \\
$2.5^{th}$-p.&0.3635 &0.2659 &0.2446 &0.3717 &0.3559 &0.2637 &0.2845 &0.2529 &0.2045 \\
$1^{st}$ Q.&0.4170 &0.3503 &0.3493 &0.4229 &0.4059 &0.3919 &0.3671 &0.3545 &0.3433 \\
median&0.4465 &0.3913 &0.3955 &0.4513 &0.4339 &0.4307 &0.4034 &0.3949 &0.3874 \\
$3^{rd}$ Q.&0.4774 &0.4298 &0.4372 &0.4804 &0.4619 &0.4655 &0.4373 &0.4306 &0.4251 \\
$97.5^{th}$-p.&0.5385 &0.4963 &0.5085 &0.5379 &0.5182 &0.5312 &0.4975 &0.4947 &0.4919 \\
max&0.6364 &0.5868 &0.6008 &0.6222 &0.6005 &0.6355 &0.5678 &0.5717 &0.5752 \\
avg. &0.4475 &0.3887 &0.3900 &0.4522 &0.4344 &0.4237 &0.4000 &0.3896 &0.3786 \\
st.dev. &0.0444 &0.0593 &0.0687 &0.0421 &0.0418 &0.0649 &0.0553 &0.0620 &0.0730 \\\hline
\multicolumn{10}{l}{$n=1000$; min-MISE bandwidths: 0.1317 (KCDFE), 0.2257 (KDE).}\\[1mm]\hline
min&0.2154 &0.1649 &0.1638 &0.2161 &0.2089 &0.1788 &0.1727 &0.1638 &0.1388 \\
$2.5^{th}$-p.&0.2303 &0.1882 &0.1949 &0.2310 &0.2233 &0.2241 &0.2000 &0.1941 &0.1882 \\
$1^{st}$ Q.&0.2403 &0.2046 &0.2123 &0.2407 &0.2333 &0.2387 &0.2151 &0.2112 &0.2076 \\
median&0.2454 &0.2130 &0.2211 &0.2457 &0.2384 &0.2446 &0.2218 &0.2185 &0.2155 \\
$3^{rd}$ Q.&0.2505 &0.2212 &0.2292 &0.2507 &0.2435 &0.2504 &0.2281 &0.2253 &0.2229 \\
$97.5^{th}$-p.&0.2604 &0.2357 &0.2444 &0.2603 &0.2530 &0.2609 &0.2398 &0.2376 &0.2359 \\
max&0.2732 &0.2516 &0.2619 &0.2728 &0.2636 &0.2753 &0.2529 &0.2573 &0.2502 \\
avg. &0.2454 &0.2128 &0.2207 &0.2457 &0.2383 &0.2441 &0.2213 &0.2179 &0.2147 \\
st.dev. &0.0077 &0.0122 &0.0125 &0.0075 &0.0076 &0.0093 &0.0100 &0.0110 &0.0120 \\\hline
\hline
\end{tabular}
\end{table}

\begin{table}[htbp]\centering
\caption{Performance of iMaxEnt bandwidth estimators: Density \#3 (Strongly skewed)}
\label{MC:Tab:SIMiMaxEntBwD3M10KSumm}
\begin{tabular}{l|*{9}{r}}\hline\hline
 & \multicolumn{3}{c}{min-Neyman Sm.} & \multicolumn{3}{c}{MEE (CUE)} & \multicolumn{3}{c}{min-CvM}\\\cline{2-10}
&$k=2$ &$k=3$ &$k=4$ &$m=2$ &$m=3$ &$m=4$ &$\alpha=0$ &$\alpha=1/4$ &$\alpha=1/2$ \\ \hline\hline
\multicolumn{10}{l}{$n=10$; min-MISE bandwidths: 0.3496 (KCDFE), 0.2355 (KDE).}\\[1mm]\hline
min&0.0033 &0.0000 &0.0000 &0.0611 &0.0418 &0.0000 &0.0007 &0.0000 &0.0000 \\
$2.5^{th}$-p.&0.2310 &0.0019 &0.0580 &0.2523 &0.1440 &0.0158 &0.0628 &0.0352 &0.0009 \\
$1^{st}$ Q.&0.5177 &0.1025 &0.1507 &0.5313 &0.3299 &0.1542 &0.2035 &0.1576 &0.0986 \\
median&0.7446 &0.2329 &0.2528 &0.7424 &0.4890 &0.2922 &0.3673 &0.3057 &0.2414 \\
$3^{rd}$ Q.&1.0080 &0.4991 &0.4251 &0.9806 &0.7002 &0.4692 &0.7009 &0.5741 &0.4881 \\
$97.5^{th}$-p.&1.5372 &1.2927 &0.9154 &1.4364 &1.1893 &0.8705 &1.3658 &1.2358 &1.0910 \\
max&2.3178 &2.1789 &1.6234 &2.0319 &1.8464 &1.4259 &2.0109 &2.0164 &1.8115 \\
avg. &0.7832 &0.3564 &0.3171 &0.7691 &0.5378 &0.3330 &0.4870 &0.4064 &0.3293 \\
st.dev. &0.3443 &0.3489 &0.2260 &0.3115 &0.2763 &0.2291 &0.3638 &0.3256 &0.2971 \\\hline
\multicolumn{10}{l}{$n=100$; min-MISE bandwidths: 0.0904 (KCDFE), 0.0796 (KDE).}\\[1mm]\hline
min&0.1361 &0.0264 &0.0335 &0.1336 &0.0861 &0.0336 &0.0000 &0.0000 &0.0000 \\
$2.5^{th}$-p.&0.2234 &0.0399 &0.0489 &0.2135 &0.1233 &0.0505 &0.0039 &0.0000 &0.0000 \\
$1^{st}$ Q.&0.2805 &0.0573 &0.0667 &0.2675 &0.1622 &0.0700 &0.0394 &0.0237 &0.0000 \\
median&0.3149 &0.0700 &0.0787 &0.2991 &0.1863 &0.0834 &0.0523 &0.0407 &0.0001 \\
$3^{rd}$ Q.&0.3522 &0.0853 &0.0922 &0.3341 &0.2134 &0.0992 &0.0655 &0.0567 &0.0229 \\
$97.5^{th}$-p.&0.4294 &0.1281 &0.1272 &0.4049 &0.2751 &0.1423 &0.0972 &0.0919 &0.0816 \\
max&0.5524 &0.2304 &0.1936 &0.5244 &0.3800 &0.2147 &0.1697 &0.1673 &0.1607 \\
avg. &0.3178 &0.0733 &0.0810 &0.3020 &0.1896 &0.0866 &0.0527 &0.0408 &0.0138 \\
st.dev. &0.0529 &0.0226 &0.0201 &0.0493 &0.0385 &0.0233 &0.0214 &0.0249 &0.0250 \\\hline
\multicolumn{10}{l}{$n=1000$; min-MISE bandwidths: 0.0338 (KCDFE), 0.0399 (KDE).}\\[1mm]\hline
min&0.1119 &0.0250 &0.0273 &0.1068 &0.0601 &0.0299 &0.0000 &0.0000 &0.0000 \\
$2.5^{th}$-p.&0.1256 &0.0291 &0.0315 &0.1206 &0.0702 &0.0343 &0.0200 &0.0103 &0.0000 \\
$1^{st}$ Q.&0.1352 &0.0322 &0.0346 &0.1297 &0.0765 &0.0378 &0.0245 &0.0208 &0.0000 \\
median&0.1405 &0.0339 &0.0363 &0.1348 &0.0800 &0.0397 &0.0266 &0.0236 &0.0000 \\
$3^{rd}$ Q.&0.1459 &0.0358 &0.0382 &0.1400 &0.0837 &0.0418 &0.0287 &0.0263 &0.0182 \\
$97.5^{th}$-p.&0.1566 &0.0395 &0.0418 &0.1501 &0.0910 &0.0460 &0.0327 &0.0327 &0.0253 \\
max&0.1791 &0.0457 &0.0472 &0.1719 &0.1058 &0.0519 &0.0379 &0.0368 &0.0366 \\
avg. &0.1406 &0.0340 &0.0364 &0.1349 &0.0802 &0.0398 &0.0265 &0.0232 &0.0084 \\
st.dev. &0.0079 &0.0027 &0.0026 &0.0076 &0.0054 &0.0030 &0.0033 &0.0055 &0.0097 \\\hline
\hline
\end{tabular}
\end{table}

\begin{table}[htbp]\centering
\caption{Performance of iMaxEnt bandwidth estimators: Density \#4 (Kurtotic unimodal)}
\label{MC:Tab:SIMiMaxEntBwD4M10KSumm}
\begin{tabular}{l|*{9}{r}}\hline\hline
 & \multicolumn{3}{c}{min-Neyman Sm.} & \multicolumn{3}{c}{MEE (CUE)} & \multicolumn{3}{c}{min-CvM}\\\cline{2-10}
&$k=2$ &$k=3$ &$k=4$ &$m=2$ &$m=3$ &$m=4$ &$\alpha=0$ &$\alpha=1/4$ &$\alpha=1/2$ \\ \hline\hline
\multicolumn{10}{l}{$n=10$; min-MISE bandwidths: 0.4362 (KCDFE), 0.2419 (KDE).}\\[1mm]\hline
min&0.0478 &0.0000 &0.0000 &0.0667 &0.0596 &0.0000 &0.0000 &0.0000 &0.0000 \\
$2.5^{th}$-p.&0.1998 &0.0682 &0.0009 &0.2558 &0.1821 &0.0481 &0.1012 &0.0703 &0.0072 \\
$1^{st}$ Q.&0.5128 &0.2729 &0.1590 &0.5968 &0.5044 &0.3983 &0.3318 &0.2468 &0.1650 \\
median&0.7414 &0.5510 &0.4418 &0.8105 &0.7274 &0.7922 &0.6980 &0.5072 &0.3404 \\
$3^{rd}$ Q.&0.9893 &0.8701 &0.8649 &1.0381 &0.9695 &1.1740 &1.0586 &0.9220 &0.7430 \\
$97.5^{th}$-p.&1.5106 &1.4689 &1.4905 &1.5072 &1.4355 &1.9674 &1.6003 &1.5006 &1.3734 \\
max&2.2317 &2.2405 &2.5601 &2.1840 &2.1644 &3.0406 &2.3337 &2.3306 &2.3325 \\
avg. &0.7682 &0.6061 &0.5437 &0.8288 &0.7468 &0.8270 &0.7248 &0.6051 &0.4786 \\
st.dev. &0.3423 &0.3915 &0.4396 &0.3218 &0.3301 &0.5223 &0.4348 &0.4167 &0.3882 \\\hline
\multicolumn{10}{l}{$n=100$; min-MISE bandwidths: 0.0993 (KCDFE), 0.0958 (KDE).}\\[1mm]\hline
min&0.1334 &0.0945 &0.0000 &0.1494 &0.1493 &0.0000 &0.0000 &0.0000 &0.0000 \\
$2.5^{th}$-p.&0.2009 &0.1715 &0.0000 &0.2233 &0.2199 &0.0001 &0.0001 &0.0000 &0.0000 \\
$1^{st}$ Q.&0.2698 &0.2479 &0.0035 &0.2976 &0.2941 &0.0138 &0.0678 &0.0646 &0.0614 \\
median&0.3113 &0.2943 &0.0091 &0.3394 &0.3362 &0.0271 &0.0868 &0.0829 &0.0797 \\
$3^{rd}$ Q.&0.3576 &0.3434 &0.0209 &0.3854 &0.3821 &0.0436 &0.1077 &0.1016 &0.0979 \\
$97.5^{th}$-p.&0.4499 &0.4377 &0.0738 &0.4726 &0.4703 &0.5694 &0.1914 &0.1654 &0.1524 \\
max&0.5760 &0.5754 &0.6515 &0.5884 &0.5876 &{\it 0.8544}$(^{*}$&0.5693 &0.5473 &0.3918 \\
avg. &0.3150 &0.2970 &0.0177 &0.3417 &0.3384 &{\it 0.0555}$(^{*}$ &0.0901 &0.0837 &0.0787 \\
st.dev. &0.0634 &0.0688 &0.0350 &0.0636 &0.0640 &{\it 0.1209}$(^{*}$&0.0455 &0.0394 &0.0369 \\\hline
\multicolumn{10}{l}{$n=1000$; min-MISE bandwidths: 0.0407 (KCDFE), 0.0539 (KDE).}\\[1mm]\hline
min&0.1071 &0.1070 &0.0000 &0.1100 &0.1100 &0.0014 &0.0000 &0.0000 &0.0000 \\
$2.5^{th}$-p.&0.1223 &0.1210 &0.0003 &0.1255 &0.1254 &0.0020 &0.0340 &0.0326 &0.0315 \\
$1^{st}$ Q.&0.1338 &0.1324 &0.0008 &0.1373 &0.1371 &0.0024 &0.0424 &0.0413 &0.0406 \\
median&0.1400 &0.1389 &0.0010 &0.1436 &0.1434 &0.0026 &0.0458 &0.0447 &0.0439 \\
$3^{rd}$ Q.&0.1466 &0.1455 &0.0013 &0.1503 &0.1502 &0.0029 &0.0490 &0.0478 &0.0471 \\
$97.5^{th}$-p.&0.1598 &0.1590 &0.0022 &0.1637 &0.1635 &0.0036 &0.0550 &0.0536 &0.0527 \\
max&0.1821 &0.1761 &0.0044 &0.1864 &0.1855 &0.0066 &0.0626 &0.0609 &0.0597 \\
avg. &0.1403 &0.1391 &0.0011 &0.1439 &0.1437 &0.0027 &0.0455 &0.0443 &0.0435 \\
st.dev. &0.0095 &0.0096 &0.0005 &0.0097 &0.0097 &0.0004 &0.0056 &0.0056 &0.0057 \\\hline
\hline
\end{tabular}
\begin{flushleft}
$(^{*}$-- without one extreme value of $\hat{b}$ ($1.62e7$).
\end{flushleft}
\end{table}

\begin{table}[htbp]\centering
\caption{Performance of iMaxEnt bandwidth estimators: Density \#5 (Outlier)}
\label{MC:Tab:SIMiMaxEntBwD5M10KSumm}
\begin{tabular}{l|*{9}{r}}\hline\hline
 & \multicolumn{3}{c}{min-Neyman Sm.} & \multicolumn{3}{c}{MEE (CUE)} & \multicolumn{3}{c}{min-CvM}\\\cline{2-10}
&$k=2$ &$k=3$ &$k=4$ &$m=2$ &$m=3$ &$m=4$ &$\alpha=0$ &$\alpha=1/4$ &$\alpha=1/2$ \\ \hline\hline
\multicolumn{10}{l}{$n=10$; min-MISE bandwidths: 0.2250 (KCDFE), 0.2457 (KDE).}\\[1mm]\hline
min&0.0640 &0.0000 &0.0000 &0.0615 &0.0396 &0.0000 &0.0035 &0.0001 &0.0000 \\
$2.5^{th}$-p.&0.1577 &0.0686 &0.0606 &0.1718 &0.1429 &0.0315 &0.1208 &0.1003 &0.0652 \\
$1^{st}$ Q.&0.2637 &0.2063 &0.1808 &0.2766 &0.2412 &0.1637 &0.2550 &0.2362 &0.2169 \\
median&0.3331 &0.2889 &0.2664 &0.3517 &0.3094 &0.2515 &0.3202 &0.3056 &0.2937 \\
$3^{rd}$ Q.&0.4203 &0.3708 &0.3506 &0.4562 &0.3944 &0.3530 &0.3957 &0.3829 &0.3699 \\
$97.5^{th}$-p.&0.7102 &0.5808 &0.5321 &0.9540 &0.8014 &0.7970 &0.6231 &0.5659 &0.5309 \\
max&2.1502 &1.9930 &1.8215 &2.9251 &3.2072 &6.3076 &2.6660 &1.5242 &0.9372 \\
avg. &0.3591 &0.2956 &0.2713 &0.3993 &0.3461 &0.3044 &0.3356 &0.3134 &0.2942 \\
st.dev. &0.1559 &0.1362 &0.1250 &0.2111 &0.1920 &0.3517 &0.1374 &0.1188 &0.1172 \\\hline
\multicolumn{10}{l}{$n=100$; min-MISE bandwidths: 0.1042 (KCDFE), 0.1418 (KDE).}\\[1mm]\hline
min&0.1103 &0.0831 &0.0018 &0.1129 &0.1116 &0.0105 &0.0000 &0.0000 &0.0000 \\
$2.5^{th}$-p.&0.1385 &0.1256 &0.0749 &0.1433 &0.1414 &0.0715 &0.1108 &0.1008 &0.0894 \\
$1^{st}$ Q.&0.1588 &0.1518 &0.1385 &0.1637 &0.1619 &0.1317 &0.1412 &0.1374 &0.1340 \\
median&0.1700 &0.1641 &0.1569 &0.1753 &0.1734 &0.1531 &0.1534 &0.1506 &0.1481 \\
$3^{rd}$ Q.&0.1822 &0.1772 &0.1717 &0.1878 &0.1862 &0.1693 &0.1652 &0.1629 &0.1609 \\
$97.5^{th}$-p.&0.2068 &0.2025 &0.1985 &0.2135 &0.2119 &0.1974 &0.1874 &0.1857 &0.1844 \\
max&0.2482 &0.2483 &0.2381 &0.2602 &0.2588 &0.2295 &0.2204 &0.2190 &0.2186 \\
avg. &0.1709 &0.1642 &0.1524 &0.1762 &0.1745 &0.1480 &0.1524 &0.1489 &0.1457 \\
st.dev. &0.0174 &0.0195 &0.0298 &0.0179 &0.0180 &0.0311 &0.0196 &0.0216 &0.0239 \\\hline
\multicolumn{10}{l}{$n=1000$; min-MISE bandwidths: 0.0496 (KCDFE), 0.0863 (KDE).}\\[1mm]\hline
min&0.0825 &0.0816 &0.0492 &0.0832 &0.0832 &0.0484 &0.0657 &0.0573 &0.0500 \\
$2.5^{th}$-p.&0.0885 &0.0877 &0.0728 &0.0892 &0.0891 &0.0700 &0.0782 &0.0759 &0.0740 \\
$1^{st}$ Q.&0.0924 &0.0917 &0.0843 &0.0930 &0.0929 &0.0819 &0.0834 &0.0820 &0.0807 \\
median&0.0944 &0.0938 &0.0894 &0.0950 &0.0949 &0.0872 &0.0859 &0.0847 &0.0835 \\
$3^{rd}$ Q.&0.0964 &0.0959 &0.0935 &0.0971 &0.0970 &0.0917 &0.0882 &0.0870 &0.0861 \\
$97.5^{th}$-p.&0.1004 &0.0999 &0.0996 &0.1011 &0.1010 &0.0985 &0.0922 &0.0913 &0.0906 \\
max&0.1055 &0.1053 &0.1057 &0.1063 &0.1062 &0.1047 &0.0975 &0.0962 &0.0960 \\
avg. &0.0944 &0.0938 &0.0885 &0.0951 &0.0950 &0.0864 &0.0857 &0.0844 &0.0832 \\
st.dev. &0.0030 &0.0031 &0.0070 &0.0030 &0.0030 &0.0074 &0.0036 &0.0039 &0.0042 \\\hline
\hline
\end{tabular}
\end{table}

\begin{table}[htbp]\centering
\caption{Performance of iMaxEnt bandwidth estimators: Density \#6 (Bimodal)}
\label{MC:Tab:SIMiMaxEntBwD6M10KSumm}
\begin{tabular}{l|*{9}{r}}\hline\hline
 & \multicolumn{3}{c}{min-Neyman Sm.} & \multicolumn{3}{c}{MEE (CUE)} & \multicolumn{3}{c}{min-CvM}\\\cline{2-10}
&$k=2$ &$k=3$ &$k=4$ &$m=2$ &$m=3$ &$m=4$ &$\alpha=0$ &$\alpha=1/4$ &$\alpha=1/2$ \\ \hline\hline
\multicolumn{10}{l}{$n=10$; min-MISE bandwidths: 0.6762 (KCDFE), 0.7477 (KDE).}\\[1mm]\hline
min&0.1650 &0.0000 &0.0000 &0.2074 &0.1339 &0.0000 &0.0299 &0.0135 &0.0000 \\
$2.5^{th}$-p.&0.5363 &0.1650 &0.1791 &0.5699 &0.4130 &0.0605 &0.3304 &0.2530 &0.0503 \\
$1^{st}$ Q.&0.8782 &0.6794 &0.4048 &0.8565 &0.7416 &0.3118 &0.8568 &0.7935 &0.7134 \\
median&1.0421 &1.0104 &0.6422 &0.9936 &0.9079 &0.5619 &1.0449 &1.0253 &1.0051 \\
$3^{rd}$ Q.&1.2075 &1.2247 &0.8900 &1.1333 &1.0648 &0.7828 &1.2094 &1.2073 &1.2034 \\
$97.5^{th}$-p.&1.5158 &1.5642 &1.3381 &1.3894 &1.3447 &1.2107 &1.5034 &1.5182 &1.5239 \\
max&1.9962 &2.0315 &2.2831 &1.7784 &1.7643 &1.9298 &2.0276 &2.0825 &2.0183 \\
avg. &1.0383 &0.9370 &0.6683 &0.9908 &0.8988 &0.5669 &1.0066 &0.9724 &0.9256 \\
st.dev. &0.2478 &0.3888 &0.3192 &0.2075 &0.2379 &0.3102 &0.2966 &0.3338 &0.3877 \\\hline
\multicolumn{10}{l}{$n=100$; min-MISE bandwidths: 0.2825 (KCDFE), 0.3207 (KDE).}\\[1mm]\hline
min&0.4091 &0.1820 &0.1426 &0.4101 &0.3491 &0.1374 &0.0000 &0.0000 &0.0000 \\
$2.5^{th}$-p.&0.4845 &0.3212 &0.2002 &0.4707 &0.4468 &0.1963 &0.2129 &0.1394 &0.0000 \\
$1^{st}$ Q.&0.5302 &0.4869 &0.2529 &0.5100 &0.4994 &0.2518 &0.3051 &0.2754 &0.2185 \\
median&0.5532 &0.5291 &0.2875 &0.5298 &0.5221 &0.2883 &0.3578 &0.3419 &0.3129 \\
$3^{rd}$ Q.&0.5753 &0.5598 &0.3288 &0.5496 &0.5436 &0.3320 &0.4084 &0.4110 &0.4142 \\
$97.5^{th}$-p.&0.6154 &0.6070 &0.4316 &0.5865 &0.5817 &0.4318 &0.4857 &0.4987 &0.5129 \\
max&0.6771 &0.6744 &0.5466 &0.6388 &0.6360 &0.5404 &0.5651 &0.5944 &0.5889 \\
avg. &0.5523 &0.5137 &0.2948 &0.5296 &0.5202 &0.2959 &0.3556 &0.3387 &0.2989 \\
st.dev. &0.0335 &0.0700 &0.0585 &0.0295 &0.0345 &0.0606 &0.0728 &0.0955 &0.1444 \\\hline
\multicolumn{10}{l}{$n=1000$; min-MISE bandwidths: 0.1270 (KCDFE), 0.1838 (KDE).}\\[1mm]\hline
min&0.2695 &0.2138 &0.1229 &0.2641 &0.2552 &0.1230 &0.1321 &0.0848 &0.0000 \\
$2.5^{th}$-p.&0.2815 &0.2660 &0.1388 &0.2760 &0.2745 &0.1387 &0.1632 &0.1509 &0.1318 \\
$1^{st}$ Q.&0.2886 &0.2836 &0.1495 &0.2828 &0.2819 &0.1494 &0.1774 &0.1685 &0.1558 \\
median&0.2923 &0.2891 &0.1554 &0.2865 &0.2857 &0.1553 &0.1848 &0.1776 &0.1670 \\
$3^{rd}$ Q.&0.2958 &0.2935 &0.1615 &0.2899 &0.2893 &0.1614 &0.1921 &0.1869 &0.1784 \\
$97.5^{th}$-p.&0.3026 &0.3010 &0.1747 &0.2965 &0.2960 &0.1744 &0.2063 &0.2046 &0.2004 \\
max&0.3109 &0.3107 &0.1955 &0.3043 &0.3042 &0.1957 &0.2299 &0.2341 &0.2294 \\
avg. &0.2922 &0.2877 &0.1557 &0.2863 &0.2855 &0.1556 &0.1847 &0.1777 &0.1669 \\
st.dev. &0.0054 &0.0088 &0.0091 &0.0052 &0.0055 &0.0091 &0.0110 &0.0137 &0.0176 \\\hline
\hline
\end{tabular}
\end{table}

\end{document}